\documentclass[prx,twocolumn,showpacs,superscriptaddress,preprintnumbers,amssymb]{revtex4-2}
\usepackage{graphicx}
\usepackage{float}
\usepackage{color}
\usepackage{dcolumn}
\usepackage{bm}
\usepackage{physics}
\usepackage{comment}
\usepackage{svg}
\usepackage{placeins}
\usepackage{hyperref}
\usepackage{soul}
\usepackage{feynmf}
\usepackage{tikz}
\usepackage[normalem]{ulem} 

\usetikzlibrary{decorations.pathmorphing}
\usetikzlibrary{arrows.meta,decorations.markings}

\newcommand{\beq}{\begin{equation}}
\newcommand{\eeq}{\end{equation}}
\newcommand{\beqn}{\begin{eqnarray}}
\newcommand{\eeqn}{\end{eqnarray}}

\newcommand{\p}{\partial}

\newcommand{\sgn}{{\rm sgn}}

\newcommand{\cL}{ {\cal L} }

\newcommand{\cZ}{ {\cal Z} }
\newcommand{\cT}{ {\cal T} }

\newcommand{\abhipurp}[1]{\textcolor{violet}{[(AS) #1]}}

\newcommand{\ZZ}{\mathbb{Z}}

\newcommand{\sfigref}[2]{Fig.~\hyperref[#1]{\ref{#1}#2}}

\begin{document}

\title{Field Theory of the Gutzwiller Projection}

\author{Abhijat Sarma}

\affiliation{Department of Physics, University of California,
Santa Barbara, CA 93106, USA}

\affiliation{Perimeter Institute for Theoretical Physics, Waterloo, Ontario, N2L 2Y5, Canada}

\author{Cenke Xu}

\affiliation{Department of Physics, University of California,
Santa Barbara, CA 93106, USA}

\author{Chong Wang}

\affiliation{Perimeter Institute for Theoretical Physics, Waterloo, Ontario, N2L 2Y5, Canada}

\begin{abstract}
The Gutzwiller projection produces a well-known ansatz for spin-liquid and quantum Hall ground-state wavefunctions that is amenable to numerical simulation, with the Laughlin wavefunction being the most celebrated example. A long-standing issue is the precise relation between the Gutzwiller projection and the closely related parton gauge theory. Here, we clarify this issue by providing a field-theoretic description of the Gutzwiller projection. In a path integral description, the universal physics of the projected state is described in terms of a gauge field supported on a temporal slab of vanishing width $\epsilon \rightarrow 0$ coupled to the bulk matter fields. We analytically derive the universal physics of various Gutzwiller projected states. Our construction is also applicable to discrete symmetries, and we use it to analyze various projected SPT states. Among our results, we show that the Gutzwiller projection is not a good ansatz for the long-distance properties of $U(1)$ Dirac spin liquids, spinon Fermi surface spin liquids or composite Fermi liquids, and that the Gutzwiller projection of a quantum spin Hall state has only quasi-long range order. In contrast, in examples where the gauge field is gapped, including Laughlin states, $\mathbb{Z}_2$ spin liquids and Luttinger liquids, we show that the slab gauge theory from Gutzwiller projection reproduces the correct universal physics of the full parton gauge theory. We also present a generalization of our field theory that describes $G$-intertwining defects, with application to self-duality defects of recent theoretical interest.

\end{abstract}

\maketitle

\tableofcontents

\section{Introduction}

The Gutzwiller projection is a celebrated ansatz, inspired by the \emph{parton} formalism, for wavefunctions of strongly correlated systems including quantum spin liquids and quantum Hall effects. Spin-liquids and quantum Hall systems have no local order, often featuring emergent excitations fundamentally different in character than the microscopic spin degrees of freedom, and are described in the low-energy (IR) limit by \emph{emergent gauge theories} \cite{Savary_2016, Broholm_2020, RevModPhys.78.17, wen2004quantum, PhysRevB.37.3774, hermele_2005, PhysRevLett.59.2095, PhysRevB.44.2664, hermele2008, Ran_2006, PhysRevB.39.11413, PhysRevLett.54.966, Bauer_2014}. For example, certain spin-liquids can be described in terms of emergent fermionic excitations $c = (c_{\uparrow}, c_{\downarrow})$ known as partons, with the microscopic spins given in terms of the partons as $\vec{S}_j = c^\dagger_j \vec{\sigma} c_j$. Such a parton decomposition necessarily introduces a gauge redundancy which is the source of the emergent dynamical gauge field \cite{Wen_2002}. In this case the gauge redundancy is $U(1)$ corresponding to $c_j \rightarrow c_j e^{i \theta_j}$. In other words, the partons have additional ``charge" degrees of freedom which are gauged away to recover the physical spin Hilbert space corresponding to one parton per site. Similarly, in quantum Hall systems the familiar notions of ``composite bosons''~\cite{ZhangHanssonKivelson1989,Read1989CompositeBoson} and ``composite fermions''~\cite{Jain1989Parton,HalperinLeeRead1993} can also be viewed as partons interacting with dynamical (typically $U(1)$) gauge fields.

Directly simulating interacting electrons is difficult in general. The Gutzwiller projection is an ansatz for ground state wavefunctions that is numerically amenable and designed to capture the universal physics of the parton gauge theory \cite{PhysRevLett.10.159, Becca_Sorella_2017, Paramekanti_2007, PhysRevB.38.931, SHIBA1987264, Ran_2007, Iqbal_2013, PhysRevB.88.060402}. In the spin liquid example, the Gutzwiller projected ansatz is $|\psi_{\text{gutz}} \rangle =\hat{P}_{n_{i,\uparrow} + n_{i,\downarrow}=1} |\psi_{0} \rangle$, where $|\psi_{0}\rangle$ is some free fermion state describing the mean-field configuration of the partons and $\hat{P}_{n_{i,\uparrow} + n_{i,\downarrow}=1}$ is the Gutzwiller projection that recovers the correct spin $1/2$ Hilbert space with one parton per site. In other words, instead of coupling the $U(1)$ symmetry to a dynamical gauge field, we just project to a fixed $U(1)$ charge sector on each site. 

The Gutzwiller projection is an infinite strength perturbation on $|\psi_{0} \rangle$, making it unclear how to analyze the projected state analytically. It was historically believed that the Gutzwiller projection should reproduce the universal physics of the parton gauge theory. This belief is based on the conceptual similarity between the two theories, as well as several successful examples including the Luttinger liquid, $\mathbb{Z}_2$ spin liquids and the celebrated Laughlin wavefunction. However, we shall argue that in general the Gutzwiller projection and the parton gauge theory describe different universal physics, especially when the underlying gauge field is gapless. The fundamental issue is that in the parton gauge theory the gauge field coupling to the partons lives in the entire spacetime, while the Gutzwiller projection is a temporal defect.

We shall argue that the Gutzwiller projection \emph{can} be accurately described by a dynamical gauge field, but this gauge field only lives on a thin \emph{temporal slab} $|\tau| \leq \epsilon\to 0^+$. This prescription is amenable to analytical treatment, allowing us to make predictions on the universal physics of the projected state. We apply our slab gauge theory to various parton mean-field states, showing that the results are in alignment with exact arguments and numerics when available. In many cases, this slab gauge theory is qualitatively different than the corresponding bulk gauge theory, explaining tensions between previous numerical results and bulk gauge theory predictions. Some well-studied such examples, mainly in two space dimensions, include the composite Fermi liquid, the spinon Fermi surface and the $U(1)$ Dirac spin liquids. 

At a more conceptual level, a variational wavefunction tries to minimize the energy density, which is typically a local operator. We can therefore expect a good trial wavefunction to correctly capture the short-distance physics (the local density matrix). However, it shall not be surprising when a trial wavefunction, even if energetically very compelling, does not produce the correct long-distance physics. A very familiar example is the BCS wavefunction for a neutral superfluid, which on its own does not capture the long-distance power-law correlation from the Goldstone mode. To capture such long-distance physics, one needs to go beyond the trial wavefunction and deploy complementary methods such as field theory. From this perspective, gapped phases are more amenable to variational approach since the short-distance behaviors already encode the universal physics.

We note that our slab field theory works regardless of the particular gauge group: we can pick any state $|\psi_0\rangle$ with global symmetry $G$ and define $|\psi_\text{gutz}\rangle = \hat{P} |\psi_0\rangle$ where $\hat{P}$ projects to a fixed $G$ charge sector on each site, even when $G$ is a discrete symmetry. We apply this to study projected $G \times K$ SPT states, deriving stat-mech models descending from the topological fixed-point action of the SPT \cite{Chen_2013}. 

In addition to being a useful ansatz for spin-liquid ground states, it has recently been observed that the Gutzwiller projection is intimately related to the quantum information theoretic problems of \emph{measurement} and \emph{decoherence}. Several works have made clear the connection between these problems and temporal defects \cite{Garratt2023,Weinstein2023,Lee_2023, Su_2024, Sarma_2025, yang2023, hauser2026}. In particular, the Gutzwiller projection can be interpreted as measuring the local $G$ charge and postselecting onto the same outcome on each site, a problem relevant to measurement-assisted state preparation studied in various works \cite{lu2022, verresen2022, lee2022, yang2023, Tantivasadakarn_2024, lu2026}. On the other hand, the infinite-time limit of charge dephasing maps to a kind of Gutzwiller projection in the doubled space (Choi) representation of the density matrix \cite{Su_2025, Sarma_2025, lu2026, sarma2026inequalitystrongweakspontaneoussymmetry, wang2026, sarma2026strongtoweakspontaneoussymmetrybreaking}. Therefore Renyi-2 observables of the fully dephased density matrix can be understood using our formalism as well. We can also consider perturbing away from the hard projection limit to understand the late but finite time behavior of the dephased density matrix. Using this perspective we are able to derive the Renyi-2 behavior of fractional quantum hall states under dephasing with only universal field theory arguments. 

The remainder of this paper is organized as follows. In Sec.~\ref{sec:eft}, we derive the slab gauge theory of the Gutzwiller projection and discuss its symmetries and their relation to confinement and deconfinement. Section \ref{sec:examples} applies this framework to a range of projected states, distinguishing examples that reproduce the long-distance predictions of full parton gauge theories from those that do not. In Sec.~\ref{sec:defect_fusion}, we formulate the projection in terms of defect fusion and extend the construction to $G$-intertwining defects, including self-duality defects. We conclude in Sec.~\ref{sec:conclusions} with open questions and future directions. Appendices \ref{app:dirac_cone} and \ref{app:proj_fermi} present the perturbative calculations for projected Dirac fermions and Fermi surfaces, respectively.

\section{Slab Gauge Theory}
\label{sec:eft}

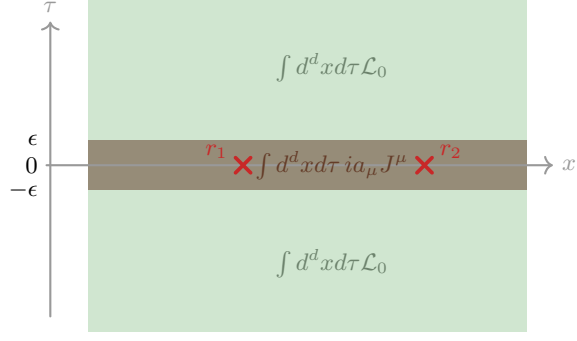
\begin{figure*}
\begin{tikzpicture}[scale=1.0, line cap=round, line join=round]

  \definecolor{slabgreen}{RGB}{208,230,208}
  \definecolor{stripgray}{RGB}{150,140,120}
  \definecolor{myred}{RGB}{200,40,40}
  \definecolor{mydarkred}{RGB}{100,60,40}
  \definecolor{mydarkgreen}{RGB}{108,130,108}

  \fill[slabgreen] (1.2,-2.2) rectangle (7.0,2.2);
  \fill[stripgray]  (1.2,-0.33) rectangle (7.0,0.33);

  \draw[gray!80, line width=0.8pt, ->] (0.7,-2.0) -- (0.7,1.9)
    node[above] {$\tau$};
  \draw[gray!80, line width=0.8pt, ->] (0.65,0) -- (7.35,0)
    node[right] {$x$};

  \node[left] at (0.65,0) {$0$};
  \node[left] at (0.65,0.33) {$\epsilon$};
  \node[left] at (0.65,-0.33) {$-\epsilon$};

  \node[mydarkred] at (4.425,0.) {$\int d^d x d\tau \,ia_\mu J^\mu$};
  \node[mydarkgreen] at (4.425,1.3) {$\int d^dx d\tau \cL_{0}$};
  \node[mydarkgreen] at (4.425,-1.3) {$\int d^dx d\tau \cL_{0}$};

  \node[myred] at (2.9,0.18) {$r_1$};
  \node[myred] at (6,0.18) {$r_2$};

  \draw[myred, line width=1.4pt] (3.15,-0.10) -- (3.35,0.10);
  \draw[myred, line width=1.4pt] (3.15, 0.10) -- (3.35,-0.10);

  \draw[myred, line width=1.4pt] (5.55,-0.10) -- (5.75,0.10);
  \draw[myred, line width=1.4pt] (5.55, 0.10) -- (5.75,-0.10);


\end{tikzpicture}
\caption{Spacetime Path Integral of the Gutzwiller Projection (figure inspired by Ref.~\cite{yang2023}). The green regions on top and bottom prepare $\langle \psi_{0}|$ and $|\psi_{0}\rangle$ respectively. Operator insertions along $\tau=0$ are denoted by a red x. We couple $J^\mu$ to a dynamical gauge field in the brown slab. The slab width $\epsilon$ is infinitesimal.}
\end{figure*}

First we define the Gutzwiller projection more generally. We begin with the theory defined on a lattice with sites labeled by $j$. Let $|\psi_{0}\rangle$ have an anomaly free (onsite) global symmetry $G$. Here we assume that $G=U(1)$, though the extension to nonabelian $G$ is straightforward. We assume translational symmetry, and without loss of generality we normal order the charge such that $|\psi_{0}\rangle$ has total charge $\hat{Q} = \sum_j \hat{Q}_j = 0$; therefore $\langle \hat{Q}_j \rangle = 0$. The effect of the Gutzwiller projection is to promote this to an operator equation. We let $\hat{P}$ denote the Gutzwiller projection, and $\hat{O}^P=\hat{P} \hat{O} \hat{P}$ denote an operator projected to the Gutzwiller Hilbert space. The Gutzwiller projection's defining characteristics are 

\begin{equation}
\label{eqn:gutz_proj_def}
    \hat{P} \hat{Q}_j = \hat{Q}_j \hat{P} = 0,\; \hat{P}^2=\hat{P}
\end{equation}
which obviously implies that $\hat{Q}_j^P = 0$. 

Suppose $\hat{O}$ is charged under a local $U(1)$ rotation (a gauge transformation), $[\hat{Q}_j, \hat{O}] = q_{O} \hat{O}$ with $q_O \neq 0$. Then $\hat{P} [\hat{Q}_j, \hat{O}] \hat{P} = q_O \hat{O}^P = [\hat{Q}_j^P, \hat{O}^P] = 0$, therefore $\hat{O}^P=0$. In other words, the Gutzwiller projection kills all operators that are not gauge-invariant. This strongly suggests the interpretation of the Gutzwiller projection as a gauge theory. Indeed we shall show how to make this connection precise.

The Gutzwiller projected state on the lattice can be represented as follows:
\begin{align}
\begin{split}
\label{eq:gutz}
&\ket{\psi_{\mathrm{gutz}}} = \lim_{g\rightarrow \infty} e^{-g \sum_j \hat{Q}_j^2} \ket{\psi_{0}}, \\
&\ket{\psi_{0}} = \int D\phi e^{- \int_{-\infty}^0 d\tau L_{0}[\phi]} \ket{\phi(\tau=0)}
\end{split}
\end{align}
where $L_{0}$ is the parton mean-field Lagrangian. The path-integral preparing the mean-field state is schematic, and should be interpreted as e.g. a coherent state path integral that may be fermionic or bosonic. This formula makes it clear that the projection is fundamentally a temporal defect, acting in the vicinity of $\tau=0$, and \textit{not} in the entire spacetime. Denoting $\hat{H}$ the mean-field Hamiltonian, a convenient regularization of the partition function $\cZ = \langle \psi_{\text{gutz}} | \psi_{\text{gutz}} \rangle$ of the Gutzwiller projected state is given by

\begin{align}
\begin{split}
\label{eq:trans}
    &\cZ = \lim_{\beta \rightarrow \infty}\text{Tr} [e^{-\beta H/2} \hat{P} e^{-\beta H/2}] \\
    \langle \psi_{\text{gutz}}|\hat{O}|\psi_{\text{gutz}}\rangle&=\langle \hat{O}(\tau=0) \rangle = \frac{1}{\cZ} \text{Tr}[e^{-\beta H/2} \hat{P} \hat{O} \hat{P} e^{-\beta H/2}]
\end{split}
\end{align}

Now we investigate the projection more closely. We perform a Hubbard-Stratonovich decomposition, introducing a Lagrange multiplier field $\theta$ to rewrite 
\begin{equation}
\label{eq:P}
    \hat{P} = \lim_{g \rightarrow \infty} e^{-g\sum_j \hat{Q}_j^2} = \int d\theta e^{-i\sum_j \theta_{j} \hat{Q}_j} = \int d\theta \hat{U}_{\theta}
\end{equation}
where $\hat U_\theta = e^{-i \sum_j \theta_j \hat{Q}_j}$ is interpreted as a gauge transformation and $\int d\theta = \Pi_j \int_{0}^{2\pi}\frac{d\theta_j}{2\pi}$. Thus, $\hat{P}$ can be viewed as summing over all spatial gauge transformations, constructing a manifestly gauge-invariant state. In other words, $\hat{U}_{\theta} \hat{P} = \hat{P}$, so the Gutzwiller wavefunction transforms trivially as $\hat{U}_{\theta} |\psi_{\text{gutz}}\rangle = |\psi_{\text{gutz}}\rangle$. The projected operator becomes
\begin{align}
\begin{split}
\label{eq:pop}
    \hat{P} \hat{O} \hat{P}= \int d\theta_+ d\theta_- \hat{U}_{\theta_+} \hat{O} \hat{U}^\dagger_{\theta_-}
\end{split}
\end{align}
It is convenient to define

\begin{equation}
\label{eq:lattice_gauge}
\bar{\theta} = \epsilon(\theta_{+} + \theta_{-}), \; \bar{a}_\tau = \theta_+ - \theta_-
\end{equation}
In terms of these fields, 
\begin{align}
\begin{split}
\label{eq:trans_gauge}
    \hat{P} \hat{O} \hat{P}\sim \int d\bar{a}_\tau d\bar{\theta} \hat{U}_{\bar{\theta}/2\epsilon+\bar{a}_\tau/2} \hat{O} \hat{U}^\dagger_{\bar{\theta}/2\epsilon-\bar{a}_\tau/2}
\end{split}
\end{align}

The interpretation of Eq.~\ref{eq:trans_gauge} is that we are undergoing time evolution from $\tau=0^-$ to $\tau=0^+$ coupled to a gauge field $\bar{a}_\mu(x)$ with $\bar{a}_j = \partial_j \bar{\theta}$. With the background gauge field set to zero, the correct continuum limit is the following:

\begin{align}
\begin{split}
\label{eq:path_int_gauge_collapsed}
&\cZ \sim \lim_{\epsilon \rightarrow 0}\int D\phi d\bar{a}_{\tau} d\bar\theta e^{- S_{\text{gutz}}} , \; \; \bar a_j = \partial_j \bar\theta, \\
&S_{\text{gutz}} = \int_{-\infty}^\infty d\tau \int d^dx  \mathcal{L}_{0}[\phi] \\&\hspace{0.8cm}+ \frac{1}{2\epsilon}\int_{-\epsilon}^{\epsilon} d\tau \int d^dx \;i\bar a_\mu(x) J^\mu(x,\tau)
\end{split}
\end{align}
The prefactors in Eq.~\ref{eq:lattice_gauge} and Eq.~\ref{eq:path_int_gauge_collapsed} are chosen such that $\bar{a}_\mu$ is $O(1)$ as $\epsilon \rightarrow 0$. We refer to this effective field theory as the ``slab gauge theory". Correlations of operators in the Gutzwiller state, i.e. $\langle \psi_{\text{gutz}}|\hat{O} | \psi_{\text{gutz}}\rangle = \langle \psi_0 | \hat{O}^P | \psi_0 \rangle$, correspond to matter field insertions at $\tau=0$ in the slab field theory. On the other hand, matter field insertions at the temporal boundary $\tau=\epsilon$ are probed by the so-called ``strange correlator"~\cite{You2014StrangeCorrelator} $\langle \psi_0 | \hat{O} | \psi_{\text{gutz}}\rangle = \langle \psi_0 | \hat{O} \hat{P} | \psi_0\rangle$, and likewise the $\tau=-\epsilon$ boundary is probed by the conjugate strange correlator with the bra and ket interchanged. 

It is sometimes helpful to rewrite the gauge field as
\begin{equation}
\label{eq:promote_gauge}
    a_\mu(x, \tau) = A_{\mu}(x,\tau)+\frac{1}{2\epsilon}\bar{a}_\mu(x) [\Theta(\tau+\epsilon)-\Theta(\tau-\epsilon)]
\end{equation}
wherein the interaction takes the standard form $\int d\tau \int d^dx i a_\mu J^\mu$ and we have restored the explicit background gauge field $A_\mu$, set by the mean-field state. We note that it is the flux of $a_\mu$, not $\bar{a}_\mu$, that is subject to Dirac quantization. From Eq.~\ref{eq:lattice_gauge} it follows that $\bar{a}_\tau$ and $\bar{\theta}$ are $d$-dimensional compact bosons, with compactification lattice
\begin{equation}
\label{eq:compact_lattice}
    (\bar{a}_\tau, \bar{\theta}) \sim (\bar{a}_\tau+2\pi, \bar{\theta}+2\pi\epsilon) \sim (\bar{a}_\tau-2\pi, \bar{\theta}+2\pi\epsilon) 
\end{equation}

Notice that in Eq.~\ref{eq:path_int_gauge_collapsed} we have kept an explicit temporal integral over a thin slab even though $\frac{1}{2\epsilon}\int_{-\epsilon}^{\epsilon}d\tau J^\mu(x,\tau) = J^\mu(x,0) + O(\epsilon)$. We keep this form due to certain subtleties regarding point-splitting which are important for properly regulating equal-time correlators. In particular, some care must be taken with the $\epsilon \rightarrow 0$ limit: though $\epsilon$ is small compared to any physical scales, it must still be large compared to the UV cutoff. In other words,
\begin{equation}
    1/\Lambda \ll \epsilon \ll t_{\text{phys}}
\end{equation}
where $\Lambda$ is the UV cutoff in the continuum formulation and $t_{\text{phys}}$ is any physical time scale that we can probe. We can verify directly that the continuum action reproduces Eq.~\ref{eq:trans_gauge} upon rediscretizing time, and it is often convenient to do so. We discretize time such that 
\begin{align}
\begin{split}
\label{eq:discrete_time_J}
    &J^{\tau}(x,\pm \epsilon) = J^{\tau}_{\pm}(x) \\
    &J^\tau(x,0) = (J^\tau_+(x) + J^\tau_-(x))/2\\
    &\partial_\tau J^\tau(x,0) = (J^\tau_+(x) - J^\tau_-(x))/2\epsilon
\end{split}
\end{align}
We can take a gauge transformation $\phi'(x,\tau) = e^{i\lambda(x,\tau)} \phi(x,\tau)$, $a'_\mu(x,\tau) = a_\mu(x,\tau) + \partial_\mu \lambda(x,\tau)$ with parameter $\lambda(x,\tau) = -\bar\theta(x)/2\epsilon \cdot [\Theta(\tau+\epsilon)-\Theta(\tau-\epsilon)]$ and integrate by parts to rewrite the interaction as 
\begin{align}
\begin{split}
&\frac{1}{2\epsilon}\int_{-\epsilon}^{\epsilon}d\tau i\bar{a}_\mu(x) J^\mu(x,\tau)\cong\frac{1}{2\epsilon}\int_{-\epsilon}^{\epsilon} d\tau i (\bar{a}_\tau + \bar{\theta}\partial_\tau) J^\tau  \\
=& i\bar{a}_\tau J^\tau(x,0) + \frac{1}{2\epsilon}i\bar{\theta} (J^\tau(x,\epsilon)-J^\tau(x,-\epsilon)) \\
=& i\theta_+(x) J^\tau_+(x) - i\theta_-(x) J^\tau_-(x)
\end{split}
\label{eq:defectaction}
\end{align}
which is just Eq.~\ref{eq:pop}.

The role of the gauge field in the slab gauge theory is clear: integrating out $\bar{a}_\tau$ enforces $J^\tau(\tau=0) = 0$. Integrating out $\bar{\theta}$ enforces $\partial_j J^j(\tau=0) = 0$, or equivalently $\partial_\tau J^\tau(\tau=0) = 0$. Indeed, these are exactly the two constraints that one expects from the projected operator algebra. By construction, $\hat{P} \hat{Q}_x\hat{P} = 0$. Likewise, $\hat{P} \partial_\tau \hat{Q}_x \hat{P} = -\hat{P} [\hat H, \hat Q_x] \hat{P} = 0$. On the other hand, higher derivatives like $\hat{P} \partial_\tau^2 \hat{Q}_x \hat{P} = \hat{P} [\hat H, [\hat H, \hat Q_x]] \hat P$ need not vanish. Similarly, because $\hat{P} \hat{Q}_x=0$, but $\hat{P} \partial_\tau \hat{Q}_x \neq 0$, we conclude that the boundaries at $\tau = \pm \epsilon$ are Dirichlet boundaries for $J^\tau$, i.e. $J^\tau(\tau=\pm \epsilon)=0$ but $\partial_\tau J^\tau(\tau = \pm \epsilon)$ is unconstrained. 

We now comment on the need to have two projections $\theta_{\pm}$ in our formulation. In principle, if we are only interested in calculating $\langle \hat{O}\rangle$ for some gauge-invariant operator $\hat{O}$, then strictly speaking we only need one projection since $\hat{P}\hat{O}\hat{P}=\hat{P}\hat{O}=\hat{O}\hat{P}$. Removing one projection is equivalent to picking a gauge that sets $\bar{a}_j=0$. However, to make the field theory perturbatively accessible (such as the perturbative calculations done in Appendices~\ref{app:dirac_cone} and \ref{app:proj_fermi}), we may  also want gauge-non-invariant quantities, such as the parton fermion Green's function $\langle c_i c_j^{\dagger}\rangle$, to satisfy desirable properties. Our regularization scheme, with two projections at $\tau=\pm\epsilon$, leads to consistent perturbative calculations, as demonstrated in Appendix~\ref{app:dirac_cone}.

When treating the theory perturbatively, we must deal with the fact that there is no bare kinetic term for the gauge field in the slab gauge theory. Instead the gauge fluctuations are controlled by the self-energy resulting from interactions with matter fields, 
\begin{equation}
\label{eq:self_energy}
    \bar \Pi^{\mu\nu}(k) = -\frac{1}{(2\epsilon)^2} \int_{-\epsilon}^{\epsilon} d\tau_1 d\tau_2 \langle J^\mu(-k,\tau_1) J^\nu(k, \tau_2)\rangle
\end{equation}
The $\bar{a}_\mu$ propagator is 
\begin{align}
\begin{split}
\label{eq:gauge_prop}
    \bar{\Delta}^{\tau\tau}(k) = -\frac{1}{\bar{\Pi}^{\tau\tau}(k)}, \; \;\bar{\Delta}^{ij}(k) = -\frac{1}{\bar\Pi_L(k)} P_L^{ij}(k)
\end{split}
\end{align}
where $P_L^{ij} = \frac{k_i k_j}{k^2}$ and $P_T^{ij} = \delta^{ij} - \frac{k_i k_j}{k^2}$ are longitudinal and transverse projectors and $\bar\Pi_L$ is the longitudinal part of $\bar\Pi^{ij}$. One can derive this form explicitly by enforcing $\bar{f}_{ij}=0$ by adding a Maxwell term $\frac{1}{4e^2}\bar{f}_{ij} \bar{f}^{ij}$ to the defect Lagrangian and taking $e=0$. This is locally equivalent to setting $\bar{a}_j = \partial_j \bar{\theta}$. 

We now briefly comment on how to generalize the slab gauge theory to the case of nonabelian $G$. Eq.~\ref{eq:pop} becomes $\hat{P}\hat{O}\hat{P} = \int dg_+ dg_- \hat{U}_{g_+} \hat{O} \hat{U}_{g_-}^\dagger$ where $dg$ is the Haar measure on $G$ and $\hat{U}_g$ is the $G$ representation carried by the matter fields. Defining $W(x) = g_+(x) g_-^{-1}(x)$ and $h(x) = g_-(x)$, $\hat{P}\hat{O}\hat{P} = \int dW dh \hat{U}_{Wh} \hat{O} \hat{U}_{h}^\dagger$. We can absorb $h$ into a spatial gauge field $\bar{a}_i = -2\epsilon ih^{-1} \partial_i h$ and locally pick $\bar{a}_\tau$ to satisfy $W(x) = e^{i \bar{a}_\tau(x)}$. Then the nonabelian slab gauge theory is given by $\cZ = \int D\phi dW dh e^{-S_{\text{gutz}}}$ with $S_{\text{gutz}}$ given as in Eq.~\ref{eq:path_int_gauge_collapsed}, but with $\bar{a}_\mu$ and $J^\mu$ understood as a nonabelian gauge field and current respectively. 

\subsection{Symmetry, symmetry breaking and (de)confinement}

We note that the action in Eq.~\eqref{eq:defectaction} is not manifestly gauge invariant in the sense of having a local redundancy in the entire spacetime. The theory is gauge-invariant only in the weaker sense that correlation functions of gauge-dependent observables at $\tau=0$ are vanishing. However, the slab gauge theory does possess several new symmetries. Firstly, the original $U(1)$ symmetry of the bulk gives rise to two $U(1)$ symmetries of the slab gauge theory. The first is a uniform rotation on the entire spacetime, implemented as 
\begin{align}
\begin{split}
\label{eq:sym_uni}
\phi'(x,\tau) = e^{i\alpha}\phi(x,\tau), \; \bar{\theta}'(x) = \bar{\theta}(x)+2\epsilon\alpha, \; \bar{a}_\tau'(x) = \bar{a}_\tau(x)
\end{split}
\end{align}
The second is an opposite rotation, call it $U(1)_o$ on each half of spacetime, implemented as 
\begin{align}
\begin{split}
\label{eq:sym_rel}
&\phi'(x,\tau) = e^{i\alpha h_{\epsilon}(\tau)}\phi(x,\tau), \; \bar{\theta}'(x) = \bar{\theta}(x), \; \bar{a}_\tau'(x) = \bar{a}_\tau(x)+2\alpha \\
&h_\epsilon(\tau) = -1+(\tau/\epsilon+1)\Theta(\tau+\epsilon) - (\tau/\epsilon-1)\Theta(\tau-\epsilon)
\end{split}
\end{align}
where $h_{\epsilon}(\tau)$ interpolates between $-1$ for $\tau<-\epsilon$ and $+1$ for $\tau>\epsilon$. 

The traditional parton gauge theory is most useful when the IR fluctuation of the gauge field is controlled in some sense. In the simplest scenario, the gauge theory is free in the IR and the theory is fully controlled --  this is the case for topologically ordered phases in $d\geq2$ as well as $U(1)$ Coulomb spin liquids in $d\geq3$. In our slab gauge theory picture, the free gauge theory phase is described by weakly fluctuating $\bar{a}_{\tau}$, with the corresponding symmetries (e.g. the opposite $U(1)_o$ described above) spontaneously broken -- this is the remnant of the familiar spontaneous breaking of $1$-form symmetries in free gauge theories. In the following we explain this correspondence in more detail.

First we consider the $U(1)_o$ or $\mathbb{Z}_{2,o}$ symmetry as a topological defect operator in the path integral. Recall that the symmetry operator is a co-dimension one operator in the path integral. But since our symmetry acts nontrivially only on the $\tau\to 0$ time slice, the corresponding symmetry defect operator lives on a co-dimension one manifold in space rather than spacetime. The effect of these symmetry defect operators is to create domain walls, across which $\bar{a}_\tau\to\bar{a}_\tau+\Delta$ for some angle $\Delta$ in either $U(1)$ or $\mathbb{Z}_2$. In the wavefunction language, they should be viewed as operators acting on $(d-1)$ manifolds, and can be viewed as $1$-form symmetries. 

We now examine operators charged under the symmetries. Consider the following ``electric'' vertex operators
\begin{equation}
W_{q, q'}(x) = e^{iq \bar{a}_\tau(x) + iq' \bar{\theta}(x)/\epsilon}
\end{equation}
This operator is well-defined when $(q,q') \in \mathbb{Z}^2$ or $(q,q') \in (\mathbb{Z}+\frac{1}{2})^2$. Microscopically this operator corresponds to modifying the projection $\hat{P} \hat{O} \hat{P}$ to $\hat{P}_x^{(q+q')} \hat{O} \hat{P}_x^{(q-q')}$, where $\hat{Q}_y \hat{P}^{(q)}_x = \hat{P}^{(q)}_x\hat{Q}_y = q \hat{P}^{(q)}_x \delta_{x,y}$. In other words, $\hat{P}_x^{(q)}$ projects to the charge $q$ sector on site $x$ and the charge neutral sector on every other site. From Eqs.~\ref{eq:sym_uni}-~\ref{eq:sym_rel} one sees that $W_{q, q'}$ carries charges $2q'$ and $2q$ under the uniform and relative $U(1)$ symmetries respectively.

In a standard gauge theory, an unbroken $1$-form symmetry leads to confinement of gauge charge, while symmetry breaking leads to deconfinement. This picture also holds in the Gutzwiller theory, as we now explain. The breaking of the symmetry is probed by the two-point correlation function $\langle W_{1,0}(x)W^{\dagger}_{1,0}(y)\rangle$. In the wavefunction language, this quantity is nothing but the normalization of the wavefunction with modified charge projection ($Q_x=1, Q_y=-1$). One can think of this as a trial excited state, with a gauge charge moved from $x$ to $y$. If the symmetry is unbroken and $\langle W_{1,0}(x)W^{\dagger}_{1,0}(y)\rangle\sim e^{-|x-y|/\xi}$, then for large $|x-y|$ such a trial excited state does not exist. In contrast, if the symmetry is broken, $\langle W_{1,0}(x)W^{\dagger}_{1,0}(y)\rangle\sim O(1)$, then the trial excited state does exist. This is the remnant of confinement vs. deconfinement at the level of wavefunctions. As a trivial example, if the parton mean field state is strictly a product state (an atomic insulator) with exactly one fermion per site, then the Gutzwiller projected state is also a product state. When formulated in the path integral approach, it is easy to see that $\bar{a}_\tau$ must be in the disordered phase and $\langle W_{1,0}(x)W^{\dagger}_{1,0}(y)=0$ for $x\neq y$, signaling charge confinement, in agreement with the above analysis.

For continuous gauge groups, the Goldstone mode on the $\tau\to0$ time slice is the remnant of the gauge boson, viewed as the Goldstone mode of the broken $1$-form symmetry.

We can also consider dual symmetries in free gauge theory. In the weakly fluctuating slab $U(1)$ gauge theory, for $d\geq 2$ there are two more symmetries associated to the gauge field, namely the vortex conservation of $\theta_+$ and $\theta_-$ with currents $J_{\pm}(x) = \frac{1}{2\pi} *_d d\theta_{\pm} = \frac{1}{4\pi}*_{d} (d\bar{\theta}/\epsilon \pm d\bar{a}_\tau)$. These are \textit{defect} $d-2$ \textit{form symmetries}, with $*_d$ denoting the $d$-dimensional Hodge star. It is useful to change basis so that the symmetries are diagonal in $\bar{a}_\tau$, $\bar{\theta}$. The first symmetry we denote $U(1)_{\text{flux}}$, with current
\begin{equation}
J_{\text{flux}}(x)=\frac{1}{4\pi \epsilon} *_{d} d\bar{\theta} = \frac{1}{4\pi \epsilon} *_d \bar{a}
\end{equation}
The reasoning for this terminology is that the charged object of this symmetry is the vortex of $\bar{\theta}$, which is nothing more than a flux tube passing through the slab. This is obvious from the symmetry operator $V_\alpha(C) = e^{i \alpha \int_C *_d J_{\text{flux}}} = e^{\frac{i\alpha}{4\pi\epsilon}\int_C \bar{a}_j dx^j}$. The second symmetry can actually be promoted to a \textit{bulk} $d$-2 form symmetry, denoted $U(1)_{\text{mon}}$. Taking $a_\mu(x,\tau)$ as Eq.~\ref{eq:promote_gauge} with $f = da$, the current is
\begin{equation}
    J_{\text{mon}}(x,\tau) = \frac{1}{2\pi}*_{d+1}f
\end{equation}
Again the terminology is chosen because the charged operators are monopoles. In the original $\theta_{\pm}$ basis, a monopole can be viewed as unequal vortices of $\theta_+$ and $\theta_-$, i.e.\ vortices of $\bar{a}_\tau$. 


We can also consider the particle-vortex duals of $\bar{a}_\tau$, $\bar{\theta}$, which we denote $\bar{b}_\tau$, $\bar{\vartheta}$, which are $d-2$ form gauge fields \footnote{When $d=2$ the dual fields are also compact bosons.}. The ``magnetic" vertex operators are
\begin{equation}
M_{m, m'}(\Sigma_{d-2}) = e^{i\int_{\Sigma_{d-2}} 2m\bar{b}_\tau + 2m'\epsilon \bar\vartheta}
\end{equation}
which insert vortices of $\bar{a}_\tau$ and $\bar{\theta}$, characterized by $\int_{S^1} d\bar{a}_\tau = 4\pi m $ and $\int_{S^1} d\bar{\theta} = 4\pi m' \epsilon$ on any $S^1$ linking $\Sigma_{d-2}$. Equivalently, the operator inserts $2\pi (m'+m)$ and $2\pi(m'-m)$ vortices of $\theta_+$ and $\theta_-$, respectively. Once again the operator is well-defined when $(m, m') \in \mathbb{Z}^2$ or $(m, m') \in (\mathbb{Z}+\frac{1}{2})^2$. Based on the preceding discussion $M_{m, m'}$ carries charge $2m$ under $U(1)_{\text{mon}}$ and charge $m'$ under $U(1)_{\text{flux}}$.

\section{Examples}
\label{sec:examples}

We will now use the slab field theory to derive the universal properties of various Gutzwiller projected states. All of our results, summarized in Table~\ref{tab:results}, are consistent with previous numerical studies.

\begin{table*}[t]
\centering
\renewcommand{\arraystretch}{1.4}
\begin{tabular}{ c|c|c}
\hline\hline
\textbf{Unprojected State} & \textbf{Projected Symmetry} & \textbf{IR Theory} \\
\hline

$N$ Flavor Luttinger Liquid
& $U(1)$
& Spin Sector \\

Fractional Quantum Spin Hall 
& $U(1)_s$
& Compact Boson CFT \\

Exciton Bose Liquid Mean-Field
& $U(1)_s$
& Mean-Field Charge Sector\\

Fermi Liquid (Generic)
& $U(1)$
& Fermi Liquid \\

Fermi Liquid (Nested)
& $U(1)/SU(2)$
& Antiferromagnet / Superconductor \\

Dirac Fermions ($d\geq2$)
& $U(1)$
& Dirac Fermions \\

$N$ Copies of Chern Insulator 
& $U(1)/SU(N)$
& $U(1)_N/SU(N)_1$ Chern-Simons Theory\\

BCS Superconductor 
& $U(1)$ 
& $\mathbb{Z}_2$ Gauge Theory \\

$\mathbb{Z}_2 \times \mathbb{Z}_2$ Cluster State ($d=1$) 
& $\mathbb{Z}_2$ 
& $\mathbb{Z}_2$ SSB \\

$\mathbb{Z}_2 \times \mathbb{Z}_2 \times \mathbb{Z}_2$ Cluster State ($d=2$)
& $\mathbb{Z}_2$
& $O(2)$ Loop Model \\
\hline\hline

\end{tabular}
\caption{Survey of Results. The second column denotes the symmetry $G$ that is Gutzwiller projected. The Luttinger liquid and Exciton Bose Liquid (EBL) mean field are (quasi)-1d systems exhibiting spin-charge separation, and the projection kills the projected sector while leaving the other untouched. The projected fractional quantum spin hall (fQSH) state is a 2D CFT, exhibiting quasi-long range superconductivity. Projecting the charge symmetry is irrelevant for generic fermi liquids and Dirac fermions in the sense that it does not modify the correlation functions of charge neutral observables in the IR. However, if there is nesting in the Fermi surface, then the projection will lead to a symmetry broken state. Projected ansatze for topologically ordered spin liquids are generically accurate. In some cases the IR behavior of the projected theory depends on non-universal (microscopic) details of the pre-projection state, such as the Luttinger parameter of the compact boson descending from the fQSH state and the loop tension of the $O(2)$ model descending from the $d=2$ cluster state.}
\label{tab:results}
\end{table*}

\subsection{Examples consistent with full gauge theory}

We first discuss examples where the Gutzwiller projected wavefunction can produce the correct universal physics predicted by the full parton theory. Intuitively, in such examples the gauge field is gapped, so restricting it to the $\tau=0$ slab does not qualitatively change the universal behaviors.

\subsubsection{Luttinger Liquids (1+1D Compact Boson CFT)}
\label{sec:1+1D}

To begin with, we consider Gutzwiller projection of free fermions in one spatial dimension. The mean-field state $\ket{\psi_{0}}$ is just the ground state of a 1D fermion nearest-neighbor tight binding model, with $N$ flavors of fermions at half filling (assuming $N$ even as in Ref.~\cite{Paramekanti_2007}). Interactions in the mean-field state can easily be incorporated, though we consider the noninteracting case for simplicity. Through non-abelian bosonization, this free-fermion state is described by the $U(N)_1 \sim U(1)_N \times SU(N)_1$ WZW model \footnote{To be precise, the $U(1)_N$ and $SU(N)_1$ WZW models separately have operators with fractional conformal spin $s=h-\bar{h}$. The physical operators of the theory are obtained by gluing $U(1)_N$ and $SU(N)_1$ fields such that the composite has half-integer spin $s\in \frac{1}{2}\mathbb{Z}$, in accordance with a local fermionic theory.}, which has central charge $c=N$. The first factor is the $U(1)_N$ WZW model describing the charge fluctuations, i.e. a compact boson CFT with action $S_0=\int dxdy \frac{K}{4\pi} (\partial_\mu \phi_c)^2$ and Luttinger parameter $K = 1/2N$. The second factor is the $SU(N)_1$ WZW model describing the spin fluctuations. We denote the charge conservation as $U(1)_c$, whose current is the vortex current $J^\mu = \frac{1}{2\pi}\epsilon ^{\mu \nu}\partial_\nu\phi_c$. 

We can exactly integrate out $\bar a_\tau$ which enforces $\partial_x \phi_c|_{\tau=0}=0$. On the other hand integration of $\bar{\theta}$ enforces $\partial_x \partial_\tau \phi_c|_{\tau=0} = 0$, or equivalently $\partial_x^2 \theta_c|_{\tau=0} = 0$ where $\theta_c$ is the dual boson to $\phi_c$. Thus neither $\phi_c$ nor $\theta_c$ can fluctuate at $\tau=0$, and the charge sector is completely gapped. We can interpret the slab field theory (of the charge sector) as two Dirichlet boundaries \footnote{More precisely, each boundary is a direct sum $\bigoplus_{\alpha} \text{Dir}(\alpha)$ of Dirichlet boundaries with $\phi_c=\alpha$.} at $\tau=\pm \epsilon$ with the vacuum sandwiched in between. This is an instance of what is known as a ``factorizing defect" in the literature \cite{popov2025}, so named because they effectively split spacetime in half as no excitation can be transmitted through them. On the Dirichlet boundaries the spatial current $J^x$ has boundary scaling dimension $\Delta_{J^x}=1$ as predicted by boundary conformal field theory, which can be verified via the strange correlator. 

Due to spin-charge separation in 1+1D the spin sector is untouched, and thus the Gutzwiller projected state will just display $SU(N)_1$ scaling of $U(1)_c$ neutral fermion bilinears, as found in Ref.~\cite{Paramekanti_2007}. It is also well-known that fully gauging the $U(1)_c$ symmetry in the entire spacetime produces the $SU(N)_1$ CFT. So in this example the Gutzwiller projection does agree with the full gauge theory, at least for the local observables. In Appendix \ref{app:dirac_cone} we derive the same conclusion without using the bosonization framework. Namely, we treat the gauge field perturbatively and find an anomalous dimension $\gamma_{\bar{\psi} T^a \psi} = -\frac{1}{N}$ for fermion bilinears (where $T^a$ is an $SU(N)$ generator), reproducing the $SU(N)_1$ scaling dimension. 

Lastly, one may wonder about the electric vertex operators $W_{q,q'}$. The winding operator $\mathcal{W}_{q} = e^{iq \theta_c}$ where $\theta_c$ is dual to $\phi_c$ creates $q$ units of $U(1)_c$ charge. Therefore $W_{q,q'}$ is equivalent to inserting winding operators $\mathcal{W}_{q+q'}$ and $\mathcal{W}_{q-q'}^\dagger$ on the temporal boundaries $\tau=\pm\epsilon$ respectively. It is easy to see that $\Delta_{\mathcal{W}_q, \text{Dir}} = \frac{q^2}{2N}$. Therefore, $\Delta_{W_{q,q'}} = \Delta_{\mathcal{W}_{q+q'}, \text{Dir}}+\Delta_{\mathcal{W}_{q-q'}, \text{Dir}} = \frac{q^2+q'^2}{N}$.

\subsubsection{1+1D Ising CFT}

The Gutzwiller projection can be applied to discrete symmetries as well, where we project onto the symmetric sector on each lattice site. Here we consider Gutzwiller projecting the $\mathbb{Z}_2$ symmetry in the Ising CFT. For convenience we take the usual transverse field Ising model as the lattice regularization, with Hamiltonian $\hat{H}_{\text{Ising}} = \sum_j \hat{Z}_j \hat{Z}_{j+1} + \hat{X}_j$. The lattice model is useful because we can directly write the Gutzwiller projection as $\hat{P} = |F\rangle \langle F|$, where $|F\rangle = \prod_j |\hat{X}_j=1\rangle$. From this we see that the projected Hilbert space is trivial as it is one-dimensional. This also agrees with full bulk-gauging: gauging the $\ZZ_2$ symmetry of the critical Ising CFT leads to a trivial vacuum due to instanton proliferation. 

From the defect point of view, $|F\rangle$ flows to the free boundary of the Ising CFT, and like the previous section we again find a factorizing defect, now with free boundaries at $\tau=\pm \epsilon$ and the vacuum sandwiched in between. Like the last section, the vertex operators $W = e^{i\bar{a}_{\tau}}\sim e^{i\bar{\theta}/\epsilon}$ also remain nontrivial; an insertion of $W_i$ is equivalent to an insertion of the spin field $\sigma$ on each temporal boundary $\tau = \pm \epsilon$. In this example the two vertex operators are identified because the charge is only defined mod 2. Therefore, $\Delta_W = 2\Delta_{\sigma, F}=1$, where $\Delta_{\sigma, F}=1/2$ is the boundary scaling dimension of $\sigma$ on the free boundary \cite{Cardy_bcft}.

\subsubsection{Topological Orders}

There are Gutzwiller projected states which are expected to realize topological orders. Since the gauge field is gapped and only has short-range correlations in such phases, restricting it to a temporal slab should not affect the universal behavior at long distance. We therefore expect the Gutzwiller ansatze for topologically ordered states to be generally accurate in capturing the universal behavior of gapped phases. This agrees with known examples, including $\mathbb{Z}_2$ spin liquids obtained from projecting BCS superconductors, and Laughlin states obtained from projecting multiple copies of integer quantum Hall states. We analyze these two examples in more detail below. 

{\textbf{$\mathbb{Z}_2$ spin liquids}:} Gutzwiller projecting the $U(1)$ symmetry of a BCS superconductor is expected to lead to a $\mathbb{Z}_2$ topological order in $d\geq2$. The logic is that the Cooper pair condensate will Higgs the dynamical gauge field introduced by the projection down to a $\mathbb{Z}_2$ gauge field, which remains in its deconfined phase. The stability of the deconfined phase depends on two facts: (1) the weakly fluctuating phase of the temporal gauge field corresponds to the ordered phase of a $d$-dimensional Ising ferromagnet, with the temporal gauge field viewed as the effective Ising spin, and this ordered phase is stable in $d\geq2$; (2) the spatial gauge field is flat $f_{ij}=0$, as a finite stiffness for $f_{ij}$ would lead to confinement of the $(2+0)d$  $\mathbb{Z}_2$ gauge field in the $\epsilon \rightarrow 0$ limit. Therefore the theory at $\tau=0$ has the same universality as the toric code, $|\psi_{\text{gutz}}\rangle \sim |\psi_{\text{TC}}\rangle$.

One may be concerned that our field theory does not account for the global holonomy sectors of the TQFT, which encode e.g. the ground-state degeneracy. In a standard TQFT one has the flatness condition $f_{ij}=0$, but here we have the stronger constraint $a_j = \partial_j \theta$. However, this is to be expected; after all, we are describing a particular wavefunction with fixed boundary conditions, and thus the global holonomy does not fluctuate. In order to probe the topological order in the wavefunction, we must puncture the slab manifold somewhere. The pure gauge spatial field can then have nontrivial holonomy around the puncture, which accounts for the topological sector. 

We can also restore the ground state degeneracy and construct the logical operators more explicitly. We place the theory on a spatial torus of linear size $L$. We begin with the $\mathbb{Z}_2$ case. Let $|\psi_0^{ij}\rangle$ denote the mean-field (BCS) ground state on the torus with a $(-1)^{i}$ twist along the $x$ cycle and a $(-1)^{j}$ twist along the $y$ cycle, i.e.\ periodic boundary conditions along $x$ if $i=0$ and antiperiodic if $i=1$, and likewise for $y$ and $j$. The projector is the same in every sector, because $\theta$ is integrated independently on each site, and we obtain four states~\cite{IvanovSenthil2002},
\begin{equation}
    |\psi_{\text{gutz}}^{ij}\rangle = \hat{P}\, |\psi_0^{ij}\rangle .
    \label{eq:torus_states}
\end{equation}
They are degenerate ground states in the following sense~\cite{BravyiHastingsMichalakis2010}: for the normalized states and any gauge-invariant local operator $\hat{O}$,
\begin{equation}
    \langle \psi_{\text{gutz}}^{kl} |\hat{O}|\psi_{\text{gutz}}^{ij}\rangle = c_{O}\,\delta_{ik}\delta_{jl} + O(e^{-L/\xi}),
    \label{eq:tqo}
\end{equation}
with $c_O$ independent of the sector. The four states are therefore orthogonal, have the same reduced density matrix on any contractible patch, and cannot be split or mixed by a local Hamiltonian, up to exponentially small corrections\footnote{We assume that each $|\psi_0^{ij}\rangle$ is gapped and satisfies $(-1)^{\hat{Q}}|\psi_0^{ij}\rangle=|\psi_0^{ij}\rangle$ for the total charge $\hat{Q}$. A sector with the opposite parity is annihilated by $\hat{P}$.}.

To derive Eq.~\ref{eq:tqo} we use the slab theory. For gauge-invariant $\hat{O}$ only $\bar{a}_\tau=\theta_+-\theta_-$ survives in $\hat{P}\hat{O}\hat{P}$, and
\begin{equation}
    \langle \psi_{\text{gutz}}^{kl}|\hat{O}|\psi_{\text{gutz}}^{ij}\rangle = \int d\bar{a}_\tau\, \langle \psi_0^{kl}|\hat{O}\,\hat{U}_{\bar{a}_\tau}|\psi_0^{ij}\rangle .
    \label{eq:ME}
\end{equation}
This is a $d$-dimensional statistical mechanics problem for $\bar{a}_\tau$, whose weight is generated by integrating out the matter fields. For the BCS state this weight has two features. First, since the pair condensate breaks $U(1)$ down to $\mathbb{Z}_2$, it is sharply peaked at $\bar{a}_\tau\in\{0,\pi\}$, so that $e^{i\bar{a}_\tau}$ is an Ising spin. Second, a domain $R$ of $\bar{a}_\tau=\pi$ inserts the fermion parity $\hat{U}_R=\prod_{x\in R}(-1)^{\hat{Q}_x}$ of $R$, whose expectation value in a gapped superconductor obeys a perimeter law, $\langle\psi_0|\hat{U}_R|\psi_0\rangle\sim e^{-\kappa|\partial R|}$. The matter fields therefore generate a ferromagnetic coupling (a stiffness) $\kappa$ for the Ising spin, which orders in $d\geq2$ when $\kappa$ is large enough. We implement the twists by reversing the sign of every term of the mean-field Hamiltonian that crosses a closed curve $\ell$, and denote the corresponding ground state by $|\psi_0^{\ell}\rangle$. The parity $\hat{U}_R$ deforms the curve, $\hat{U}_R|\psi_0^{\ell}\rangle \propto |\psi_0^{\ell+\partial R}\rangle$ with curves added mod 2, while $\hat{P}\hat{U}_R=\hat{P}$. Hence $\hat{P}|\psi_0^{\ell}\rangle$ depends, up to a phase, only on the homology class of $\ell$, which is what $(i,j)$ labels. In the language of Sec.~\ref{sec:defect_fusion}, the twist is a symmetry defect that is absorbed by the Gutzwiller defect, and it is therefore a topological line of the slab theory. Consider first $(k,l)=(i,j)$ in Eq.~\ref{eq:ME}. If all domain walls are contractible, $\hat{U}_{\bar{a}_\tau}$ is a product of the parities of discs, and $\ell$ can be moved away from these discs and from the support of $\hat{O}$. Because the mean-field state is gapped, its reduced density matrix away from $\ell$ is that of the untwisted state up to exponentially small corrections, so the weight of such a configuration does not depend on $(i,j)$. This fails only when the discs obstruct every curve in the class of $\ell$, which requires domain walls of total length at least $2L$, and such configurations are exponentially suppressed in the ordered phase. For $(k,l)\neq(i,j)$, $\hat{U}_{\bar{a}_\tau}$ can deform the curve of the ket but cannot change its class, so every term in Eq.~\ref{eq:ME} is proportional to $\langle\psi_0^{\ell'}|\hat{O}|\psi_0^{\ell+\partial R}\rangle$ with $\ell'$ and $\ell+\partial R$ in different classes. In other words, a mismatch between the twists of the bra and the ket imposes an antiperiodic boundary condition on the Ising spin around the corresponding cycle. Every configuration then contains a domain wall of length at least $L$ that wraps the torus, and the matrix element is exponentially small in $L$. In the disordered phase the same expressions give normalized overlaps that approach unity, so the four states collapse onto a single state, as expected for a confined phase.

On the ground state manifold, let $\hat{M}^{ij}$ permute the normalized states, $\hat{M}^{ij}|\psi_{\text{gutz}}^{kl}\rangle = |\psi_{\text{gutz}}^{i+k,j+l}\rangle$ with the addition taken mod $2$, i.e.\ $\hat{M}^{x}=\hat{M}^{10}$ inserts a $\pi$ holonomy along $x$ and $\hat{M}^{y}=\hat{M}^{01}$ does the same along $y$. In the slab theory $\hat{M}^{x}$ is a closed twist line wrapping the $y$ cycle, and an open twist line ends on a pair of $\pi$ fluxes (visons). Also, let $\hat{V}^{x/y} = e^{i \oint_{C_{x/y}} a_j dx^j}$ denote a spatial Wilson loop along an $x/y$ cycle, where $a_j$ is the full connection seen by the partons, namely the flat background of the mean-field state plus the pure gauge fluctuation $\partial_j\bar{\theta}/2\epsilon$. Since the latter has no holonomy, $\hat{V}^{x}=(-1)^{i}$ and $\hat{V}^{y}=(-1)^{j}$ on $|\psi_{\text{gutz}}^{ij}\rangle$. Then we have
\begin{equation}
    \hat{M}^{x} \hat{V}^{x} = - \hat{V}^{x} \hat{M}^{x}, \qquad \hat{M}^{y} \hat{V}^{y} = - \hat{V}^{y} \hat{M}^{y}
    \label{eq:Z2_algebra}
\end{equation}
on the ground state manifold, while $\hat{M}^{x}$ commutes with $\hat{V}^{y}$ and $\hat{M}^{y}$ with $\hat{V}^{x}$, i.e.\ a twist line anticommutes with the Wilson loop that crosses it. These are the exactly analogous to the logical operators of the toric code.

{\textbf{Laughlin state}: As another example, one can consider $N$ Chern insulators with Chern number $C=1$, with $U(1) \times SU(N)$ symmetry. Integrating out the matter fields gives a level $N$ Chern-Simons term for the background $U(1)$ gauge field and a level $1$ CS term for the background $SU(N)$ gauge field. Therefore if we project either symmetry, the corresponding background gauge field is promoted to a dynamical gauge field on the slab, giving us $U(1)_N$ or $SU(N)_1$ CS theory. Like the $\mathbb{Z}_2$ example above, the spatial gauge fields remain deconfined as $\epsilon \rightarrow 0$ because $f_{ij}=0$ identically, while the thin slab limit of course does not modify the anyon structure of the theory. Therefore the theory near $\tau=0$ has the same universality as the corresponding CS theory, $|\psi_{\text{gutz}}\rangle \sim |\psi_{CS}\rangle$. In the continuum Landau level, projecting out the $SU(N)$ index leads to the celebrated Laughlin wavefunction for fractional quantum Hall state~\cite{Laughlin1983}. This explains the success of the Gutzwiller projection in capturing topological orders, for example in making accurate measurements of the topological entanglement entropy \cite{Zhang_2011, Zhang_2013}.

Let us analyze the $U(1)_N$ example in more detail. The topological part of the effective action for the background gauge field is 
\begin{equation}
\label{eq:CS_action}
    S_{top}[A] = i\frac{N}{4\pi}\int d\tau \int d^dx\epsilon^{\mu\nu\rho} A_\mu \partial_\nu A_\rho
\end{equation}
Plugging in Eq.~\ref{eq:promote_gauge}, we find the topological action
\begin{equation}
\label{eq:CS_slab}
    S_{top}[\bar{a}] = i\frac{N}{4\pi}\frac{1}{2\epsilon} \int d^dx 2\bar a_\tau \bar f_{xy}
\end{equation}
Without any operator insertions, we sum over configurations with $\bar{f}_{xy}=0$, and thus the topological action is automatically zero in the absence of operator insertions. Integrating out the matter fields also generates a stiffness for $\bar{a}_\tau$ and $\bar{\theta}$ that is not captured by the topological term.
Now consider a vortex of $\bar{a}_\tau$, denoted as $M_{1,0}$. The CS response attaches an overall $U(1)$ charge $N$ to the vortex, so to make the operator local we should attach $W_{0,N}=e^{iN\bar{\theta}/\epsilon}$ to it. The combined operator can then be added to the action as a legitimate perturbation. Likewise, we can also attach the vortex of $\bar{\theta}$, denoted as $M_{0,1}$ with $W_{N,0}=e^{iN\bar{a}_\tau}$ to make it a local perturbation. These perturbations, with sufficient strength, can lead to a fully gapped theory on the slab. The ``deconfined'' operator is $W_{1,0}M_{0,1/N}$ (as the $N$'th root of the perturbation $W_{N,0}M_{0,1}$), which is nothing but a gauge charge attached with a $2\pi/N$ flux. The fractional flux can be detected by the spatial Wilson loop $V_{2\pi p}(C)=e^{ip\int_C a_j(x,\tau)dx^j}=e^{i\frac{p}{2\epsilon} \int_C d\bar{\theta}}$. Defining $\tilde W_q = W_{q,0}M_{0,q/N}$, we have
\begin{equation}
\label{eq:braiding_phase}
\langle V_{2\pi p} (C) \tilde{W}_q(x) \rangle = e^{\frac{2\pi i p q}{N} \text{Link}(C, x)} \langle V_{2\pi p}(C)\rangle \langle \tilde{W}_q(x)\rangle
\end{equation}
This is the Chern-Simons braiding phase between spatial and temporal Wilson loops. 

To understand the ground state degeneracy and logical operators, we need to consider projecting mean field states with various boundary conditions (holonomy along $\hat{x}$ and $\hat{y}$ loops). This can be done most easily by introducing dynamical holonomy in the path integral. Namely, we further couple the system to a dynamical flat spatial gauge field $a_{x,y}$ that lives on the entire spacetime -- the only physical degrees of freedom are the holonomies $\theta_{x,y}=\int_{\mathcal{C}_{x,y}}\vec{a}\cdot d\vec{l}$. Choosing a gauge where $a_{x,y}=\theta_{x,y}/L$, the CS response term leads to
\begin{equation}
    S[\theta_{x,y}]=\frac{iN}{2\pi}\int d\tau \theta_x\partial_\tau\theta_y
\end{equation}
This leads to the familiar commutation relation between the holonomies $\hat{M}^{x,y}=e^{i\theta_{x,y}}$:
\begin{equation}
    \hat{M}^{x} \hat{M}^{y} = e^{2\pi i/N}\, \hat{M}^{y} \hat{M}^{x},
    \label{eq:ZN_algebra}
\end{equation}
which spans the $N$-dimensional ground state subspace. Our analysis agrees with the explicit construction of $N$-fold degenerate Laughlin wavefunctions on torus~\cite{HaldaneRezayi1985}.

\subsubsection{Discrete SPTs}
\label{sec:disc_spt}

Here we consider projecting mixed-SPT states, i.e. an SPT protected by $G\times K$ where defects of $K$ are decorated with $G$ charge and $G$ and $K$ are discrete. Generally, based on the decorated domain wall picture, projection onto the $G$ singlet state on each site gives a state that has nontrivial correlations of $K$ charge \cite{Tantivasadakarn_2024, zhang2024, lu2026}. 

For concreteness, we consider the $\mathbb{Z}^{(a)}_2 \times \mathbb{Z}^{(b)}_2$ cluster state in $d=1$ defined on a periodic chain, stabilized by $K_j|\psi_{\mathrm{cluster}}\rangle = |\psi_{\mathrm{cluster}}\rangle$ for $K_j = Z_{j-1}X_j Z_{j+1}$, with the two $\mathbb{Z}_2$ symmetries generated by $U^{(a)}=\Pi_{j}X_{2j}$ and $U^{(b)} = \Pi_{j} X_{2j+1}$. Consider the state $|\psi_{\mathrm{gutz}}\rangle = \hat{P}_{X_{2j}=1}|\psi_{\mathrm{cluster}}\rangle$. It is easy to see that $K_{2j}$ commutes with the projection, and in the projected subspace $K_{2j} = Z_{j-1}Z_{j+1}=1$ when acting on $|\psi_{\mathrm{gutz}}\rangle$. Therefore $Z_{1}=Z_3=Z_5=\dots$, and thus $|\psi_{\mathrm{gutz}}\rangle$ has perfect long range order of $\mathbb{Z}_2^{(b)}$, i.e. $\langle Z_{2j+1} Z_{2k+1} \rangle = 1$. 

We can understand this SSB from the path integral directly. We pick a triangulation of spacetime with variables $g_i = (g_{i,a}, g_{i,b}) \in \{0, 1\}^2$ defined at each vertex $i$ representing $\mathbb{Z}_2^{(a)}\times \mathbb{Z}_2^{(b)}$ elements, with $\delta g_{ij} = g_j - g_i$. The SPT action can be schematically represented as $S = i\pi\int \delta g_{a} \cup \delta g_b$, where $\delta g_{a/b}$ are $\mathbb{Z}_2$ 1-coboundaries. Because $\delta g_{a} \cup \delta g_b = \delta (g_a \cup \delta g_b)$ is a coboundary, the action is trivial on any closed spacetime manifold, $S=0$. 

Now we consider the Gutzwiller theory defined on a spacetime manifold $\mathbb{R} \times \Sigma$ where $\Sigma$ is closed. We can imagine cutting the spacetime at $\tau=0$ into two half-spacetimes $I^\pm \times \Sigma$ with $I^- = (-\infty, -\epsilon]$ and $I^+ = [\epsilon, \infty)$. We denote $g_{a/b}^{\pm} = g_{a/b}(\pm \epsilon)$ the matter fields on the top and bottom boundaries. The action reduces to the boundary contributions, 
\begin{equation}
S = i\pi \int (g_{a}^+ \cup \delta g_b^+ - g_a^- \cup \delta g_b^-)
\end{equation} 
The bulk path integral then enforces continuity of $g_{a/b}$, gluing $g_{a/b}^+ = g_{a/b}^-$ and recovering $S=0$. This continuity can be understood as arising from an overlap $\langle g^+_a g^+_b | g^-_a g^-_b\rangle = \delta_{g^+_a g^-_a} \delta_{g^+_b g^-_b}$ coming from resolutions of identity inserted at $\tau=\pm \epsilon$. In the Gutzwiller path integral on the other hand, the projection modifies the gluing so that we no longer necessarily need $g_{a/b}$ continuous at $\tau=0$. In particular, the overlap becomes $\langle g^+_a g^+_b | \hat{P} | g^-_a g^-_b\rangle = \frac{1}{2}\delta_{g^+_b g^-_b}$, so $g_a$ is free to be discontinuous at $\tau=0$. Defining $g^0_b = g_b(\tau=0) = g_b^+ = g_b^-$, after integrating out the bulk the partition function instead reduces to 
\begin{equation}
\cZ \sim \int dg_{a}^{+} dg_a^- dg_b^0 e^{-i\pi\int (g_a^+ - g_a^-) \cup \delta g_b^0}
\end{equation}
Obviously integrating out $g_{a}^{\pm}$ then gives $\delta g^0_b = 0$, which simply means that there are no $\mathbb{Z}_2^{(b)}$ domain walls at $\tau=0$, i.e. there is perfect long range order. 

The above example is again consistent with the full bulk gauge theory: the cluster state SPT is characterized by the topological response term $S_{top}=i\pi\int a\cup b$ where $a,b$ are the background $\mathbb{Z}_2$ gauge fields. Now gauging the $\mathbb{Z}_2^{(a)}$ symmetry means integrating over $a$, which sets $b=0$ due to the topological term. In other words, the $\mathbb{Z}_2^{(b)}$ symmetry is spontaneously broken.

This example can be generalized to higher dimensions. For example, the $d=2$ cluster state protected by $\mathbb{Z}_2^{(a)} \times\mathbb{Z}_2^{(b)} \times\mathbb{Z}_2^{(c)}$ is described by fixed point action $S = i\pi \int \delta g_a \cup \delta g_b \cup \delta g_c$. After cutting and integrating out the $a$ variables we find $\delta g_b^0 \cup \delta g_c^0 = 0$. This constraint means that the $\mathbb{Z}_2^{(b)}$ and $\mathbb{Z}_2^{(c)}$ domain walls, which are closed two dimensional surfaces, are not allowed to intersect at $\tau=0$. After integrating out the bulk and analyzing the remaining degrees of freedom on the timeslice $\tau=0$, these domain walls become closed nonintersecting loops. Since each loop can correspond either to $\mathbb{Z}_2^{(b)}$ or $\mathbb{Z}_2^{(c)}$, the remaining 2+0D stat mech model reduces to what is known as the classical $O(2)$ loop model. This result was derived for a particular microscopic realization of this SPT in \cite{zhang2024}. The loop tension, a free parameter of this model, is nonuniversal, controlled by non-topological contributions to the action.

In the full gauge theory picture, gauging the $\mathbb{Z}_2^{(a)}$ symmetry leads to a gapped $\mathbb{Z}_2$ topological order, with the $\mathbb{Z}_2^{(b)}\times \mathbb{Z}_2^{(c)}$ symmetry acting projectively on the $\mathbb{Z}_2$ flux anyon. In the above wavefunction picture, this corresponds to the gapped phase in the $O(2)$ loop model.

\subsection{Examples deviating from full gauge theory}

We now discuss examples where the Gutzwiller projected wavefunction leads to long-distance physics that deviates from the expectations from full parton gauge theories. Intuitively, in such examples the gauge field remains gapless, which leads to nontrivial dynamical consequences. Therefore it matters if this gauge field can propagate in the entire spacetime or is restricted only to the $\tau=0$ slab.

\subsubsection{Free photons}
\label{sec:photon}

The discrepancy between the full gauge theory and the slab gauge theory can already be seen from the simplest gauge theory: the free $U(1)$ Maxwell photon theory. This theory emerges in a $U(1)$ spin liquid in $d\geq 3$ where the charged fields are all gapped. The standard Maxwell action is
\begin{equation}
    S=\frac{1}{4e^2}\int d\tau d^dx f_{\mu\nu}^2.
\end{equation}
The most relevant operators are the field strength $f_{\mu\nu}$, having power-law correlation functions with scaling dimension
\begin{equation}
    \Delta_{f}=\frac{d+1}{2}.
\end{equation}

In contrast, our slab version is described by a massless free scalar $\phi=\bar{a}_\tau$ defined in space:
\begin{equation}
    S_{\rm slab}=\frac{K}{2}\int d^dx(\partial\phi)^2.
\end{equation}
The most relevant local operator is $\partial\phi$, with scaling dimension
\begin{equation}
    \Delta_{d\phi}=\frac{d}{2},
\end{equation}
which is clearly different from the bulk gauge theory. The two theories even differ kinematically: in the $(3+1)d$ bulk gauge theory, the field strength $f_{\mu\nu}$ contains both a vector (electric field) and a pseudo-vector (magnetic field), assuming standard implementation of reflection or time-reversal symmetry. In the slab theory, however, there is only vector operator $\partial\phi$, and no pseudo-vector among the lightest operators.

The above discussions can be generalized to non-abelian continuous gauge groups (like $SU(2)$) in $d\geq 4$, where the bulk gauge theory is again free, and the slab theory is the free Goldstone scalar theory of the corresponding gauge group. 

The difference between the bulk and slab theories becomes even more dramatic as we approach some lower critical dimension. For example, for non-abelian gauge theories in $(3+1)d$, the gauge bosons are gapped in the IR, but the slab theory can remain gapless since a $3d$ non-abelian Goldstone boson is stable. Another example is $U(1)$ gauge theory in $(2+1)d$, which is nothing but a dual description of a superfluid. The slab version, however, is a free scalar theory in $2d$ with only algebraic quasi-long-range order. We shall discuss more about the last example in the following.

\subsubsection{From Quantum Spin Hall to Superconductor}
\label{sec:fqsh}

The fractional quantum spin Hall (fQSH) state in 2+1D arises when spin-up electrons form a $\nu=+1/m$ fractional quantum Hall (fQH) or fractional Chern insulator (fCI) state, while spin-down electrons form the time reversed analogue, i.e. the $\nu=-1/m$ FCI state. The $m=1$ case corresponds to the familiar integer quantum spin Hall state. The fQSH state is topological phase enriched by $(U(1)_c \times U(1)_s) \rtimes \mathbb{Z}_2^T$ symmetry. We seek to understand the effect of Gutzwiller projecting out the spin sector, $|\psi_{\text{gutz}} \rangle = \hat{P}_{n_{\uparrow}=n_{\downarrow}}|\psi_{\text{fQSH}}\rangle$. The universal IR physics of the fQSH state can be captured by $U(1)_{m} \times U(1)_{-m}$ Chern-Simons (CS) theory:
\begin{equation}
\label{eq:fqsh_bulk_action}
    S_{\text{bulk}} = i\frac{m}{4\pi}\int a_{\uparrow} \wedge da_{\uparrow} - a_{\downarrow} \wedge da_{\downarrow}
\end{equation}
where $U(1)_{c}$ and $U(1)_s$ have currents $J^\mu_{c/s} = \frac{1}{2\pi}\epsilon^{\mu\nu\rho} \partial_\nu (a_{\rho,\uparrow} \pm a_{\rho,\downarrow})$. 

In the full parton gauge theory, gauging the $U(1)_s$ symmetry means coupling another dynamical $U(1)$ gauge field $a$ through $S_{\rm gauge}=\frac{i}{2\pi}\int a\wedge(da_{\uparrow}-da_{\downarrow})$, which upon integrating out $a$ identifies the gauge fields $a_{\uparrow}-a_{\downarrow}$. The CS terms then cancel out and the theory is described by a free Maxwell photon, which is nothing but the Goldstone mode of a superfluid that spontaneously breaks the $U(1)_c$ symmetry. The Gutzwiller projection, however, behaves differently.

To make progress, we will utilize the same cutting and gluing trick that we utilized for the discrete SPT example in Sec.~\ref{sec:disc_spt}. We cut the spacetime open at $\tau=0$, giving two Chern-Simons theories defined on the half-spacetimes $I^{\pm} \times \Sigma$. Each half-spacetime theory, labelled by $b\in\{+,-\}$, has a gapless right-moving and left-moving chiral $U(1)_m$ WZW field at the boundary, which we denote $\phi^b_{R/L}$. Near the boundary the CS fields are identified with the boundary fields as 
\begin{align}
\begin{split}
    &a_{\uparrow}(\epsilon) \sim d\phi^+_R, \qquad a_{\uparrow}(-\epsilon) \sim d\phi^-_L\\
    &a_{\downarrow}(\epsilon) \sim d\phi^+_L, \qquad a_{\downarrow}(-\epsilon) \sim d\phi^-_R
\end{split}
\end{align}
with boundary actions at $\tau=\pm \epsilon$
\begin{align}
\begin{split}
    S_{\text{bdry}}^{\pm} = \frac{m}{4\pi}&\int d^2x [i\p_y \phi_L^{\pm} \p_x \phi_L^{\pm} - i\p_y \phi_R^{\pm} \p_x \phi_R^{\pm} \\
    &+ v_1((\p_x \phi_L^{\pm})^2 + (\p_x \phi_R^{\pm})^2) -2v_2\p_x\phi_L^{\pm} \p_x\phi_R^{\pm}]
\end{split}
\end{align}
where $v_1, v_2$ descend from non-topological terms in the bulk action. The action of the matter sector on the defect is then
\begin{equation}
\label{eq:fqsh_bdry_sum}
    S = S_{\text{bdry}}^+ + S_{\text{bdry}}^-
\end{equation}

In order to recover the original fQSH state we must sew together the two boundary theories. Continuity of $a_{\uparrow/\downarrow}$ at $\tau=0$ identifies $d\phi^+_R = d\phi^-_L$ and $d\phi^+_L = d\phi^-_R$. The two pairs of left and right movers gap each other out, giving the original gapped fQSH state. 

Like the discrete SPT example, the Gutzwiller projection enforces a different sewing condition between the two half-spacetimes. In particular, since $J_s^\tau = \frac{1}{2\pi}(f_{xy,\uparrow}-f_{xy,\downarrow})$ and $f_{xy,\uparrow/\downarrow}$ generate gauge transformations of $a_{\uparrow/\downarrow}$ with opposite signs due to the opposite Chern-Simons levels, a $U(1)_s$ gauge transformation with parameter $\theta$ acts uniformly on both species as
\begin{equation}
\label{eq:u1_s_gauge_trans}
    a_{\uparrow/\downarrow} \rightarrow a_{\uparrow/\downarrow} + \frac{d\theta}{m}
\end{equation}
The overlap that controls the continuity around $\tau=0$ in the path-integral is modified to 
\begin{equation}
\label{eq:gauge_overlap}
\langle a_{\uparrow/\downarrow}(\epsilon) | \hat{P} | a_{\uparrow/\downarrow}(-\epsilon) \rangle = \int d\theta_+ d\theta_-\langle a_{\uparrow/\downarrow}(\epsilon) | \hat{U}_{\theta_+} \hat{U}_{\theta_-}^\dagger| a_{\uparrow/\downarrow}(-\epsilon) \rangle
\end{equation}
where $|a_{\uparrow/\downarrow}\rangle$ denote field configurations in the temporal gauge, for which the unprojected overlap enforces continuity. Therefore using Eqs~\ref{eq:u1_s_gauge_trans}-~\ref{eq:gauge_overlap}, the sewing constraint becomes 
\begin{equation}
d\phi_{R/L}^{+} - \frac{d\theta_{+}}{m} = d\phi_{L/R}^{-} - \frac{d\theta_{-}}{m}
\end{equation}

This implies that one nonchiral WZW field survives on the defect. In particular, defining the nonchiral fields
\begin{align}
\begin{split}
    &\phi_\uparrow = \phi_R^+ - \phi_L^-\\
    &\phi_\downarrow = \phi_L^+ - \phi_R^-
\end{split}
\end{align}
the sewing constraints become
\begin{align}
\begin{split}
\label{eq:fqsh_identify}
    d\phi_c \equiv d\phi_\uparrow = d\phi_\downarrow = \frac{d\bar{a}_\tau}{m} \\
\end{split}
\end{align}

This fixes the anomaly structure of the boundary fields. In particular, the $U(1)_{\text{space}}$ spatial rotation symmetry of the boundary theory is fractionalized, with boundary operators inherited from the bulk anyons carrying conformal spin $s=h-\bar{h} = \frac{q_\uparrow^2-q_{\downarrow}^2}{2m}$ for integers $q_\uparrow, q_\downarrow$. Finally, substituting Eq.~\ref{eq:fqsh_identify} into Eq.~\ref{eq:fqsh_bdry_sum} and integrating out the remaining boundary fields, the action becomes
\begin{align}
\begin{split}
&S  =\frac{K_{\text{eff}}}{4\pi} \int d^2x (\partial_j \phi_c)^2, \qquad K_\text{eff} = m \sqrt{\frac{v_1-v_2}{v_1+v_2}}
\end{split}
\end{align}
where we also rescaled space to make the kinetic term homogeneous. 

The $U(1)_c$ charge density on the defect is just the vortex density $J^\tau_c =\frac{2}{2\pi} \epsilon^{ij} \partial_i \partial_j \phi_c$. The elementary winding operator $\mathcal{W}$ carries charge $2$ and is identified with the Cooper pair operator $c_{i,\uparrow} c_{i,\downarrow}$. Its scaling dimension is $\Delta_c = \frac{m}{2}\sqrt{\frac{v_1-v_2}{v_1+v_2}}$, and thus the projected fQSH state has quasi-long range superconductivity characterized by $\langle c_{i,\uparrow} c_{i,\downarrow} c^\dagger_{j,\uparrow} c^\dagger_{j,\downarrow} \rangle \sim \frac{1}{|i-j|^{2\Delta_c}}$. Crucially, the values of $v_1, v_2$ are inherited from the bulk theory, and are not renormalized by the Gutzwiller projection. In particular, $\partial_x \phi_L \partial_x \phi_R$ is produced by an interaction between the two spin species in the bulk. Thus if the realization of the fQSH state has no such interaction, or in other words $|\psi_\text{fQSH}\rangle = |\psi_{\text{fQH}}\rangle \otimes |\psi_{\text{fQH}}^*\rangle$ for some microscopic realization $|\psi_{\text{fQH}}\rangle$ of $U(1)_m$ Chern-Simons theory, then $v_2=0$ and $\Delta_c = \frac{m}{2}$. More generally, our argument implies that the Luttinger parameter of the Gutzwiller projected state is half the Luttinger parameter that one would measure when placing the fQSH state on a manifold with boundary, following from Eq.~\ref{eq:fqsh_bdry_sum}. 


Now we consider large but finite $g$ where we slightly perturb away from the hard projection limit $g=\infty$ in Eq.~\ref{eq:gutz}. Applying the HS transformation of Eq.~\ref{eq:P} at finite $g$ it follows that perturbing away from $g=\infty$ corresponds to adding $-\lambda \cos(\bar{a}_\tau)=-\lambda \cos(m\phi_c)$ to the action, with $\lambda \sim e^{-2g}$. Indeed this operator explicitly breaks the relative $U(1)_s$ symmetry Eq.~\ref{eq:sym_rel} between the two half-spacetimes that is only present at $g=\infty$. This vertex operator has dimension $\Delta_{v} = \frac{m}{2} \sqrt{\frac{v_1+v_2}{v_1-v_2}}$. Thus $\lambda$ is irrelevant if $\Delta_{v} > 2$. For $v_2=0$, $\lambda$ is irrelevant when $m > 4$. This means that for some value of $0\ll g <\infty $ there is a KT transition from short-range to power-law superconducting correlations on the defect when $m>4$, though for $m \leq 4$ short-range correlations persist for all $g<\infty$. This implies that the $1/m$ fractional quantum Hall states with $m>4$ display Renyi-2 strong to weak spontaneous symmetry breaking (SWSSB) at a finite strength of density dephasing \footnote{The doubled space representation of the dephased quantum Hall state is exactly the quantum spin Hall state with an attractive density-density interaction between the left and right space supported on a temporal defect, which becomes the Gutzwiller projection in the infinite dephasing limit. The doubled space representation before dephasing is a tensor product state, and therefore $v_2=0$.}.

These results are in agreement with Laughlin wavefunction based arguments  \cite{Girvin_1987, Kane_1991, Lu_2023, Sarma_2025, wang2025fractionalquantumhallstates}, and the prediction of $\Delta_c = \frac{m}{2}$ for $v_2=0$ has been verified numerically \cite{Sarma_2025, lu2026}. Our field theoretic argument is universal and thus holds even on lattice model realizations where the Laughlin wavefunction is not a valid approximation to the ground state wavefunction. Our field theory arguments also imply nontrivial predictions that cannot be argued via the Laughlin wavefunction alone: for example, the spatial $U(1)_c$ currents $J^j_c$ have scaling dimension $\Delta_{J^j_c}=1$, consistent with being the conserved current of a global symmetry in a 2D CFT. On the other hand, the $U(1)_c$ charge density $J^\tau_c=n_{\uparrow}+n_{\downarrow}$ has short-range correlations. 

\subsubsection{Exciton Bose Liquid}

Another example where the Gutzwiller projection does not give the same answer as the bulk gauge field picture is the Exciton Bose Liquid (EBL) \cite{Tay_2011, Paramekanti_2002}, which has the nice feature of being exactly solvable. The EBL state is captured by a bulk gauge theory with two flavors of bosonic partons $b_{x}, b_{y}$ corresponding to the parton decomposition $b_i = b_{i,x} b_{i,y}$. Before introducing the gauge field, the EBL mean-field state $\ket{\psi_{0}}$ consists of $b_x$ hopping freely in the $x$ direction but not in the $y$ direction and vice-versa for $b_y$. There are $U(1)_c \times U(1)_s$ symmetries, rotating $b_{x}$ and $b_{y}$ in the same and opposite directions respectively. The EBL theory is obtained by coupling $U(1)_s$ to a dynamical bulk gauge field, which is exactly solvable through bosonization thanks to the quasi-1D character of the bosonic partons. 

Based on the incorrect bulk gauge theory picture of the Gutzwiller projection, this would lead one to believe that $\ket{\psi_{\mathrm{gutz}}} = \hat{P}_{n_x = n_y} \ket{\psi_{0}}$ would capture the same universal physics. In fact, this is not the case, as shown in \cite{Tay_2011}. They solve the constraint $n_x = n_y$ exactly in bosonization and show that a different field theory is obtained than the EBL theory, and that this field theory gives the same scaling as the numerically extracted scaling from the Gutzwiller wavefunction. However, to obtain this result, they enforce the constraint $n_x = n_y$ in the entire spacetime, and not only on the $|\tau|<\epsilon$ slab. Thankfully this subtlety does not affect the result in this case. The parton mean-field Lagrangian is given by 

\begin{align}
\begin{split}
\mathcal{L}_{0}=&\frac{v}{2\pi} [g^{-1}(\partial_x \theta_x)^2 + g(\partial_x \phi_x)^2] + \frac{i}{\pi} (\partial_x \theta_x)(\partial_\tau \phi_x) \\
+ &\frac{v}{2\pi} [g^{-1}(\partial_y \theta_y)^2 + g(\partial_y \phi_y)^2] + \frac{i}{\pi} (\partial_y \theta_y)(\partial_\tau \phi_y) 
\end{split}
\end{align}

\noindent where $n_j \sim \partial_j \theta_j$. To properly treat the Gutzwiller projection, we rewrite $\theta_x = \partial_y \vartheta + \partial_x\varphi$, $\theta_y = \partial_x \vartheta - \partial_y\varphi$. Viewing $\vec{\theta} = (\theta_y, \theta_x)$ as a 2D vector field at a given timeslice, this is just the standard Helmholtz decomposition $\vec{\theta} = \vec{\nabla} \vartheta + \vec{z} \times \vec{\nabla} \varphi$, and thus introduces no redundancies. We also integrate out $\phi_j$ to find

\begin{align}
\begin{split}
\mathcal{L}_{0}=&\frac{v}{2 \pi g}[2(\partial_x \partial_y \vartheta)^2 + (\partial_x^2 \varphi)^2 + (\partial_y^2 \varphi)^2 \\ & \hspace{1cm}+ 2(\partial_x \partial_y \vartheta) (\partial_x^2 \varphi - \partial_y^2 \varphi)]\\
+& \frac{1}{2\pi vg} [(\partial_\tau\partial_x\vartheta)^2 + (\partial_\tau \partial_y \vartheta)^2 + (\partial_\tau\partial_x\varphi)^2 + (\partial_\tau \partial_y \varphi)^2 \\& \hspace{1cm}+ 2((\partial_\tau \partial_y \vartheta)(\partial_\tau\partial_x \varphi) - (\partial_\tau \partial_x \vartheta)(\partial_\tau\partial_y \varphi))]
\end{split}
\end{align}

The $U(1)_s$ current is $J^0 = \partial_j \partial^j\varphi$, $J^x = -\partial_\tau \partial_x \varphi$, $J^y = -\partial_\tau \partial_y \varphi$. We can integrate out $\bar a_\tau$ and $\bar \theta$ which enforces $\partial_j \partial^j \varphi|_{\tau=0}= 0$ and $\partial_\tau \partial_j \partial^j \varphi|_{\tau=0} = 0$. Expanding in Fourier modes, this kills every term in $\cL_{0}$ that includes $\varphi$ at $\tau=0$, recovering the action describing $\vartheta$ fluctuations derived in Ref.~\cite{Tay_2011}. Thanks to the quasi-1d nature of the EBL mean-field the result is qualitatively similar to the Luttinger liquid case - the projection kills the entire projected sector while leaving the remaining sector untouched.

\subsubsection{Spinon Fermi Surface and composite Fermi liquid}

We consider now the Gutzwiller projection of the charge $U(1)$ symmetry of the spinon Fermi surface, $|\psi_{\text{gutz}}\rangle = \hat{P}_{n_{\uparrow} + n_{\downarrow} = \bar{n}} |\psi_{0}\rangle$. Coupling a spinon Fermi surface to a bulk gauge field leads to Landau damping of the gauge fluctuations, producing a non-Fermi liquid fixed point which displays logarithmic corrections to the electron self-energy \cite{Nayak_1994, Polchinski_1994, Nave_2007, Lee_2009, Anakru_2025}. A similar theory with a Fermi surface coupled to $U(1)$ gauge field also describes the composite Fermi liquid phase in quantum Hall systems, with additional Chern-Simons terms for the gauge field that is expected to not change the IR behavior. On the other hand, the Gutzwiller projection of a generic (i.e., without nesting) spinon (or composite fermion) Fermi sea may not display such non-Fermi liquid behavior. 

In Appendix \ref{app:proj_fermi} we show that the slab gauge theory leads to a Fermi liquid in the IR. In particular, perturbative treatment shows that the gauge fluctuations do not modify the leading power law of spin-spin correlation functions, though they renormalize the amplitude. This is consistent with the numerics of Refs.~\cite{Anakru_2025, Motrunich_2005} which find Fermi liquid scaling of correlation functions in the projected Fermi surface wavefunction. Therefore the Gutzwiller projected Fermi surface does not faithfully describe, in terms of long-distance behaviors, the non-Fermi liquid phase of spinons or composite fermions.

Our perturbative result can be intuitively understood as follows. In the standard treatment of the gauged Fermi surface problem, the non-Fermi liquid behavior comes from fluctuations of the spatial gauge field ${a}_{i}$ (the temporal gauge field becomes unimportant in the IR due to Debye screening). In our slab theory, the spatial gauge field is (1) flat, and (2) fluctuating only on a temporal slab. The slab geometry means that we are mostly considering fluctuations in the $q\to0$ limit, keeping $\omega$ finite, where the gauge field dynamics is dominated by the diamagnetic term $(n/2m)a_i^2$ and behaves as a gapped field. It is therefore natural to expect Fermi liquid behavior.

On the other hand, suppose there is perfect nesting, such as the nearest-neighbor square lattice tight-binding model at half-filling, with nesting vector $Q=(\pi, \pi)$. We will show that the projected state develops N\'eel order for $d\geq2$ spatial dimensions. We take $\rho=\bar{\psi}\psi$ the local density, with $J^\tau=\rho-1$, and $\vec{S}=\frac{1}{2}\bar{\psi} \vec{\sigma} \psi$. First, we rewrite the projection with a different Hubbard-Stratonovich decomposition, utilizing the identity $(\rho_j-1)^2 = 1-\frac{4}{3}\vec{S}^2_j$. We also keep the projection strength $g$ finite for the moment.
\begin{align}
\begin{split}
    \hat{P} = e^{-g \sum_j (\rho_j-1)^2} \sim e^{\frac{4g}{3}\sum_j \vec{S}_j^2} =\int d\vec{n} e^{-\sum_j (\frac{3}{16g} \vec{n}_j^2 - \vec{n}_j \cdot \vec{S}_j)} 
\end{split}
\end{align}
Taking $\vec{n}_{\pm}$ the HS fields for the left / right projection respectively, we define $\vec{N} = \vec{n}_+ + \vec{n}_-$ and $\vec{M} = \epsilon(\vec{n}_+ - \vec{n}_-)$, the continuum action is
\begin{align}
\begin{split}
    S_{\text{gutz}} = S_0 + &\frac{1}{2\epsilon}\int_{-\epsilon}^{\epsilon}d\tau\int d^dx [\frac{3}{32g}(\vec{N}(x)^2+\vec{M}(x)^2/\epsilon^2) \\&- \vec{N}(x) \cdot \vec{S}(x,\tau) - \vec{M}(x) \cdot \partial_\tau \vec{S}(x,\tau)]
\end{split}
\end{align}
The purpose of this rewriting is to integrate out the fermions and obtain an effective action for $\vec{N}$, which we analyze to leading order in $\vec{N}$. It is necessary to keep $g$ finite because the saddle point equation for $\vec{N}$ is $\vec{N} = \frac{16g}{3} \vec{S}(\tau=0)$, and so $\vec{N}^2 \ll 1$ is small for small $g$ and the expansion in powers of $\vec{N}$ is controlled. The saddle point equation for $\vec{M}$ is $\vec{M} = \epsilon^2\frac{16g}{3} \partial_\tau \vec{S}(\tau=0)$. At present we are interested in the equal-time correlations of $\vec{S}(\tau=0)$, so we focus on the $\vec{N}$ sector. We integrate out the fermions to find
\begin{align}
\begin{split}
   S_{eff}[\vec{N}]=\frac{1}{2}\int_{k} [\frac{3}{16g}-\frac{1}{3}\chi_0(k)]\vec{N}(-k) \cdot \vec{N}(k)
\end{split}
\end{align}
to leading order, where 
\begin{equation}
\chi_0(k) = \frac{1}{(2\epsilon)^2}\int_{-\epsilon}^{\epsilon} d\tau_1 d\tau_2 \langle \vec{S}(-k,\tau_1)\cdot \vec{S}(k,\tau_2)\rangle_0
\end{equation}
is the (suitably regulated) equal-time spin structure factor in the unprojected state. There is an instability to $\vec{N}$ ordering at $k$ when $\frac{3}{16g}-\frac{1}{3}\chi_0(k) < 0$. Since there is Fermi surface nesting, $\chi_0(k)$ is maximal at $k=Q$. In fact, near $k=Q$ it has a cusp singularity,
\begin{equation}
    \chi_0(Q+q) =\chi_0(Q)-\alpha |q|
\end{equation}
with $\chi_0(Q), \alpha>0$. Since $\chi_0(Q)$ is maximal, we can imagine tuning $g$ up from $g=0$ to $g\gtrsim g_c = \frac{9}{16\chi_0(Q)}$, at which point an instability towards N\'eel order develops. Thus, we can expand about a symmetry broken saddle 
\begin{equation}
    \vec{N}(r) = (-1)^{|r|}(\vec{N}_0\sqrt{1-|\vec{\pi}|^2} + \vec{\pi}), \; \vec{\pi} \cdot \vec{N}_0=0
\end{equation}
where $|\vec{N}_0| = \rho_0 + \delta \rho$. The amplitude fluctuations $\delta \rho$ have a mass of order $1/g_c-1/g$, and therefore do not destabilize the saddle. Thus we need only examine the transverse fluctuations given by $\vec{\pi}$. This gives us an effective action for $\vec{\pi}$:
\begin{align}
\begin{split}
\label{eqn:trans_action}
    S_{eff}[\vec{\pi}] &= \frac{\rho_0^2}{2} \int \frac{d^d q}{(2\pi)^d} [\chi_0(Q)-\chi_0(Q+q)] |\vec{\pi}(q)|^2 \\
    &= \frac{\rho_0^2}{2} \int \frac{d^d q}{(2\pi)^d} \alpha|q| |\vec{\pi}(q)|^2 
\end{split}
\end{align}
Then, the transverse fluctuations to leading order are given by
\begin{equation}
    \langle |\vec{\pi}|^2\rangle \sim \int \frac{d^dq}{(2\pi)^d} \frac{1}{|q|}
\end{equation}
which has no IR divergence for $d\geq2$. Thus, the saddle point is stable, and the system spontaneously develops N\'eel order at $\tau=0$. Now as we take $g\rightarrow\infty$, we expect the phase to remain stable, although the perturbative calculation becomes uncontrolled. In particular, the $\alpha$ appearing in Eq.~\ref{eqn:trans_action} will be nonperturbatively renormalized, but we expect the form of the effective action to remain the same. For $d=2$ though the defect problem is effectively 2+0 dimensional, the system circumvents the Mermin-Wagner theorem thanks to the nonlocal kernel $\chi_0(k)$ supplied by the gapless 2+1D bulk. This argument breaks down for a generic Fermi surface because there is no longer an isolated saddle. In particular, $\chi_0(k)$ is maximized for $|k|=2k_F$, and therefore fluctuations of the direction of $k$ destabilize the symmetry broken saddle. Our analysis agrees with previous numerical study of projected nested Fermi surface on the square lattice~\cite{Li2013UniformRVB}. A similar argument implies that Gutzwiller projecting the $SU(2)$ flavor symmetry of a spinon Fermi surface leads to a long range superconductor. This is used to study the strong-to-weak symmetry breaking properties of a dephased Fermi sea in Ref.~\cite{sarma2026strongtoweakspontaneoussymmetrybreaking}.

\subsubsection{Dirac Fermions}

A paradigmatic example of an algebraic spin liquid is $QED_3 / QCD_3$ \cite{hermele_2005, Ran_2006}, describing $N$ flavors of Dirac fermions in 2+1D coupled to a bulk gauge field. In the large $N$ limit, the theory flows to a nontrivial interacting CFT with algebraically decaying correlations that are enhanced compared to the mean-field state, with anomalous dimensions given by certain functions of $N$ and the gauge group. For $U(1)$ Dirac spin liquids, existing numerics suggest (but do not prove) that this picture holds until some critical value $N_c<4$, with $N=4$ being a well-studied example relevant for spin-$1/2$ systems on triangular and kagome lattices~\cite{Song2019,Song2020,KagomeDirac2017,TriangularDirac2019,TriangularDirac2024}.

The mean-field state is simply $N$ free Dirac fermions in the IR. We consider the effect of Gutzwiller projecting the charge $U(1)$ symmetry, which leads to a slab version of $QED_3$. Previous numerical studies on this state \cite{Su_2025, Paramekanti_2007, Li_2023, RogersOnTS, PhysRevB.103.195140} extracted various conflicting exponents. In Appendix \ref{app:dirac_cone} we study the slab version of $QED_3$ which properly captures the projected Dirac fermion state. Unlike the bulk gauge theory, it does not flow to an interacting CFT. We find no anomalous dimension, only scale dependent subleading contributions. This explains the tension between exponents extracted in previous works, as the exponent one will extract is effectively scale-dependent and non-universal, with correlations tending towards free-fermion in the IR. Consequently, the Gutzwiller projected Dirac fermions could appear critical for any flavor number $N$, even though the true field theory is critical only for sufficiently large $N$.

More generally, we can have the following scaling argument that applies when the parton mean field is described in the IR by some conformal field theory $S^{(d+1)}_{\rm CFT}$. Consider the defect theory, schematically written as
\begin{equation}
    S_{\rm CFT}^{(d+1)}+\int_{\tau\to0} d^dx\left[\frac{K}{2}(\partial\bar{a}_\tau)^2+\bar{a}_\tau\rho\right],
\end{equation}
where $\rho$ is the $U(1)$ charge density from the bulk CFT. Now treating the bulk CFT and the boundary free scalar as fixed point and the gauge coupling as perturbation, we have $\Delta_\rho=d$ for any $(d+1)$ dimensional CFT and $\Delta_{\bar{a}_\tau}=d/2-1$ from free scalar. So the scaling dimension of the gauge coupling term is $\frac{3}{2}d-1$, and is irrelevant when $d>2$. For free Dirac fermions at $d=2$, the gauge coupling is marginal at leading order, and our calculation in Appendix~\ref{app:dirac_cone} shows that it becomes irrelevant in higher order, similar to ordinary QED in $(3+1)d$. At $d=1$, the gauge coupling is relevant and the free theory eventually flows to the $SU(N)_1$ CFT, as explained in Sec.~\ref{sec:1+1D} and Appendix~\ref{app:dirac_cone}.

\section{Defect Fusion and Intertwining Defects}
\label{sec:defect_fusion}
\begin{figure*}[t]
\centering

\begin{minipage}{0.32\textwidth}
\centering
\begin{tikzpicture}[
    scale=0.52,
    transform shape,
    line cap=round,
    line join=round,
    >={Stealth[length=2.5mm,width=1.5mm]}
]

\draw[
    black,
    line width=1.1pt,
    postaction={decorate},
    decoration={
        markings,
        mark=at position 0.76 with {\arrow{>}}
    }
]
(0,-1.85) -- (0,2.15);

\draw[
    red,
    line width=1.1pt,
    postaction={decorate},
    decoration={
        markings,
        mark=at position 0.47 with {\arrow{>}}
    }
]
(3.15,0) -- (0,0);

\fill[black] (0,0) circle (2.6pt);

\node[red, font=\Large] at (2.05,0.53) {$U_g$};
\node[font=\Large] at (0,-2.20) {$\mathcal{D_{\text{gutz}}}$};

\end{tikzpicture}
\end{minipage}%
\hfill
\begin{minipage}{0.32\textwidth}
\centering
\begin{tikzpicture}[
    scale=0.52,
    transform shape,
    line cap=round,
    line join=round,
    >={Stealth[length=2.5mm,width=1.5mm]}
]

\draw[
    black,
    line width=1.1pt,
    postaction={decorate},
    decoration={
        markings,
        mark=at position 0.76 with {\arrow{>}}
    }
]
(0,-1.85) -- (0,2.15);

\draw[
    red,
    line width=1.1pt,
    postaction={decorate},
    decoration={
        markings,
        mark=at position 0.47 with {\arrow{>}}
    }
]
(0,0) -- (-3.15,0);

\fill[black] (0,0) circle (2.6pt);

\node[red, font=\Large] at (-2.05,0.53) {$U_g$};
\node[font=\Large] at (0,-2.20) {$\mathcal{D}_{\text{gutz}}$};

\end{tikzpicture}
\end{minipage}%
\hfill
\begin{minipage}{0.32\textwidth}
\centering
\begin{tikzpicture}[
    scale=0.52,
    transform shape,
    line cap=round,
    line join=round,
    >={Stealth[length=2.5mm,width=1.5mm]}
]

\draw[
    black,
    line width=1.1pt,
    postaction={decorate},
    decoration={
        markings,
        mark=at position 0.76 with {\arrow{>}}
    }
]
(0,-1.85) -- (0,2.15);

\draw[
    blue,
    line width=1.1pt,
    postaction={decorate},
    decoration={
        markings,
        mark=at position 0.47 with {\arrow{>}}
    }
]
(0,0) -- (-3.15,0);

\draw[
    red,
    line width=1.1pt,
    postaction={decorate},
    decoration={
        markings,
        mark=at position 0.47 with {\arrow{>}}
    }
]
(3.15,0) -- (0,0);

\fill[black] (0,0) circle (2.6pt);

\node[blue, font=\Large] at (-2.05,0.53) {$V_g$};
\node[red, font=\Large] at (2.05,0.53) {$U_g$};
\node[font=\Large] at (0,-2.18) {$\mathcal{D}_{\text{intwn}}$};

\end{tikzpicture}
\end{minipage}
\caption{Fusion of Gutzwiller and $G$-Intertwining Defects \cite{wei2026}. These fusion properties are explicit in the UV and preserved by RG flow, and therefore must hold in the IR.}
\label{fig:fusion}
\end{figure*}
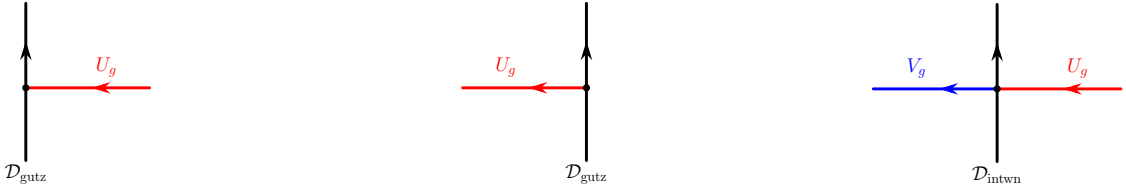

The Gutzwiller projection is a non-invertible defect, and it is useful to think about its properties from the perspective of defect fusion. In particular, we can reinterpret the defining characteristics of the Gutzwiller projection, Eq.~\ref{eqn:gutz_proj_def}, in terms of defect fusion. We denote $U_g$ the topological $G$ symmetry defects, which form a fusion category. The defects $U_g(\Sigma)$ are defined for any codimension-one cycle $\Sigma$, with $\p \Sigma=0$. They satisfy the $G$ fusion rule $U_g \otimes U_h \cong U_{gh}$. An abstract Gutzwiller defect $D_{\text{gutz}}(\eta)$ is defined for any codimension-one chain $\eta$ which need not be closed, $\p \eta \neq 0$. This is because the Gutzwiller projection is on-site and therefore can terminate. The Gutzwiller projection property $\hat{Q}_j \hat{P} = \hat{P} \hat{Q}_j = 0$ means that $U_g$ defects can terminate on $D_{\text{gutz}}$, so $U_g(\Sigma)D_{\text{gutz}}(\eta)$ is well defined for $\p\Sigma \neq 0$ if $\p \Sigma \subset \eta$. Equivalently, the Gutzwiller defect satisfies the fusion rule $U_g \otimes D_{\text{gutz}} \cong D_{\text{gutz}} \otimes U_g \cong D_{\text{gutz}}$. Although the Gutzwiller defect is not necessarily topological, the fusion rule is still meaningful because $U_g$ are. Further, the Gutzwiller defect satisfies $D_\text{gutz}(\eta) D_{\text{gutz}}(\eta') = D_{\text{gutz}}(\eta \cup \eta')$.

The fusion rule has a striking consequence: the Gutzwiller defect naturally mixes twisted sectors of the theory. To illustrate the point, we focus on 1+1D spacetime, with a Gutzwiller line $D_\text{gutz}(\eta)$ inserted. A symmetry line $U_g(\gamma)$ generically terminates on conjugate twist operators $\mu$, $\bar{\mu}$ inserted at the two endpoints of $\gamma$. Denoting $\p\gamma = \{x, y\}$, $U_g(\gamma)\mu(x)\bar{\mu}(y)$ is well-defined. However, when the Gutzwiller defect is present, we can consider $\gamma$ configurations with one endpoint on the Gutzwiller line, leaving an unpaired twist operator $\mu$ at the other endpoint. For example, with $y \in \eta$, $D_{\text{gutz}}(\eta) U_g(\gamma) \mu(x)$ is well-defined. Thus, denoting $\cZ_{\text{gutz}} = \int D\phi D_{\text{gutz}}(\eta)e^{-S}$ the defect partition function on a torus, generically $\langle U_g(\gamma)\mu(x)\rangle_{\text{gutz}} \neq 0$, which is what we mean when we say that the Gutzwiller defect mixes twisted sectors. 

We can generalize this property of the Gutzwiller defect to construct a UV regularization of so-called ``$G$-intertwining" defects \cite{wei2026}. These arise naturally as symmetry preserving defects at $\tau=0$ when we have a $G$ symmetry that acts differently on $\tau<0$ and $\tau>0$. Unlike the Gutzwiller defect, $G$-intertwining defects are well-defined even when $G$ is anomalous. In particular, let $U_g$, $V_g$ denote the $G$ symmetry defects for $\tau<0$, $\tau>0$ respectively (which by assumption have the same anomaly). Then a $G$-intertwining defect $D_{\text{intwn}}$ is characterized by the fusion rule $V_g\otimes D_{\text{intwn}} \cong D_{\text{intwn}} \otimes U_g$, or equivalently, $U_g$ and $V_g$ can terminate on $D_{\text{intwn}}$ at a shared topological junction. Note that $G$-symmetric defects are a subset of $G$-intertwining defects with $V_g = U_g$. Like the Gutzwiller defect, a $G$-intertwining defect may or may not be topological. If $D_{\text{intwn}}$ is a topological intertwining defect, it will automatically scatter incoming local operators into twisted sector operators attached to a line ending at $D_{\text{intwn}}$ with perfect transmission \cite{ueda2026}. In particular, let $A=\bar{D}_{\text{intwn}}\otimes D_{\text{intwn}}$, $A' = D_{\text{intwn}}\otimes \bar{D}_{\text{intwn}}$. If a local operator is charged under a defect in $A$, then upon dragging the local operator through $D_{\text{intwn}}$, it becomes attached to a string $L \in A'$ terminating on $D_{\text{intwn}}$ \cite{wei2026}.

Based off of what we know of the Gutzwiller projection, it is easy to construct a slab field theory that is a UV regularization of a $G$-intertwining defect. For the moment, suppose that $G$ is anomaly free. In this case $G$ is onsite and we can define $\hat{Q}_{j,<}$ the $U_g$ charge on site $j$ and $\hat{Q}_{j,>}$ the $V_g$ charge on site $j$. Then we can define $\hat{P}_{\text{intwn}}$ to be an ``intertwining operator" if it satisfies \begin{equation}
\label{eqn:int_op_def}
    \hat{Q}_{>,j} \hat{P}_{\text{intwn}} = \hat{P}_{\text{intwn}} \hat{Q}_{<, j}
\end{equation} 
and an obvious realization is
\begin{equation}
\hat{P}_{\text{intwn}}  = \int d\theta e^{-i\theta_j \hat Q_{j,>}}e^{i\theta_j \hat Q_{j,<}}
\end{equation}
The partition function is then $\cZ = \langle\psi_0|\hat{P}_{\text{intwn}} |\psi_0\rangle$. From here, the derivation is slightly different from the Gutzwiller projection. In particular, an operator insertion at $\tau=0$ should be interpreted as 
\begin{equation}
    \langle \hat{O}(\tau=0)\rangle = \frac{\text{Tr} [e^{-\beta H/2}\int d\theta e^{-i\theta_j \hat{Q}_{j,>}}\hat{O} e^{i\theta_j \hat{Q}_{j,<}}e^{-\beta H/2}]}{\text{Tr} [e^{-\beta H/2}\int d\theta e^{-i\theta_j \hat{Q}_{j,>}}e^{i\theta_j \hat{Q}_{j,<}}e^{-\beta H/2}]}
\end{equation}
Thus, unlike the Gutzwiller case, there is only one compact boson introduced on the defect. This can be intuited from the defect fusion rule: the Gutzwiller defect can absorb a symmetry defect from either side, while the intertwining defect can only transmute an incoming $U_g$ defect into an outgoing $V_g$ defect (Fig.~\ref{fig:fusion}). The continuum limit then follows directly. In particular, let $J^\mu_{>}$ be the $G$ current for $\tau>0$, and likewise for $J^\mu_{<}$. Then, $\cZ_{\text{intwn}} \sim \lim_{\epsilon \rightarrow 0} \int D\phi d\bar{a}_{\tau} e^{-S_{\text{intwn}}}$, with 

\begin{equation}
\label{eq:intertwine_action}
S_{\text{intwn}} = S_0 + \frac{1}{\epsilon}\int_{0}^\epsilon d\tau \int d^dxi\bar{a}_\tau(x) (J_>^\tau(x,\tau) - J^\tau_<(x, -\tau))
\end{equation}
where $\bar{a}_\tau$ has the standard compactification $\bar{a}_\tau \sim \bar{a}_\tau + 2\pi$. 

Although we assumed $G$ is anomaly free to arrive at the continuum action Eq.~\ref{eq:intertwine_action}, this continuum action is consistent even when $G$ is anomalous. In that case $\hat{P}_{\text{intwn}}$ must be realized as some non-onsite matrix product operator.

\subsection{Self-Duality Defects}

Such intertwining defects arise naturally in theories with a self-duality. Let $\cT$ be a 1+1D QFT with global $G$-symmetry, and symmetry lines $U_g$. Suppose there is an anomaly-free subgroup $A \subset G$ and $\cT$ is self-dual under gauging $A$, i.e. there exists an isomorphism $\chi: \cT/A \rightarrow \cT$. Then $\cT/A$ has a dual (or ``quantum") global $G$ symmetry, with symmetry lines $\hat{V}_g$. Then one can construct a non-invertible topological defect $D_{\text{dual}}$ in the following manner. One first gauges $A$ on the half space $\tau>0$ with Dirichlet boundary conditions for the gauge field at $\tau=0$, producing a topological interface between $\cT$ and $\cT/G$, and then applies $\chi$ to $\tau>0$, restoring the $\tau>0$ theory to $\cT$. Defects obtained in this manner are known as self-duality defects. Then, defining $V_g = \chi(\hat{V}_g)$, the self-duality defect is naturally a $G$-intertwining defect with $U_g \otimes D_{\text{dual}} \cong D_{\text{dual}} \otimes V_g$. Further, $\bar{D}_{\text{dual}} \otimes D_{\text{dual}} = A$ \footnote{We slightly abuse notation, using $A$ to refer both to the subgroup of $G$ as well as the direct sum $\bigoplus_{a \in A} U_a$ of all $A$ defects.}. Therefore self-duality defects are always topological intertwining defects. The converse is also true; if $D_{\text{intwn}}$ is a topological intertwining defect, then the theory is self-dual under gauging $A=\bar{D}_{\text{intwn}} \otimes D_{\text{intwn}}$ \cite{wei2026}. Thus, self-duality defects are exactly equivalent to topological intertwining defects in 1+1D. The half-space gauging construction also extends to any even spacetime dimension \cite{Choi_2022}, although in that case $A$ must be a $q$-form symmetry with $q=\frac{d-2}{2}$.

Self-duality defects have attracted much attention recently \cite{Choi_2022, Kaidi_2022, choi2023, antinucci2023}. One intriguing application is the problem of scattering a Weyl fermion off of a massive monopole in 3+1D \cite{vanbeest2023, tamargo2026, wei2026, loladze2026, ueda2026}. Indeed, although the incoming fermion is created by a local operator, the outgoing fermion is attached to a symmetry defect that terminates on the monopole worldline, and thus lives in a twisted sector. In the $s$-wave channel this problem can be reduced to an effective 1+1D problem of $N$ free Dirac fermions moving on the half-plane, with some boundary condition capturing the effect of the monopole \cite{vanbeest2023}. If we specify a non-anomalous $G=U(1)^N = U(1)_1 \times \dots \times U(1)_N$ symmetry to be preserved by the boundary, then the boundary state is uniquely determined \cite{vanbeest2023} when we also impose conformal invariance. Upon unfolding, this becomes a theory of $N$ Weyl fermions on the whole plane with a topological $G$-intertwining defect $D$ at $\tau=0$. Explicit expressions for $U_g$, $V_g$ can be found in \cite{wei2026}. Since $D$ is the unique conformal defect with the $G$-intertwining property in this theory, Eq.~\ref{eq:intertwine_action} is guaranteed to flow to $D$ in the IR, realizing a class of noninvertible topological defects of Weyl fermions. In this case our construction reduces to the rotor regularization used in \cite{loladze2026}. Our construction also reproduces the self-duality defects of the $U(1)_N$ WZW model \cite{Choi_2022}.

\section{Conclusions}
\label{sec:conclusions}

In this work we present a novel field theory that captures the universal physics of the Gutzwiller projection and is amenable to perturbative treatment. The theory consists of coupling the bulk matter fields to a gauge field supported on a temporal slab $|\tau| \leq \epsilon$ of vanishing width. We are able to reproduce various known analytical results while also making new predictions. In particular, we show that the Gutzwiller projection does not modify the IR correlations of fermi liquids and Dirac fermions ($d\geq2$), contrary to expectations from field theory. We also show analytically that the Gutzwiller projection of a fractional quantum spin hall insulator is described by a particular compact boson CFT, and that our theory can be successfully applied to states with discrete global symmetry such as $G\times K$ SPTs. Lastly, we construct a similar field theory of so-called $G$-intertwining defects, giving a UV regularization of many self-duality defects, including many of those recently studied in Weyl fermions with application to the monopole scattering paradox. Our results open up a new route towards understanding the universal behavior of projected wavefunctions and $G$-intertwining defects, with applications across condensed matter and quantum information. 

We end with some open directions to be explored in the future:
\begin{itemize}
 \item \textit{From UV to IR}: We have shown that for a large class of Gutzwiller projected wave functions, the true long-distance behavior is not properly captured by the trial wave function. Suppose a Gutzwiller wave function (for example, the projected Fermi sea) produces very good ground state energy, it seems physically quite reasonable to say that the wave function correctly captures the short-distance physics, while the long-distance physics should be described by the corresponding gauge theory (for example, a Fermi surface coupled to $U(1)$ gauge field). Is there a way to make this kind of UV-IR correspondence more precise? Recall in the BCS superconductor, we have the key notion of order parameter that bridges between the UV BCS trial wave function and the IR Ginzburg-Landau field theory. What is the analogue of the order parameter in the more complicated, highly entangled, quantum phases?

 \item \textit{Ground-state-ability}: Given a Gutzwiller wave function, can we always realize it as the exact ground state of some local Hamiltonian? For those that are consistent with a field description, one naturally expects the answer to be positive, since a field theory respecting time-translation symmetry does admit a Hamiltonian description. Again the Laughlin state serves as the best understood example in this category. However, for those wave functions that do not agree with any known field theory (with time-translation symmetry) at long distance, such as the projected Dirac fermions in $2d$, the answer is not obvious.
\end{itemize}

\section{Acknowledgements}

A.S. is grateful to Roger Melko for hosting him at Perimeter Institute where this work was initiated. The authors are grateful to Matthew Fisher, Tarun Grover and Zhehao Zhang for helpful discussions. Research at Perimeter Institute is supported in part by the Government of Canada through the Department of Innovation, Science and Economic Development; and by the Province of Ontario through the Ministry of Colleges, Universities, Research Excellence and Security. Sarma and Xu are supported by the Simons Foundation International through the Simons Investigator program.

\bibliography{bibliography}

\begin{thebibliography}{89}%
\makeatletter
\providecommand \@ifxundefined [1]{%
 \@ifx{#1\undefined}
}%
\providecommand \@ifnum [1]{%
 \ifnum #1\expandafter \@firstoftwo
 \else \expandafter \@secondoftwo
 \fi
}%
\providecommand \@ifx [1]{%
 \ifx #1\expandafter \@firstoftwo
 \else \expandafter \@secondoftwo
 \fi
}%
\providecommand \natexlab [1]{#1}%
\providecommand \enquote  [1]{``#1''}%
\providecommand \bibnamefont  [1]{#1}%
\providecommand \bibfnamefont [1]{#1}%
\providecommand \citenamefont [1]{#1}%
\providecommand \href@noop [0]{\@secondoftwo}%
\providecommand \href [0]{\begingroup \@sanitize@url \@href}%
\providecommand \@href[1]{\@@startlink{#1}\@@href}%
\providecommand \@@href[1]{\endgroup#1\@@endlink}%
\providecommand \@sanitize@url [0]{\catcode `\\12\catcode `\$12\catcode `\&12\catcode `\#12\catcode `\^12\catcode `\_12\catcode `\%12\relax}%
\providecommand \@@startlink[1]{}%
\providecommand \@@endlink[0]{}%
\providecommand \url  [0]{\begingroup\@sanitize@url \@url }%
\providecommand \@url [1]{\endgroup\@href {#1}{\urlprefix }}%
\providecommand \urlprefix  [0]{URL }%
\providecommand \Eprint [0]{\href }%
\providecommand \doibase [0]{https://doi.org/}%
\providecommand \selectlanguage [0]{\@gobble}%
\providecommand \bibinfo  [0]{\@secondoftwo}%
\providecommand \bibfield  [0]{\@secondoftwo}%
\providecommand \translation [1]{[#1]}%
\providecommand \BibitemOpen [0]{}%
\providecommand \bibitemStop [0]{}%
\providecommand \bibitemNoStop [0]{.\EOS\space}%
\providecommand \EOS [0]{\spacefactor3000\relax}%
\providecommand \BibitemShut  [1]{\csname bibitem#1\endcsname}%
\let\auto@bib@innerbib\@empty
\bibitem [{\citenamefont {Savary}\ and\ \citenamefont {Balents}(2016)}]{Savary_2016}%
  \BibitemOpen
  \bibfield  {author} {\bibinfo {author} {\bibfnamefont {L.}~\bibnamefont {Savary}}\ and\ \bibinfo {author} {\bibfnamefont {L.}~\bibnamefont {Balents}},\ }\bibfield  {title} {\bibinfo {title} {Quantum spin liquids: a review},\ }\href {https://doi.org/10.1088/0034-4885/80/1/016502} {\bibfield  {journal} {\bibinfo  {journal} {Reports on Progress in Physics}\ }\textbf {\bibinfo {volume} {80}},\ \bibinfo {pages} {016502} (\bibinfo {year} {2016})}\BibitemShut {NoStop}%
\bibitem [{\citenamefont {Broholm}\ \emph {et~al.}(2020)\citenamefont {Broholm}, \citenamefont {Cava}, \citenamefont {Kivelson}, \citenamefont {Nocera}, \citenamefont {Norman},\ and\ \citenamefont {Senthil}}]{Broholm_2020}%
  \BibitemOpen
  \bibfield  {author} {\bibinfo {author} {\bibfnamefont {C.}~\bibnamefont {Broholm}}, \bibinfo {author} {\bibfnamefont {R.~J.}\ \bibnamefont {Cava}}, \bibinfo {author} {\bibfnamefont {S.~A.}\ \bibnamefont {Kivelson}}, \bibinfo {author} {\bibfnamefont {D.~G.}\ \bibnamefont {Nocera}}, \bibinfo {author} {\bibfnamefont {M.~R.}\ \bibnamefont {Norman}},\ and\ \bibinfo {author} {\bibfnamefont {T.}~\bibnamefont {Senthil}},\ }\bibfield  {title} {\bibinfo {title} {Quantum spin liquids},\ }\href {https://doi.org/10.1126/science.aay0668} {\bibfield  {journal} {\bibinfo  {journal} {Science}\ }\textbf {\bibinfo {volume} {367}},\ \bibinfo {pages} {eaay0668} (\bibinfo {year} {2020})},\ \Eprint {https://arxiv.org/abs/https://www.science.org/doi/pdf/10.1126/science.aay0668} {https://www.science.org/doi/pdf/10.1126/science.aay0668} \BibitemShut {NoStop}%
\bibitem [{\citenamefont {Lee}\ \emph {et~al.}(2006)\citenamefont {Lee}, \citenamefont {Nagaosa},\ and\ \citenamefont {Wen}}]{RevModPhys.78.17}%
  \BibitemOpen
  \bibfield  {author} {\bibinfo {author} {\bibfnamefont {P.~A.}\ \bibnamefont {Lee}}, \bibinfo {author} {\bibfnamefont {N.}~\bibnamefont {Nagaosa}},\ and\ \bibinfo {author} {\bibfnamefont {X.-G.}\ \bibnamefont {Wen}},\ }\bibfield  {title} {\bibinfo {title} {Doping a mott insulator: Physics of high-temperature superconductivity},\ }\href {https://doi.org/10.1103/RevModPhys.78.17} {\bibfield  {journal} {\bibinfo  {journal} {Rev. Mod. Phys.}\ }\textbf {\bibinfo {volume} {78}},\ \bibinfo {pages} {17} (\bibinfo {year} {2006})}\BibitemShut {NoStop}%
\bibitem [{\citenamefont {Wen}(2004)}]{wen2004quantum}%
  \BibitemOpen
  \bibfield  {author} {\bibinfo {author} {\bibfnamefont {X.-G.}\ \bibnamefont {Wen}},\ }\href@noop {} {\emph {\bibinfo {title} {Quantum Field Theory of Many-Body Systems: From the Origin of Sound to an Origin of Light and Electrons}}}\ (\bibinfo  {publisher} {Oxford University Press},\ \bibinfo {year} {2004})\BibitemShut {NoStop}%
\bibitem [{\citenamefont {Affleck}\ and\ \citenamefont {Marston}(1988)}]{PhysRevB.37.3774}%
  \BibitemOpen
  \bibfield  {author} {\bibinfo {author} {\bibfnamefont {I.}~\bibnamefont {Affleck}}\ and\ \bibinfo {author} {\bibfnamefont {J.~B.}\ \bibnamefont {Marston}},\ }\bibfield  {title} {\bibinfo {title} {Large-n limit of the heisenberg-hubbard model: Implications for high-${T}_{c}$ superconductors},\ }\href {https://doi.org/10.1103/PhysRevB.37.3774} {\bibfield  {journal} {\bibinfo  {journal} {Phys. Rev. B}\ }\textbf {\bibinfo {volume} {37}},\ \bibinfo {pages} {3774} (\bibinfo {year} {1988})}\BibitemShut {NoStop}%
\bibitem [{\citenamefont {Hermele}\ \emph {et~al.}(2005)\citenamefont {Hermele}, \citenamefont {Senthil},\ and\ \citenamefont {Fisher}}]{hermele_2005}%
  \BibitemOpen
  \bibfield  {author} {\bibinfo {author} {\bibfnamefont {M.}~\bibnamefont {Hermele}}, \bibinfo {author} {\bibfnamefont {T.}~\bibnamefont {Senthil}},\ and\ \bibinfo {author} {\bibfnamefont {M.~P.~A.}\ \bibnamefont {Fisher}},\ }\bibfield  {title} {\bibinfo {title} {Algebraic spin liquid as the mother of many competing orders},\ }\bibfield  {journal} {\bibinfo  {journal} {Physical Review B}\ }\textbf {\bibinfo {volume} {72}},\ \href {https://doi.org/10.1103/physrevb.72.104404} {10.1103/physrevb.72.104404} (\bibinfo {year} {2005})\BibitemShut {NoStop}%
\bibitem [{\citenamefont {Kalmeyer}\ and\ \citenamefont {Laughlin}(1987)}]{PhysRevLett.59.2095}%
  \BibitemOpen
  \bibfield  {author} {\bibinfo {author} {\bibfnamefont {V.}~\bibnamefont {Kalmeyer}}\ and\ \bibinfo {author} {\bibfnamefont {R.~B.}\ \bibnamefont {Laughlin}},\ }\bibfield  {title} {\bibinfo {title} {Equivalence of the resonating-valence-bond and fractional quantum hall states},\ }\href {https://doi.org/10.1103/PhysRevLett.59.2095} {\bibfield  {journal} {\bibinfo  {journal} {Phys. Rev. Lett.}\ }\textbf {\bibinfo {volume} {59}},\ \bibinfo {pages} {2095} (\bibinfo {year} {1987})}\BibitemShut {NoStop}%
\bibitem [{\citenamefont {Wen}(1991)}]{PhysRevB.44.2664}%
  \BibitemOpen
  \bibfield  {author} {\bibinfo {author} {\bibfnamefont {X.~G.}\ \bibnamefont {Wen}},\ }\bibfield  {title} {\bibinfo {title} {Mean-field theory of spin-liquid states with finite energy gap and topological orders},\ }\href {https://doi.org/10.1103/PhysRevB.44.2664} {\bibfield  {journal} {\bibinfo  {journal} {Phys. Rev. B}\ }\textbf {\bibinfo {volume} {44}},\ \bibinfo {pages} {2664} (\bibinfo {year} {1991})}\BibitemShut {NoStop}%
\bibitem [{\citenamefont {Hermele}\ \emph {et~al.}(2008)\citenamefont {Hermele}, \citenamefont {Ran}, \citenamefont {Lee},\ and\ \citenamefont {Wen}}]{hermele2008}%
  \BibitemOpen
  \bibfield  {author} {\bibinfo {author} {\bibfnamefont {M.}~\bibnamefont {Hermele}}, \bibinfo {author} {\bibfnamefont {Y.}~\bibnamefont {Ran}}, \bibinfo {author} {\bibfnamefont {P.~A.}\ \bibnamefont {Lee}},\ and\ \bibinfo {author} {\bibfnamefont {X.-G.}\ \bibnamefont {Wen}},\ }\bibfield  {title} {\bibinfo {title} {Properties of an algebraic spin liquid on the kagome lattice},\ }\href {https://doi.org/10.1103/PhysRevB.77.224413} {\bibfield  {journal} {\bibinfo  {journal} {Phys. Rev. B}\ }\textbf {\bibinfo {volume} {77}},\ \bibinfo {pages} {224413} (\bibinfo {year} {2008})}\BibitemShut {NoStop}%
\bibitem [{\citenamefont {Ran}\ and\ \citenamefont {gang Wen}(2006)}]{Ran_2006}%
  \BibitemOpen
  \bibfield  {author} {\bibinfo {author} {\bibfnamefont {Y.}~\bibnamefont {Ran}}\ and\ \bibinfo {author} {\bibfnamefont {X.}~\bibnamefont {gang Wen}},\ }\href {https://arxiv.org/abs/cond-mat/0609620} {\bibinfo {title} {Continuous quantum phase transitions beyond landau's paradigm in a large-n spin model}} (\bibinfo {year} {2006}),\ \Eprint {https://arxiv.org/abs/cond-mat/0609620} {arXiv:cond-mat/0609620 [cond-mat.str-el]} \BibitemShut {NoStop}%
\bibitem [{\citenamefont {Wen}\ \emph {et~al.}(1989)\citenamefont {Wen}, \citenamefont {Wilczek},\ and\ \citenamefont {Zee}}]{PhysRevB.39.11413}%
  \BibitemOpen
  \bibfield  {author} {\bibinfo {author} {\bibfnamefont {X.~G.}\ \bibnamefont {Wen}}, \bibinfo {author} {\bibfnamefont {F.}~\bibnamefont {Wilczek}},\ and\ \bibinfo {author} {\bibfnamefont {A.}~\bibnamefont {Zee}},\ }\bibfield  {title} {\bibinfo {title} {Chiral spin states and superconductivity},\ }\href {https://doi.org/10.1103/PhysRevB.39.11413} {\bibfield  {journal} {\bibinfo  {journal} {Phys. Rev. B}\ }\textbf {\bibinfo {volume} {39}},\ \bibinfo {pages} {11413} (\bibinfo {year} {1989})}\BibitemShut {NoStop}%
\bibitem [{\citenamefont {Affleck}(1985)}]{PhysRevLett.54.966}%
  \BibitemOpen
  \bibfield  {author} {\bibinfo {author} {\bibfnamefont {I.}~\bibnamefont {Affleck}},\ }\bibfield  {title} {\bibinfo {title} {Large-$n$ limit of $\mathrm{SU}(n)$ quantum "spin" chains},\ }\href {https://doi.org/10.1103/PhysRevLett.54.966} {\bibfield  {journal} {\bibinfo  {journal} {Phys. Rev. Lett.}\ }\textbf {\bibinfo {volume} {54}},\ \bibinfo {pages} {966} (\bibinfo {year} {1985})}\BibitemShut {NoStop}%
\bibitem [{\citenamefont {Bauer}\ \emph {et~al.}(2014)\citenamefont {Bauer}, \citenamefont {Cincio}, \citenamefont {Keller}, \citenamefont {Dolfi}, \citenamefont {Vidal}, \citenamefont {Trebst},\ and\ \citenamefont {Ludwig}}]{Bauer_2014}%
  \BibitemOpen
  \bibfield  {author} {\bibinfo {author} {\bibfnamefont {B.}~\bibnamefont {Bauer}}, \bibinfo {author} {\bibfnamefont {L.}~\bibnamefont {Cincio}}, \bibinfo {author} {\bibfnamefont {B.}~\bibnamefont {Keller}}, \bibinfo {author} {\bibfnamefont {M.}~\bibnamefont {Dolfi}}, \bibinfo {author} {\bibfnamefont {G.}~\bibnamefont {Vidal}}, \bibinfo {author} {\bibfnamefont {S.}~\bibnamefont {Trebst}},\ and\ \bibinfo {author} {\bibfnamefont {A.}~\bibnamefont {Ludwig}},\ }\bibfield  {title} {\bibinfo {title} {Chiral spin liquid and emergent anyons in a kagome lattice mott insulator},\ }\bibfield  {journal} {\bibinfo  {journal} {Nature Communications}\ }\textbf {\bibinfo {volume} {5}},\ \href {https://doi.org/10.1038/ncomms6137} {10.1038/ncomms6137} (\bibinfo {year} {2014})\BibitemShut {NoStop}%
\bibitem [{\citenamefont {Wen}(2002)}]{Wen_2002}%
  \BibitemOpen
  \bibfield  {author} {\bibinfo {author} {\bibfnamefont {X.-G.}\ \bibnamefont {Wen}},\ }\bibfield  {title} {\bibinfo {title} {Quantum orders and symmetric spin liquids},\ }\bibfield  {journal} {\bibinfo  {journal} {Physical Review B}\ }\textbf {\bibinfo {volume} {65}},\ \href {https://doi.org/10.1103/physrevb.65.165113} {10.1103/physrevb.65.165113} (\bibinfo {year} {2002})\BibitemShut {NoStop}%
\bibitem [{\citenamefont {Zhang}\ \emph {et~al.}(1989)\citenamefont {Zhang}, \citenamefont {Hansson},\ and\ \citenamefont {Kivelson}}]{ZhangHanssonKivelson1989}%
  \BibitemOpen
  \bibfield  {author} {\bibinfo {author} {\bibfnamefont {S.~C.}\ \bibnamefont {Zhang}}, \bibinfo {author} {\bibfnamefont {T.~H.}\ \bibnamefont {Hansson}},\ and\ \bibinfo {author} {\bibfnamefont {S.}~\bibnamefont {Kivelson}},\ }\bibfield  {title} {\bibinfo {title} {Effective-field-theory model for the fractional quantum {Hall} effect},\ }\href {https://doi.org/10.1103/PhysRevLett.62.82} {\bibfield  {journal} {\bibinfo  {journal} {Phys. Rev. Lett.}\ }\textbf {\bibinfo {volume} {62}},\ \bibinfo {pages} {82} (\bibinfo {year} {1989})}\BibitemShut {NoStop}%
\bibitem [{\citenamefont {Read}(1989)}]{Read1989CompositeBoson}%
  \BibitemOpen
  \bibfield  {author} {\bibinfo {author} {\bibfnamefont {N.}~\bibnamefont {Read}},\ }\bibfield  {title} {\bibinfo {title} {Order parameter and {Ginzburg-Landau} theory for the fractional quantum {Hall} effect},\ }\href {https://doi.org/10.1103/PhysRevLett.62.86} {\bibfield  {journal} {\bibinfo  {journal} {Phys. Rev. Lett.}\ }\textbf {\bibinfo {volume} {62}},\ \bibinfo {pages} {86} (\bibinfo {year} {1989})}\BibitemShut {NoStop}%
\bibitem [{\citenamefont {Jain}(1989)}]{Jain1989Parton}%
  \BibitemOpen
  \bibfield  {author} {\bibinfo {author} {\bibfnamefont {J.~K.}\ \bibnamefont {Jain}},\ }\bibfield  {title} {\bibinfo {title} {Incompressible quantum {Hall} states},\ }\href {https://doi.org/10.1103/PhysRevB.40.8079} {\bibfield  {journal} {\bibinfo  {journal} {Phys. Rev. B}\ }\textbf {\bibinfo {volume} {40}},\ \bibinfo {pages} {8079} (\bibinfo {year} {1989})}\BibitemShut {NoStop}%
\bibitem [{\citenamefont {Halperin}\ \emph {et~al.}(1993)\citenamefont {Halperin}, \citenamefont {Lee},\ and\ \citenamefont {Read}}]{HalperinLeeRead1993}%
  \BibitemOpen
  \bibfield  {author} {\bibinfo {author} {\bibfnamefont {B.~I.}\ \bibnamefont {Halperin}}, \bibinfo {author} {\bibfnamefont {P.~A.}\ \bibnamefont {Lee}},\ and\ \bibinfo {author} {\bibfnamefont {N.}~\bibnamefont {Read}},\ }\bibfield  {title} {\bibinfo {title} {Theory of the half-filled {Landau} level},\ }\href {https://doi.org/10.1103/PhysRevB.47.7312} {\bibfield  {journal} {\bibinfo  {journal} {Phys. Rev. B}\ }\textbf {\bibinfo {volume} {47}},\ \bibinfo {pages} {7312} (\bibinfo {year} {1993})}\BibitemShut {NoStop}%
\bibitem [{\citenamefont {Gutzwiller}(1963)}]{PhysRevLett.10.159}%
  \BibitemOpen
  \bibfield  {author} {\bibinfo {author} {\bibfnamefont {M.~C.}\ \bibnamefont {Gutzwiller}},\ }\bibfield  {title} {\bibinfo {title} {Effect of correlation on the ferromagnetism of transition metals},\ }\href {https://doi.org/10.1103/PhysRevLett.10.159} {\bibfield  {journal} {\bibinfo  {journal} {Phys. Rev. Lett.}\ }\textbf {\bibinfo {volume} {10}},\ \bibinfo {pages} {159} (\bibinfo {year} {1963})}\BibitemShut {NoStop}%
\bibitem [{\citenamefont {Becca}\ and\ \citenamefont {Sorella}(2017)}]{Becca_Sorella_2017}%
  \BibitemOpen
  \bibfield  {author} {\bibinfo {author} {\bibfnamefont {F.}~\bibnamefont {Becca}}\ and\ \bibinfo {author} {\bibfnamefont {S.}~\bibnamefont {Sorella}},\ }\href@noop {} {\emph {\bibinfo {title} {Quantum Monte Carlo Approaches for Correlated Systems}}}\ (\bibinfo  {publisher} {Cambridge University Press},\ \bibinfo {year} {2017})\BibitemShut {NoStop}%
\bibitem [{\citenamefont {Paramekanti}\ and\ \citenamefont {Marston}(2007)}]{Paramekanti_2007}%
  \BibitemOpen
  \bibfield  {author} {\bibinfo {author} {\bibfnamefont {A.}~\bibnamefont {Paramekanti}}\ and\ \bibinfo {author} {\bibfnamefont {J.~B.}\ \bibnamefont {Marston}},\ }\bibfield  {title} {\bibinfo {title} {Su(n) quantum spin models: a variational wavefunction study},\ }\href {https://doi.org/10.1088/0953-8984/19/12/125215} {\bibfield  {journal} {\bibinfo  {journal} {Journal of Physics: Condensed Matter}\ }\textbf {\bibinfo {volume} {19}},\ \bibinfo {pages} {125215} (\bibinfo {year} {2007})}\BibitemShut {NoStop}%
\bibitem [{\citenamefont {Gros}(1988)}]{PhysRevB.38.931}%
  \BibitemOpen
  \bibfield  {author} {\bibinfo {author} {\bibfnamefont {C.}~\bibnamefont {Gros}},\ }\bibfield  {title} {\bibinfo {title} {Superconductivity in correlated wave functions},\ }\href {https://doi.org/10.1103/PhysRevB.38.931} {\bibfield  {journal} {\bibinfo  {journal} {Phys. Rev. B}\ }\textbf {\bibinfo {volume} {38}},\ \bibinfo {pages} {931} (\bibinfo {year} {1988})}\BibitemShut {NoStop}%
\bibitem [{\citenamefont {Shiba}\ and\ \citenamefont {Yokoyama}(1987)}]{SHIBA1987264}%
  \BibitemOpen
  \bibfield  {author} {\bibinfo {author} {\bibfnamefont {H.}~\bibnamefont {Shiba}}\ and\ \bibinfo {author} {\bibfnamefont {H.}~\bibnamefont {Yokoyama}},\ }\bibfield  {title} {\bibinfo {title} {Variational monte carlo studies of highly correlated electron systems},\ }\href {https://doi.org/https://doi.org/10.1016/0378-4363(87)90206-3} {\bibfield  {journal} {\bibinfo  {journal} {Physica B+C}\ }\textbf {\bibinfo {volume} {148}},\ \bibinfo {pages} {264} (\bibinfo {year} {1987})}\BibitemShut {NoStop}%
\bibitem [{\citenamefont {Ran}\ \emph {et~al.}(2007)\citenamefont {Ran}, \citenamefont {Hermele}, \citenamefont {Lee},\ and\ \citenamefont {Wen}}]{Ran_2007}%
  \BibitemOpen
  \bibfield  {author} {\bibinfo {author} {\bibfnamefont {Y.}~\bibnamefont {Ran}}, \bibinfo {author} {\bibfnamefont {M.}~\bibnamefont {Hermele}}, \bibinfo {author} {\bibfnamefont {P.~A.}\ \bibnamefont {Lee}},\ and\ \bibinfo {author} {\bibfnamefont {X.-G.}\ \bibnamefont {Wen}},\ }\bibfield  {title} {\bibinfo {title} {Projected-wave-function study of the spin-1/2 heisenberg model on the kagomé lattice},\ }\bibfield  {journal} {\bibinfo  {journal} {Physical Review Letters}\ }\textbf {\bibinfo {volume} {98}},\ \href {https://doi.org/10.1103/physrevlett.98.117205} {10.1103/physrevlett.98.117205} (\bibinfo {year} {2007})\BibitemShut {NoStop}%
\bibitem [{\citenamefont {Iqbal}\ \emph {et~al.}(2013)\citenamefont {Iqbal}, \citenamefont {Becca}, \citenamefont {Sorella},\ and\ \citenamefont {Poilblanc}}]{Iqbal_2013}%
  \BibitemOpen
  \bibfield  {author} {\bibinfo {author} {\bibfnamefont {Y.}~\bibnamefont {Iqbal}}, \bibinfo {author} {\bibfnamefont {F.}~\bibnamefont {Becca}}, \bibinfo {author} {\bibfnamefont {S.}~\bibnamefont {Sorella}},\ and\ \bibinfo {author} {\bibfnamefont {D.}~\bibnamefont {Poilblanc}},\ }\bibfield  {title} {\bibinfo {title} {Gapless spin-liquid phase in the kagome spin-1/2 heisenberg antiferromagnet},\ }\bibfield  {journal} {\bibinfo  {journal} {Physical Review B}\ }\textbf {\bibinfo {volume} {87}},\ \href {https://doi.org/10.1103/physrevb.87.060405} {10.1103/physrevb.87.060405} (\bibinfo {year} {2013})\BibitemShut {NoStop}%
\bibitem [{\citenamefont {Hu}\ \emph {et~al.}(2013)\citenamefont {Hu}, \citenamefont {Becca}, \citenamefont {Parola},\ and\ \citenamefont {Sorella}}]{PhysRevB.88.060402}%
  \BibitemOpen
  \bibfield  {author} {\bibinfo {author} {\bibfnamefont {W.-J.}\ \bibnamefont {Hu}}, \bibinfo {author} {\bibfnamefont {F.}~\bibnamefont {Becca}}, \bibinfo {author} {\bibfnamefont {A.}~\bibnamefont {Parola}},\ and\ \bibinfo {author} {\bibfnamefont {S.}~\bibnamefont {Sorella}},\ }\bibfield  {title} {\bibinfo {title} {Direct evidence for a gapless ${Z}_{2}$ spin liquid by frustrating n\'eel antiferromagnetism},\ }\href {https://doi.org/10.1103/PhysRevB.88.060402} {\bibfield  {journal} {\bibinfo  {journal} {Phys. Rev. B}\ }\textbf {\bibinfo {volume} {88}},\ \bibinfo {pages} {060402} (\bibinfo {year} {2013})}\BibitemShut {NoStop}%
\bibitem [{\citenamefont {Chen}\ \emph {et~al.}(2013)\citenamefont {Chen}, \citenamefont {Gu}, \citenamefont {Liu},\ and\ \citenamefont {Wen}}]{Chen_2013}%
  \BibitemOpen
  \bibfield  {author} {\bibinfo {author} {\bibfnamefont {X.}~\bibnamefont {Chen}}, \bibinfo {author} {\bibfnamefont {Z.-C.}\ \bibnamefont {Gu}}, \bibinfo {author} {\bibfnamefont {Z.-X.}\ \bibnamefont {Liu}},\ and\ \bibinfo {author} {\bibfnamefont {X.-G.}\ \bibnamefont {Wen}},\ }\bibfield  {title} {\bibinfo {title} {Symmetry protected topological orders and the group cohomology of their symmetry group},\ }\bibfield  {journal} {\bibinfo  {journal} {Physical Review B}\ }\textbf {\bibinfo {volume} {87}},\ \href {https://doi.org/10.1103/physrevb.87.155114} {10.1103/physrevb.87.155114} (\bibinfo {year} {2013})\BibitemShut {NoStop}%
\bibitem [{\citenamefont {Garratt}\ \emph {et~al.}(2023)\citenamefont {Garratt}, \citenamefont {Weinstein},\ and\ \citenamefont {Altman}}]{Garratt2023}%
  \BibitemOpen
  \bibfield  {author} {\bibinfo {author} {\bibfnamefont {S.~J.}\ \bibnamefont {Garratt}}, \bibinfo {author} {\bibfnamefont {Z.}~\bibnamefont {Weinstein}},\ and\ \bibinfo {author} {\bibfnamefont {E.}~\bibnamefont {Altman}},\ }\bibfield  {title} {\bibinfo {title} {Measurements conspire nonlocally to restructure critical quantum states},\ }\href {https://doi.org/10.1103/PhysRevX.13.021026} {\bibfield  {journal} {\bibinfo  {journal} {Phys. Rev. X}\ }\textbf {\bibinfo {volume} {13}},\ \bibinfo {pages} {021026} (\bibinfo {year} {2023})}\BibitemShut {NoStop}%
\bibitem [{\citenamefont {{Weinstein}}\ \emph {et~al.}(2023)\citenamefont {{Weinstein}}, \citenamefont {{Sajith}}, \citenamefont {{Altman}},\ and\ \citenamefont {{Garratt}}}]{Weinstein2023}%
  \BibitemOpen
  \bibfield  {author} {\bibinfo {author} {\bibfnamefont {Z.}~\bibnamefont {{Weinstein}}}, \bibinfo {author} {\bibfnamefont {R.}~\bibnamefont {{Sajith}}}, \bibinfo {author} {\bibfnamefont {E.}~\bibnamefont {{Altman}}},\ and\ \bibinfo {author} {\bibfnamefont {S.~J.}\ \bibnamefont {{Garratt}}},\ }\bibfield  {title} {\bibinfo {title} {{Nonlocality and entanglement in measured critical quantum Ising chains}},\ }\href {https://doi.org/10.1103/PhysRevB.107.245132} {\bibfield  {journal} {\bibinfo  {journal} {\prb}\ }\textbf {\bibinfo {volume} {107}},\ \bibinfo {eid} {245132} (\bibinfo {year} {2023})},\ \Eprint {https://arxiv.org/abs/2301.08268} {arXiv:2301.08268 [cond-mat.stat-mech]} \BibitemShut {NoStop}%
\bibitem [{\citenamefont {Lee}\ \emph {et~al.}(2023)\citenamefont {Lee}, \citenamefont {Jian},\ and\ \citenamefont {Xu}}]{Lee_2023}%
  \BibitemOpen
  \bibfield  {author} {\bibinfo {author} {\bibfnamefont {J.~Y.}\ \bibnamefont {Lee}}, \bibinfo {author} {\bibfnamefont {C.-M.}\ \bibnamefont {Jian}},\ and\ \bibinfo {author} {\bibfnamefont {C.}~\bibnamefont {Xu}},\ }\bibfield  {title} {\bibinfo {title} {Quantum criticality under decoherence or weak measurement},\ }\href@noop {} {\bibfield  {journal} {\bibinfo  {journal} {PRX quantum}\ }\textbf {\bibinfo {volume} {4}},\ \bibinfo {pages} {030317} (\bibinfo {year} {2023})}\BibitemShut {NoStop}%
\bibitem [{\citenamefont {Su}\ \emph {et~al.}(2024)\citenamefont {Su}, \citenamefont {Myerson-Jain},\ and\ \citenamefont {Xu}}]{Su_2024}%
  \BibitemOpen
  \bibfield  {author} {\bibinfo {author} {\bibfnamefont {K.}~\bibnamefont {Su}}, \bibinfo {author} {\bibfnamefont {N.}~\bibnamefont {Myerson-Jain}},\ and\ \bibinfo {author} {\bibfnamefont {C.}~\bibnamefont {Xu}},\ }\bibfield  {title} {\bibinfo {title} {{Conformal field theories generated by Chern insulators under decoherence or measurement}},\ }\href {https://doi.org/10.1103/PhysRevB.109.035146} {\bibfield  {journal} {\bibinfo  {journal} {Phys. Rev. B}\ }\textbf {\bibinfo {volume} {109}},\ \bibinfo {pages} {035146} (\bibinfo {year} {2024})},\ \Eprint {https://arxiv.org/abs/2305.13410} {arXiv:2305.13410 [cond-mat.str-el]} \BibitemShut {NoStop}%
\bibitem [{\citenamefont {Sarma}\ \emph {et~al.}(2025)\citenamefont {Sarma}, \citenamefont {Bao}, \citenamefont {Myerson-Jain}, \citenamefont {Kiely},\ and\ \citenamefont {Xu}}]{Sarma_2025}%
  \BibitemOpen
  \bibfield  {author} {\bibinfo {author} {\bibfnamefont {A.}~\bibnamefont {Sarma}}, \bibinfo {author} {\bibfnamefont {Y.}~\bibnamefont {Bao}}, \bibinfo {author} {\bibfnamefont {N.}~\bibnamefont {Myerson-Jain}}, \bibinfo {author} {\bibfnamefont {T.}~\bibnamefont {Kiely}},\ and\ \bibinfo {author} {\bibfnamefont {C.}~\bibnamefont {Xu}},\ }\bibfield  {title} {\bibinfo {title} {Effective conformal field theory generated from pure and dephased chern insulator},\ }\href {https://doi.org/10.1103/nn2m-w4vk} {\bibfield  {journal} {\bibinfo  {journal} {Phys. Rev. B}\ }\textbf {\bibinfo {volume} {112}},\ \bibinfo {pages} {085130} (\bibinfo {year} {2025})}\BibitemShut {NoStop}%
\bibitem [{\citenamefont {Yang}\ \emph {et~al.}(2023)\citenamefont {Yang}, \citenamefont {Mao},\ and\ \citenamefont {Jian}}]{yang2023}%
  \BibitemOpen
  \bibfield  {author} {\bibinfo {author} {\bibfnamefont {Z.}~\bibnamefont {Yang}}, \bibinfo {author} {\bibfnamefont {D.}~\bibnamefont {Mao}},\ and\ \bibinfo {author} {\bibfnamefont {C.-M.}\ \bibnamefont {Jian}},\ }\bibfield  {title} {\bibinfo {title} {Entanglement in a one-dimensional critical state after measurements},\ }\href {https://doi.org/10.1103/PhysRevB.108.165120} {\bibfield  {journal} {\bibinfo  {journal} {Phys. Rev. B}\ }\textbf {\bibinfo {volume} {108}},\ \bibinfo {pages} {165120} (\bibinfo {year} {2023})}\BibitemShut {NoStop}%
\bibitem [{\citenamefont {Hauser}\ \emph {et~al.}(2026)\citenamefont {Hauser}, \citenamefont {Su}, \citenamefont {Ha}, \citenamefont {Lloyd}, \citenamefont {Kiely}, \citenamefont {Vasseur}, \citenamefont {Gopalakrishnan}, \citenamefont {Xu},\ and\ \citenamefont {Fisher}}]{hauser2026}%
  \BibitemOpen
  \bibfield  {author} {\bibinfo {author} {\bibfnamefont {J.}~\bibnamefont {Hauser}}, \bibinfo {author} {\bibfnamefont {K.}~\bibnamefont {Su}}, \bibinfo {author} {\bibfnamefont {H.}~\bibnamefont {Ha}}, \bibinfo {author} {\bibfnamefont {J.}~\bibnamefont {Lloyd}}, \bibinfo {author} {\bibfnamefont {T.~G.}\ \bibnamefont {Kiely}}, \bibinfo {author} {\bibfnamefont {R.}~\bibnamefont {Vasseur}}, \bibinfo {author} {\bibfnamefont {S.}~\bibnamefont {Gopalakrishnan}}, \bibinfo {author} {\bibfnamefont {C.}~\bibnamefont {Xu}},\ and\ \bibinfo {author} {\bibfnamefont {M.~P.~A.}\ \bibnamefont {Fisher}},\ }\href {https://arxiv.org/abs/2602.16045} {\bibinfo {title} {Strong-to-weak symmetry breaking in open quantum systems: From discrete particles to continuum hydrodynamics}} (\bibinfo {year} {2026}),\ \Eprint {https://arxiv.org/abs/2602.16045} {arXiv:2602.16045 [quant-ph]} \BibitemShut {NoStop}%
\bibitem [{\citenamefont {Lu}\ \emph {et~al.}(2022)\citenamefont {Lu}, \citenamefont {Lessa}, \citenamefont {Kim},\ and\ \citenamefont {Hsieh}}]{lu2022}%
  \BibitemOpen
  \bibfield  {author} {\bibinfo {author} {\bibfnamefont {T.-C.}\ \bibnamefont {Lu}}, \bibinfo {author} {\bibfnamefont {L.~A.}\ \bibnamefont {Lessa}}, \bibinfo {author} {\bibfnamefont {I.~H.}\ \bibnamefont {Kim}},\ and\ \bibinfo {author} {\bibfnamefont {T.~H.}\ \bibnamefont {Hsieh}},\ }\bibfield  {title} {\bibinfo {title} {Measurement as a shortcut to long-range entangled quantum matter},\ }\href {https://doi.org/10.1103/PRXQuantum.3.040337} {\bibfield  {journal} {\bibinfo  {journal} {PRX Quantum}\ }\textbf {\bibinfo {volume} {3}},\ \bibinfo {pages} {040337} (\bibinfo {year} {2022})}\BibitemShut {NoStop}%
\bibitem [{\citenamefont {Verresen}\ \emph {et~al.}(2022)\citenamefont {Verresen}, \citenamefont {Tantivasadakarn},\ and\ \citenamefont {Vishwanath}}]{verresen2022}%
  \BibitemOpen
  \bibfield  {author} {\bibinfo {author} {\bibfnamefont {R.}~\bibnamefont {Verresen}}, \bibinfo {author} {\bibfnamefont {N.}~\bibnamefont {Tantivasadakarn}},\ and\ \bibinfo {author} {\bibfnamefont {A.}~\bibnamefont {Vishwanath}},\ }\href {https://arxiv.org/abs/2112.03061} {\bibinfo {title} {Efficiently preparing schr\"odinger's cat, fractons and non-abelian topological order in quantum devices}} (\bibinfo {year} {2022}),\ \Eprint {https://arxiv.org/abs/2112.03061} {arXiv:2112.03061 [quant-ph]} \BibitemShut {NoStop}%
\bibitem [{\citenamefont {Lee}\ \emph {et~al.}(2022)\citenamefont {Lee}, \citenamefont {Ji}, \citenamefont {Bi},\ and\ \citenamefont {Fisher}}]{lee2022}%
  \BibitemOpen
  \bibfield  {author} {\bibinfo {author} {\bibfnamefont {J.~Y.}\ \bibnamefont {Lee}}, \bibinfo {author} {\bibfnamefont {W.}~\bibnamefont {Ji}}, \bibinfo {author} {\bibfnamefont {Z.}~\bibnamefont {Bi}},\ and\ \bibinfo {author} {\bibfnamefont {M.~P.~A.}\ \bibnamefont {Fisher}},\ }\href {https://arxiv.org/abs/2208.11699} {\bibinfo {title} {Decoding measurement-prepared quantum phases and transitions: from ising model to gauge theory, and beyond}} (\bibinfo {year} {2022}),\ \Eprint {https://arxiv.org/abs/2208.11699} {arXiv:2208.11699 [cond-mat.str-el]} \BibitemShut {NoStop}%
\bibitem [{\citenamefont {Tantivasadakarn}\ \emph {et~al.}(2024)\citenamefont {Tantivasadakarn}, \citenamefont {Thorngren}, \citenamefont {Vishwanath},\ and\ \citenamefont {Verresen}}]{Tantivasadakarn_2024}%
  \BibitemOpen
  \bibfield  {author} {\bibinfo {author} {\bibfnamefont {N.}~\bibnamefont {Tantivasadakarn}}, \bibinfo {author} {\bibfnamefont {R.}~\bibnamefont {Thorngren}}, \bibinfo {author} {\bibfnamefont {A.}~\bibnamefont {Vishwanath}},\ and\ \bibinfo {author} {\bibfnamefont {R.}~\bibnamefont {Verresen}},\ }\bibfield  {title} {\bibinfo {title} {Long-range entanglement from measuring symmetry-protected topological phases},\ }\href {https://doi.org/10.1103/PhysRevX.14.021040} {\bibfield  {journal} {\bibinfo  {journal} {Phys. Rev. X}\ }\textbf {\bibinfo {volume} {14}},\ \bibinfo {pages} {021040} (\bibinfo {year} {2024})}\BibitemShut {NoStop}%
\bibitem [{\citenamefont {Lu}\ \emph {et~al.}(2026)\citenamefont {Lu}, \citenamefont {Silveira},\ and\ \citenamefont {You}}]{lu2026}%
  \BibitemOpen
  \bibfield  {author} {\bibinfo {author} {\bibfnamefont {S.}~\bibnamefont {Lu}}, \bibinfo {author} {\bibfnamefont {L.~Q.}\ \bibnamefont {Silveira}},\ and\ \bibinfo {author} {\bibfnamefont {Y.}~\bibnamefont {You}},\ }\href {https://arxiv.org/abs/2602.23484} {\bibinfo {title} {Nonequilibrium topological response under charge dephasing}} (\bibinfo {year} {2026}),\ \Eprint {https://arxiv.org/abs/2602.23484} {arXiv:2602.23484 [cond-mat.str-el]} \BibitemShut {NoStop}%
\bibitem [{\citenamefont {Su}\ \emph {et~al.}(2025)\citenamefont {Su}, \citenamefont {Sarma}, \citenamefont {Bintz}, \citenamefont {Kiely}, \citenamefont {Bao}, \citenamefont {Fisher},\ and\ \citenamefont {Xu}}]{Su_2025}%
  \BibitemOpen
  \bibfield  {author} {\bibinfo {author} {\bibfnamefont {K.}~\bibnamefont {Su}}, \bibinfo {author} {\bibfnamefont {A.}~\bibnamefont {Sarma}}, \bibinfo {author} {\bibfnamefont {M.}~\bibnamefont {Bintz}}, \bibinfo {author} {\bibfnamefont {T.}~\bibnamefont {Kiely}}, \bibinfo {author} {\bibfnamefont {Y.}~\bibnamefont {Bao}}, \bibinfo {author} {\bibfnamefont {M.~P.}\ \bibnamefont {Fisher}},\ and\ \bibinfo {author} {\bibfnamefont {C.}~\bibnamefont {Xu}},\ }\bibfield  {title} {\bibinfo {title} {Spin liquid and superconductivity emerging from steady states and measurements},\ }\bibfield  {journal} {\bibinfo  {journal} {Physical Review Letters}\ }\textbf {\bibinfo {volume} {135}},\ \href {https://doi.org/10.1103/pjs3-14cc} {10.1103/pjs3-14cc} (\bibinfo {year} {2025})\BibitemShut {NoStop}%
\bibitem [{\citenamefont {Sarma}\ and\ \citenamefont {Xu}(2026)}]{sarma2026inequalitystrongweakspontaneoussymmetry}%
  \BibitemOpen
  \bibfield  {author} {\bibinfo {author} {\bibfnamefont {A.}~\bibnamefont {Sarma}}\ and\ \bibinfo {author} {\bibfnamefont {C.}~\bibnamefont {Xu}},\ }\href {https://arxiv.org/abs/2603.24671} {\bibinfo {title} {Inequality for strong-weak spontaneous symmetry breaking in fermionic open quantum systems}} (\bibinfo {year} {2026}),\ \Eprint {https://arxiv.org/abs/2603.24671} {arXiv:2603.24671 [quant-ph]} \BibitemShut {NoStop}%
\bibitem [{\citenamefont {Wang}\ \emph {et~al.}(2026)\citenamefont {Wang}, \citenamefont {Kiely}, \citenamefont {Tell}, \citenamefont {Obermeyer}, \citenamefont {Barendregt}, \citenamefont {Bojović}, \citenamefont {Preiss}, \citenamefont {Sarma}, \citenamefont {Franz}, \citenamefont {Fisher}, \citenamefont {Xu},\ and\ \citenamefont {Bloch}}]{wang2026}%
  \BibitemOpen
  \bibfield  {author} {\bibinfo {author} {\bibfnamefont {S.}~\bibnamefont {Wang}}, \bibinfo {author} {\bibfnamefont {T.~G.}\ \bibnamefont {Kiely}}, \bibinfo {author} {\bibfnamefont {D.}~\bibnamefont {Tell}}, \bibinfo {author} {\bibfnamefont {J.}~\bibnamefont {Obermeyer}}, \bibinfo {author} {\bibfnamefont {M.}~\bibnamefont {Barendregt}}, \bibinfo {author} {\bibfnamefont {P.}~\bibnamefont {Bojović}}, \bibinfo {author} {\bibfnamefont {P.~M.}\ \bibnamefont {Preiss}}, \bibinfo {author} {\bibfnamefont {A.}~\bibnamefont {Sarma}}, \bibinfo {author} {\bibfnamefont {T.}~\bibnamefont {Franz}}, \bibinfo {author} {\bibfnamefont {M.~P.~A.}\ \bibnamefont {Fisher}}, \bibinfo {author} {\bibfnamefont {C.}~\bibnamefont {Xu}},\ and\ \bibinfo {author} {\bibfnamefont {I.}~\bibnamefont {Bloch}},\ }\href {https://arxiv.org/abs/2604.16137} {\bibinfo {title} {Observation of strong-to-weak spontaneous symmetry breaking in a dephased fermi gas}} (\bibinfo {year} {2026}),\ \Eprint {https://arxiv.org/abs/2604.16137} {arXiv:2604.16137
  [cond-mat.quant-gas]} \BibitemShut {NoStop}%
\bibitem [{\citenamefont {Sarma}(2026)}]{sarma2026strongtoweakspontaneoussymmetrybreaking}%
  \BibitemOpen
  \bibfield  {author} {\bibinfo {author} {\bibfnamefont {A.}~\bibnamefont {Sarma}},\ }\href {https://arxiv.org/abs/2609.37965} {\bibinfo {title} {Strong-to-weak spontaneous symmetry breaking of dephased fermions}} (\bibinfo {year} {2026}),\ \Eprint {https://arxiv.org/abs/2609.37965} {arXiv:2609.37965 [cond-mat.str-el]} \BibitemShut {NoStop}%
\bibitem [{\citenamefont {You}\ \emph {et~al.}(2014)\citenamefont {You}, \citenamefont {Bi}, \citenamefont {Rasmussen}, \citenamefont {Slagle},\ and\ \citenamefont {Xu}}]{You2014StrangeCorrelator}%
  \BibitemOpen
  \bibfield  {author} {\bibinfo {author} {\bibfnamefont {Y.-Z.}\ \bibnamefont {You}}, \bibinfo {author} {\bibfnamefont {Z.}~\bibnamefont {Bi}}, \bibinfo {author} {\bibfnamefont {A.}~\bibnamefont {Rasmussen}}, \bibinfo {author} {\bibfnamefont {K.}~\bibnamefont {Slagle}},\ and\ \bibinfo {author} {\bibfnamefont {C.}~\bibnamefont {Xu}},\ }\bibfield  {title} {\bibinfo {title} {Wave function and strange correlator of short-range entangled states},\ }\href {https://doi.org/10.1103/PhysRevLett.112.247202} {\bibfield  {journal} {\bibinfo  {journal} {Phys. Rev. Lett.}\ }\textbf {\bibinfo {volume} {112}},\ \bibinfo {pages} {247202} (\bibinfo {year} {2014})},\ \Eprint {https://arxiv.org/abs/1312.0626} {arXiv:1312.0626 [cond-mat.str-el]} \BibitemShut {NoStop}%
\bibitem [{Note1()}]{Note1}%
  \BibitemOpen
  \bibinfo {note} {When $d=2$ the dual fields are also compact bosons.}\BibitemShut {Stop}%
\bibitem [{Note2()}]{Note2}%
  \BibitemOpen
  \bibinfo {note} {To be precise, the $U(1)_N$ and $SU(N)_1$ WZW models separately have operators with fractional conformal spin $s=h-\protect \bar {h}$. The physical operators of the theory are obtained by gluing $U(1)_N$ and $SU(N)_1$ fields such that the composite has half-integer spin $s\in \protect \frac {1}{2}\protect \mathbb {Z}$, in accordance with a local fermionic theory.}\BibitemShut {Stop}%
\bibitem [{Note3()}]{Note3}%
  \BibitemOpen
  \bibinfo {note} {More precisely, each boundary is a direct sum $\DOTSB \bigoplus@ \slimits@ _{\alpha } \protect \text {Dir}(\alpha )$ of Dirichlet boundaries with $\phi _c=\alpha $.}\BibitemShut {Stop}%
\bibitem [{\citenamefont {Popov}\ and\ \citenamefont {Wang}(2025)}]{popov2025}%
  \BibitemOpen
  \bibfield  {author} {\bibinfo {author} {\bibfnamefont {F.~K.}\ \bibnamefont {Popov}}\ and\ \bibinfo {author} {\bibfnamefont {Y.}~\bibnamefont {Wang}},\ }\href {https://arxiv.org/abs/2504.06203} {\bibinfo {title} {Factorizing defects from generalized pinning fields}} (\bibinfo {year} {2025}),\ \Eprint {https://arxiv.org/abs/2504.06203} {arXiv:2504.06203 [hep-th]} \BibitemShut {NoStop}%
\bibitem [{\citenamefont {Cardy}(1989)}]{Cardy_bcft}%
  \BibitemOpen
  \bibfield  {author} {\bibinfo {author} {\bibfnamefont {J.~L.}\ \bibnamefont {Cardy}},\ }\bibfield  {title} {\bibinfo {title} {Boundary conditions, fusion rules and the verlinde formula},\ }\href {https://doi.org/https://doi.org/10.1016/0550-3213(89)90521-X} {\bibfield  {journal} {\bibinfo  {journal} {Nuclear Physics B}\ }\textbf {\bibinfo {volume} {324}},\ \bibinfo {pages} {581} (\bibinfo {year} {1989})}\BibitemShut {NoStop}%
\bibitem [{\citenamefont {Ivanov}\ and\ \citenamefont {Senthil}(2002)}]{IvanovSenthil2002}%
  \BibitemOpen
  \bibfield  {author} {\bibinfo {author} {\bibfnamefont {D.~A.}\ \bibnamefont {Ivanov}}\ and\ \bibinfo {author} {\bibfnamefont {T.}~\bibnamefont {Senthil}},\ }\bibfield  {title} {\bibinfo {title} {Projected wave functions for fractionalized phases of quantum spin systems},\ }\href {https://doi.org/10.1103/PhysRevB.66.115111} {\bibfield  {journal} {\bibinfo  {journal} {Phys. Rev. B}\ }\textbf {\bibinfo {volume} {66}},\ \bibinfo {pages} {115111} (\bibinfo {year} {2002})}\BibitemShut {NoStop}%
\bibitem [{\citenamefont {Bravyi}\ \emph {et~al.}(2010)\citenamefont {Bravyi}, \citenamefont {Hastings},\ and\ \citenamefont {Michalakis}}]{BravyiHastingsMichalakis2010}%
  \BibitemOpen
  \bibfield  {author} {\bibinfo {author} {\bibfnamefont {S.}~\bibnamefont {Bravyi}}, \bibinfo {author} {\bibfnamefont {M.~B.}\ \bibnamefont {Hastings}},\ and\ \bibinfo {author} {\bibfnamefont {S.}~\bibnamefont {Michalakis}},\ }\bibfield  {title} {\bibinfo {title} {Topological quantum order: Stability under local perturbations},\ }\href {https://doi.org/10.1063/1.3490195} {\bibfield  {journal} {\bibinfo  {journal} {J. Math. Phys.}\ }\textbf {\bibinfo {volume} {51}},\ \bibinfo {pages} {093512} (\bibinfo {year} {2010})}\BibitemShut {NoStop}%
\bibitem [{Note4()}]{Note4}%
  \BibitemOpen
  \bibinfo {note} {We assume that each $|\psi _0^{ij}\rangle $ is gapped and satisfies $(-1)^{\protect \hat {Q}}|\psi _0^{ij}\rangle =|\psi _0^{ij}\rangle $ for the total charge $\protect \hat {Q}$. A sector with the opposite parity is annihilated by $\protect \hat {P}$.}\BibitemShut {Stop}%
\bibitem [{\citenamefont {Laughlin}(1983)}]{Laughlin1983}%
  \BibitemOpen
  \bibfield  {author} {\bibinfo {author} {\bibfnamefont {R.~B.}\ \bibnamefont {Laughlin}},\ }\bibfield  {title} {\bibinfo {title} {Anomalous quantum {Hall} effect: An incompressible quantum fluid with fractionally charged excitations},\ }\href {https://doi.org/10.1103/PhysRevLett.50.1395} {\bibfield  {journal} {\bibinfo  {journal} {Phys. Rev. Lett.}\ }\textbf {\bibinfo {volume} {50}},\ \bibinfo {pages} {1395} (\bibinfo {year} {1983})}\BibitemShut {NoStop}%
\bibitem [{\citenamefont {Zhang}\ \emph {et~al.}(2011)\citenamefont {Zhang}, \citenamefont {Grover},\ and\ \citenamefont {Vishwanath}}]{Zhang_2011}%
  \BibitemOpen
  \bibfield  {author} {\bibinfo {author} {\bibfnamefont {Y.}~\bibnamefont {Zhang}}, \bibinfo {author} {\bibfnamefont {T.}~\bibnamefont {Grover}},\ and\ \bibinfo {author} {\bibfnamefont {A.}~\bibnamefont {Vishwanath}},\ }\bibfield  {title} {\bibinfo {title} {Topological entanglement entropy of ${\mathbb{z}}_{2}$ spin liquids and lattice laughlin states},\ }\href {https://doi.org/10.1103/PhysRevB.84.075128} {\bibfield  {journal} {\bibinfo  {journal} {Phys. Rev. B}\ }\textbf {\bibinfo {volume} {84}},\ \bibinfo {pages} {075128} (\bibinfo {year} {2011})}\BibitemShut {NoStop}%
\bibitem [{\citenamefont {Zhang}\ and\ \citenamefont {Vishwanath}(2013)}]{Zhang_2013}%
  \BibitemOpen
  \bibfield  {author} {\bibinfo {author} {\bibfnamefont {Y.}~\bibnamefont {Zhang}}\ and\ \bibinfo {author} {\bibfnamefont {A.}~\bibnamefont {Vishwanath}},\ }\bibfield  {title} {\bibinfo {title} {Establishing non-abelian topological order in gutzwiller-projected chern insulators via entanglement entropy and modular $\mathcal{S}$-matrix},\ }\href {https://doi.org/10.1103/PhysRevB.87.161113} {\bibfield  {journal} {\bibinfo  {journal} {Phys. Rev. B}\ }\textbf {\bibinfo {volume} {87}},\ \bibinfo {pages} {161113} (\bibinfo {year} {2013})}\BibitemShut {NoStop}%
\bibitem [{\citenamefont {Haldane}\ and\ \citenamefont {Rezayi}(1985)}]{HaldaneRezayi1985}%
  \BibitemOpen
  \bibfield  {author} {\bibinfo {author} {\bibfnamefont {F.~D.~M.}\ \bibnamefont {Haldane}}\ and\ \bibinfo {author} {\bibfnamefont {E.~H.}\ \bibnamefont {Rezayi}},\ }\bibfield  {title} {\bibinfo {title} {Periodic {Laughlin-Jastrow} wave functions for the fractional quantized {Hall} effect},\ }\href {https://doi.org/10.1103/PhysRevB.31.2529} {\bibfield  {journal} {\bibinfo  {journal} {Phys. Rev. B}\ }\textbf {\bibinfo {volume} {31}},\ \bibinfo {pages} {2529} (\bibinfo {year} {1985})}\BibitemShut {NoStop}%
\bibitem [{\citenamefont {Zhang}\ \emph {et~al.}(2024)\citenamefont {Zhang}, \citenamefont {Li},\ and\ \citenamefont {Lu}}]{zhang2024}%
  \BibitemOpen
  \bibfield  {author} {\bibinfo {author} {\bibfnamefont {Z.}~\bibnamefont {Zhang}}, \bibinfo {author} {\bibfnamefont {Y.}~\bibnamefont {Li}},\ and\ \bibinfo {author} {\bibfnamefont {T.-C.}\ \bibnamefont {Lu}},\ }\href {https://arxiv.org/abs/2411.05004} {\bibinfo {title} {Long-range entanglement from spontaneous non-onsite symmetry breaking}} (\bibinfo {year} {2024}),\ \Eprint {https://arxiv.org/abs/2411.05004} {arXiv:2411.05004 [cond-mat.str-el]} \BibitemShut {NoStop}%
\bibitem [{Note5()}]{Note5}%
  \BibitemOpen
  \bibinfo {note} {The doubled space representation of the dephased quantum Hall state is exactly the quantum spin Hall state with an attractive density-density interaction between the left and right space supported on a temporal defect, which becomes the Gutzwiller projection in the infinite dephasing limit. The doubled space representation before dephasing is a tensor product state, and therefore $v_2=0$.}\BibitemShut {Stop}%
\bibitem [{\citenamefont {Girvin}\ and\ \citenamefont {MacDonald}(1987)}]{Girvin_1987}%
  \BibitemOpen
  \bibfield  {author} {\bibinfo {author} {\bibfnamefont {S.~M.}\ \bibnamefont {Girvin}}\ and\ \bibinfo {author} {\bibfnamefont {A.~H.}\ \bibnamefont {MacDonald}},\ }\bibfield  {title} {\bibinfo {title} {Off-diagonal long-range order, oblique confinement, and the fractional quantum hall effect},\ }\href {https://doi.org/10.1103/PhysRevLett.58.1252} {\bibfield  {journal} {\bibinfo  {journal} {Phys. Rev. Lett.}\ }\textbf {\bibinfo {volume} {58}},\ \bibinfo {pages} {1252} (\bibinfo {year} {1987})}\BibitemShut {NoStop}%
\bibitem [{\citenamefont {Kane}\ \emph {et~al.}(1991)\citenamefont {Kane}, \citenamefont {Kivelson}, \citenamefont {Lee},\ and\ \citenamefont {Zhang}}]{Kane_1991}%
  \BibitemOpen
  \bibfield  {author} {\bibinfo {author} {\bibfnamefont {C.~L.}\ \bibnamefont {Kane}}, \bibinfo {author} {\bibfnamefont {S.}~\bibnamefont {Kivelson}}, \bibinfo {author} {\bibfnamefont {D.~H.}\ \bibnamefont {Lee}},\ and\ \bibinfo {author} {\bibfnamefont {S.~C.}\ \bibnamefont {Zhang}},\ }\bibfield  {title} {\bibinfo {title} {General validity of jastrow-laughlin wave functions},\ }\href {https://doi.org/10.1103/PhysRevB.43.3255} {\bibfield  {journal} {\bibinfo  {journal} {Phys. Rev. B}\ }\textbf {\bibinfo {volume} {43}},\ \bibinfo {pages} {3255} (\bibinfo {year} {1991})}\BibitemShut {NoStop}%
\bibitem [{\citenamefont {Lu}\ \emph {et~al.}(2023)\citenamefont {Lu}, \citenamefont {Zhang}, \citenamefont {Vijay},\ and\ \citenamefont {Hsieh}}]{Lu_2023}%
  \BibitemOpen
  \bibfield  {author} {\bibinfo {author} {\bibfnamefont {T.-C.}\ \bibnamefont {Lu}}, \bibinfo {author} {\bibfnamefont {Z.}~\bibnamefont {Zhang}}, \bibinfo {author} {\bibfnamefont {S.}~\bibnamefont {Vijay}},\ and\ \bibinfo {author} {\bibfnamefont {T.~H.}\ \bibnamefont {Hsieh}},\ }\bibfield  {title} {\bibinfo {title} {Mixed-state long-range order and criticality from measurement and feedback},\ }\bibfield  {journal} {\bibinfo  {journal} {PRX Quantum}\ }\textbf {\bibinfo {volume} {4}},\ \href {https://doi.org/10.1103/prxquantum.4.030318} {10.1103/prxquantum.4.030318} (\bibinfo {year} {2023})\BibitemShut {NoStop}%
\bibitem [{\citenamefont {Wang}\ \emph {et~al.}(2025)\citenamefont {Wang}, \citenamefont {Fan}, \citenamefont {Wang}, \citenamefont {Garratt},\ and\ \citenamefont {Altman}}]{wang2025fractionalquantumhallstates}%
  \BibitemOpen
  \bibfield  {author} {\bibinfo {author} {\bibfnamefont {Z.}~\bibnamefont {Wang}}, \bibinfo {author} {\bibfnamefont {R.}~\bibnamefont {Fan}}, \bibinfo {author} {\bibfnamefont {T.}~\bibnamefont {Wang}}, \bibinfo {author} {\bibfnamefont {S.~J.}\ \bibnamefont {Garratt}},\ and\ \bibinfo {author} {\bibfnamefont {E.}~\bibnamefont {Altman}},\ }\href {https://arxiv.org/abs/2510.08490} {\bibinfo {title} {Fractional quantum hall states under density decoherence}} (\bibinfo {year} {2025}),\ \Eprint {https://arxiv.org/abs/2510.08490} {arXiv:2510.08490 [cond-mat.str-el]} \BibitemShut {NoStop}%
\bibitem [{\citenamefont {Tay}\ and\ \citenamefont {Motrunich}(2011)}]{Tay_2011}%
  \BibitemOpen
  \bibfield  {author} {\bibinfo {author} {\bibfnamefont {T.}~\bibnamefont {Tay}}\ and\ \bibinfo {author} {\bibfnamefont {O.~I.}\ \bibnamefont {Motrunich}},\ }\bibfield  {title} {\bibinfo {title} {Failure of gutzwiller-type wave function to capture gauge fluctuations: Case study in the exciton bose liquid context},\ }\bibfield  {journal} {\bibinfo  {journal} {Physical Review B}\ }\textbf {\bibinfo {volume} {83}},\ \href {https://doi.org/10.1103/physrevb.83.235122} {10.1103/physrevb.83.235122} (\bibinfo {year} {2011})\BibitemShut {NoStop}%
\bibitem [{\citenamefont {Paramekanti}\ \emph {et~al.}(2002)\citenamefont {Paramekanti}, \citenamefont {Balents},\ and\ \citenamefont {Fisher}}]{Paramekanti_2002}%
  \BibitemOpen
  \bibfield  {author} {\bibinfo {author} {\bibfnamefont {A.}~\bibnamefont {Paramekanti}}, \bibinfo {author} {\bibfnamefont {L.}~\bibnamefont {Balents}},\ and\ \bibinfo {author} {\bibfnamefont {M.~P.~A.}\ \bibnamefont {Fisher}},\ }\bibfield  {title} {\bibinfo {title} {Ring exchange, the exciton bose liquid, and bosonization in two dimensions},\ }\bibfield  {journal} {\bibinfo  {journal} {Physical Review B}\ }\textbf {\bibinfo {volume} {66}},\ \href {https://doi.org/10.1103/physrevb.66.054526} {10.1103/physrevb.66.054526} (\bibinfo {year} {2002})\BibitemShut {NoStop}%
\bibitem [{\citenamefont {Nayak}\ and\ \citenamefont {Wilczek}(1994)}]{Nayak_1994}%
  \BibitemOpen
  \bibfield  {author} {\bibinfo {author} {\bibfnamefont {C.}~\bibnamefont {Nayak}}\ and\ \bibinfo {author} {\bibfnamefont {F.}~\bibnamefont {Wilczek}},\ }\bibfield  {title} {\bibinfo {title} {Non-fermi liquid fixed point in 2 + 1 dimensions},\ }\href {https://doi.org/10.1016/0550-3213(94)90477-4} {\bibfield  {journal} {\bibinfo  {journal} {Nuclear Physics B}\ }\textbf {\bibinfo {volume} {417}},\ \bibinfo {pages} {359–373} (\bibinfo {year} {1994})}\BibitemShut {NoStop}%
\bibitem [{\citenamefont {Polchinski}(1994)}]{Polchinski_1994}%
  \BibitemOpen
  \bibfield  {author} {\bibinfo {author} {\bibfnamefont {J.}~\bibnamefont {Polchinski}},\ }\bibfield  {title} {\bibinfo {title} {Low-energy dynamics of the spinon-gauge system},\ }\href {https://doi.org/10.1016/0550-3213(94)90449-9} {\bibfield  {journal} {\bibinfo  {journal} {Nuclear Physics B}\ }\textbf {\bibinfo {volume} {422}},\ \bibinfo {pages} {617–633} (\bibinfo {year} {1994})}\BibitemShut {NoStop}%
\bibitem [{\citenamefont {Nave}\ \emph {et~al.}(2007)\citenamefont {Nave}, \citenamefont {Lee},\ and\ \citenamefont {Lee}}]{Nave_2007}%
  \BibitemOpen
  \bibfield  {author} {\bibinfo {author} {\bibfnamefont {C.~P.}\ \bibnamefont {Nave}}, \bibinfo {author} {\bibfnamefont {S.-S.}\ \bibnamefont {Lee}},\ and\ \bibinfo {author} {\bibfnamefont {P.~A.}\ \bibnamefont {Lee}},\ }\bibfield  {title} {\bibinfo {title} {Susceptibility of a spinon fermi surface coupled to a u(1) gauge field},\ }\bibfield  {journal} {\bibinfo  {journal} {Physical Review B}\ }\textbf {\bibinfo {volume} {76}},\ \href {https://doi.org/10.1103/physrevb.76.165104} {10.1103/physrevb.76.165104} (\bibinfo {year} {2007})\BibitemShut {NoStop}%
\bibitem [{\citenamefont {Lee}(2009)}]{Lee_2009}%
  \BibitemOpen
  \bibfield  {author} {\bibinfo {author} {\bibfnamefont {S.-S.}\ \bibnamefont {Lee}},\ }\bibfield  {title} {\bibinfo {title} {Low-energy effective theory of fermi surface coupled with u(1) gauge field in 2+1 dimensions},\ }\bibfield  {journal} {\bibinfo  {journal} {Physical Review B}\ }\textbf {\bibinfo {volume} {80}},\ \href {https://doi.org/10.1103/physrevb.80.165102} {10.1103/physrevb.80.165102} (\bibinfo {year} {2009})\BibitemShut {NoStop}%
\bibitem [{\citenamefont {Anakru}\ \emph {et~al.}(2025)\citenamefont {Anakru}, \citenamefont {Gattu}, \citenamefont {Balram}, \citenamefont {Wu}, \citenamefont {Kumar}, \citenamefont {Bi},\ and\ \citenamefont {Jain}}]{Anakru_2025}%
  \BibitemOpen
  \bibfield  {author} {\bibinfo {author} {\bibfnamefont {A.}~\bibnamefont {Anakru}}, \bibinfo {author} {\bibfnamefont {M.}~\bibnamefont {Gattu}}, \bibinfo {author} {\bibfnamefont {A.~C.}\ \bibnamefont {Balram}}, \bibinfo {author} {\bibfnamefont {X.-C.}\ \bibnamefont {Wu}}, \bibinfo {author} {\bibfnamefont {P.}~\bibnamefont {Kumar}}, \bibinfo {author} {\bibfnamefont {Z.}~\bibnamefont {Bi}},\ and\ \bibinfo {author} {\bibfnamefont {J.~K.}\ \bibnamefont {Jain}},\ }\bibfield  {title} {\bibinfo {title} {Emergent gauge field in composite-fermion metals: A large-scale microscopic study},\ }\href {https://doi.org/10.1103/cjys-fx46} {\bibfield  {journal} {\bibinfo  {journal} {Phys. Rev. Lett.}\ }\textbf {\bibinfo {volume} {135}},\ \bibinfo {pages} {246503} (\bibinfo {year} {2025})}\BibitemShut {NoStop}%
\bibitem [{\citenamefont {Motrunich}(2005)}]{Motrunich_2005}%
  \BibitemOpen
  \bibfield  {author} {\bibinfo {author} {\bibfnamefont {O.~I.}\ \bibnamefont {Motrunich}},\ }\bibfield  {title} {\bibinfo {title} {Variational study of triangular lattice spin-$1/2$ model with ring exchanges and spin liquid state in {$\kappa$-(ET)$_2$Cu$_2$(CN)$_3$}},\ }\href {https://doi.org/10.1103/PhysRevB.72.045105} {\bibfield  {journal} {\bibinfo  {journal} {Phys. Rev. B}\ }\textbf {\bibinfo {volume} {72}},\ \bibinfo {pages} {045105} (\bibinfo {year} {2005})}\BibitemShut {NoStop}%
\bibitem [{\citenamefont {Li}(2013)}]{Li2013UniformRVB}%
  \BibitemOpen
  \bibfield  {author} {\bibinfo {author} {\bibfnamefont {T.}~\bibnamefont {Li}},\ }\bibfield  {title} {\bibinfo {title} {Antiferromagnetic long-range order in the uniform resonating valence bond state on the square lattice},\ }\href {https://doi.org/10.1209/0295-5075/103/57002} {\bibfield  {journal} {\bibinfo  {journal} {EPL (Europhysics Letters)}\ }\textbf {\bibinfo {volume} {103}},\ \bibinfo {pages} {57002} (\bibinfo {year} {2013})},\ \Eprint {https://arxiv.org/abs/1101.0193} {arXiv:1101.0193 [cond-mat.str-el]} \BibitemShut {NoStop}%
\bibitem [{\citenamefont {{Song}}\ \emph {et~al.}(2019)\citenamefont {{Song}}, \citenamefont {{Wang}}, \citenamefont {{Vishwanath}},\ and\ \citenamefont {{He}}}]{Song2019}%
  \BibitemOpen
  \bibfield  {author} {\bibinfo {author} {\bibfnamefont {X.-Y.}\ \bibnamefont {{Song}}}, \bibinfo {author} {\bibfnamefont {C.}~\bibnamefont {{Wang}}}, \bibinfo {author} {\bibfnamefont {A.}~\bibnamefont {{Vishwanath}}},\ and\ \bibinfo {author} {\bibfnamefont {Y.-C.}\ \bibnamefont {{He}}},\ }\bibfield  {title} {\bibinfo {title} {{Unifying description of competing orders in two-dimensional quantum magnets}},\ }\href {https://doi.org/10.1038/s41467-019-11727-3} {\bibfield  {journal} {\bibinfo  {journal} {Nature Communications}\ }\textbf {\bibinfo {volume} {10}},\ \bibinfo {eid} {4254} (\bibinfo {year} {2019})},\ \Eprint {https://arxiv.org/abs/1811.11186} {arXiv:1811.11186 [cond-mat.str-el]} \BibitemShut {NoStop}%
\bibitem [{\citenamefont {Song}\ \emph {et~al.}(2020)\citenamefont {Song}, \citenamefont {He}, \citenamefont {Vishwanath},\ and\ \citenamefont {Wang}}]{Song2020}%
  \BibitemOpen
  \bibfield  {author} {\bibinfo {author} {\bibfnamefont {X.-Y.}\ \bibnamefont {Song}}, \bibinfo {author} {\bibfnamefont {Y.-C.}\ \bibnamefont {He}}, \bibinfo {author} {\bibfnamefont {A.}~\bibnamefont {Vishwanath}},\ and\ \bibinfo {author} {\bibfnamefont {C.}~\bibnamefont {Wang}},\ }\bibfield  {title} {\bibinfo {title} {From spinon band topology to the symmetry quantum numbers of monopoles in dirac spin liquids},\ }\href {https://doi.org/10.1103/PhysRevX.10.011033} {\bibfield  {journal} {\bibinfo  {journal} {Phys. Rev. X}\ }\textbf {\bibinfo {volume} {10}},\ \bibinfo {pages} {011033} (\bibinfo {year} {2020})}\BibitemShut {NoStop}%
\bibitem [{\citenamefont {He}\ \emph {et~al.}(2017)\citenamefont {He}, \citenamefont {Zaletel}, \citenamefont {Oshikawa},\ and\ \citenamefont {Pollmann}}]{KagomeDirac2017}%
  \BibitemOpen
  \bibfield  {author} {\bibinfo {author} {\bibfnamefont {Y.-C.}\ \bibnamefont {He}}, \bibinfo {author} {\bibfnamefont {M.~P.}\ \bibnamefont {Zaletel}}, \bibinfo {author} {\bibfnamefont {M.}~\bibnamefont {Oshikawa}},\ and\ \bibinfo {author} {\bibfnamefont {F.}~\bibnamefont {Pollmann}},\ }\bibfield  {title} {\bibinfo {title} {Signatures of dirac cones in a dmrg study of the kagome heisenberg model},\ }\href {https://doi.org/10.1103/PhysRevX.7.031020} {\bibfield  {journal} {\bibinfo  {journal} {Phys. Rev. X}\ }\textbf {\bibinfo {volume} {7}},\ \bibinfo {pages} {031020} (\bibinfo {year} {2017})}\BibitemShut {NoStop}%
\bibitem [{\citenamefont {Hu}\ \emph {et~al.}(2019)\citenamefont {Hu}, \citenamefont {Zhu}, \citenamefont {Eggert},\ and\ \citenamefont {He}}]{TriangularDirac2019}%
  \BibitemOpen
  \bibfield  {author} {\bibinfo {author} {\bibfnamefont {S.}~\bibnamefont {Hu}}, \bibinfo {author} {\bibfnamefont {W.}~\bibnamefont {Zhu}}, \bibinfo {author} {\bibfnamefont {S.}~\bibnamefont {Eggert}},\ and\ \bibinfo {author} {\bibfnamefont {Y.-C.}\ \bibnamefont {He}},\ }\bibfield  {title} {\bibinfo {title} {Dirac spin liquid on the spin-$1/2$ triangular heisenberg antiferromagnet},\ }\href {https://doi.org/10.1103/PhysRevLett.123.207203} {\bibfield  {journal} {\bibinfo  {journal} {Phys. Rev. Lett.}\ }\textbf {\bibinfo {volume} {123}},\ \bibinfo {pages} {207203} (\bibinfo {year} {2019})}\BibitemShut {NoStop}%
\bibitem [{\citenamefont {Wietek}\ \emph {et~al.}(2024)\citenamefont {Wietek}, \citenamefont {Capponi},\ and\ \citenamefont {L{\"a}uchli}}]{TriangularDirac2024}%
  \BibitemOpen
  \bibfield  {author} {\bibinfo {author} {\bibfnamefont {A.}~\bibnamefont {Wietek}}, \bibinfo {author} {\bibfnamefont {S.}~\bibnamefont {Capponi}},\ and\ \bibinfo {author} {\bibfnamefont {A.~M.}\ \bibnamefont {L{\"a}uchli}},\ }\bibfield  {title} {\bibinfo {title} {Quantum electrodynamics in $2+1$ dimensions as the organizing principle of a triangular lattice antiferromagnet},\ }\href {https://doi.org/10.1103/PhysRevX.14.021010} {\bibfield  {journal} {\bibinfo  {journal} {Phys. Rev. X}\ }\textbf {\bibinfo {volume} {14}},\ \bibinfo {pages} {021010} (\bibinfo {year} {2024})}\BibitemShut {NoStop}%
\bibitem [{\citenamefont {Li}\ \emph {et~al.}(2023)\citenamefont {Li}, \citenamefont {von Delft},\ and\ \citenamefont {Tu}}]{Li_2023}%
  \BibitemOpen
  \bibfield  {author} {\bibinfo {author} {\bibfnamefont {J.-W.}\ \bibnamefont {Li}}, \bibinfo {author} {\bibfnamefont {J.}~\bibnamefont {von Delft}},\ and\ \bibinfo {author} {\bibfnamefont {H.-H.}\ \bibnamefont {Tu}},\ }\bibfield  {title} {\bibinfo {title} {U(1)-symmetric gaussian fermionic projected entangled paired states and their gutzwiller projection},\ }\bibfield  {journal} {\bibinfo  {journal} {Physical Review B}\ }\textbf {\bibinfo {volume} {107}},\ \href {https://doi.org/10.1103/physrevb.107.085148} {10.1103/physrevb.107.085148} (\bibinfo {year} {2023})\BibitemShut {NoStop}%
\bibitem [{\citenamefont {Rogers}()}]{RogersOnTS}%
  \BibitemOpen
  \bibfield  {author} {\bibinfo {author} {\bibfnamefont {E.}~\bibnamefont {Rogers}},\ }\bibinfo {title} {On the su(2) theory of the t-j model}\BibitemShut {NoStop}%
\bibitem [{\citenamefont {Ferrari}\ \emph {et~al.}(2021)\citenamefont {Ferrari}, \citenamefont {Parola},\ and\ \citenamefont {Becca}}]{PhysRevB.103.195140}%
  \BibitemOpen
\bibfield  {title} {  }\bibfield  {author} {\bibinfo {author} {\bibfnamefont {F.}~\bibnamefont {Ferrari}}, \bibinfo {author} {\bibfnamefont {A.}~\bibnamefont {Parola}},\ and\ \bibinfo {author} {\bibfnamefont {F.}~\bibnamefont {Becca}},\ }\bibfield  {title} {\bibinfo {title} {Gapless spin liquids in disguise},\ }\href {https://doi.org/10.1103/PhysRevB.103.195140} {\bibfield  {journal} {\bibinfo  {journal} {Phys. Rev. B}\ }\textbf {\bibinfo {volume} {103}},\ \bibinfo {pages} {195140} (\bibinfo {year} {2021})}\BibitemShut {NoStop}%
\bibitem [{\citenamefont {Wei}\ and\ \citenamefont {Zheng}(2026)}]{wei2026}%
  \BibitemOpen
  \bibfield  {author} {\bibinfo {author} {\bibfnamefont {P.}~\bibnamefont {Wei}}\ and\ \bibinfo {author} {\bibfnamefont {Y.}~\bibnamefont {Zheng}},\ }\href {https://arxiv.org/abs/2605.19363} {\bibinfo {title} {Non-invertible symmetries in weyl fermions, and applications to fermion-boundary scattering problem}} (\bibinfo {year} {2026}),\ \Eprint {https://arxiv.org/abs/2605.19363} {arXiv:2605.19363 [hep-th]} \BibitemShut {NoStop}%
\bibitem [{\citenamefont {Ueda}\ \emph {et~al.}(2026)\citenamefont {Ueda}, \citenamefont {Linden}, \citenamefont {Lootens}, \citenamefont {Haegeman}, \citenamefont {Fendley},\ and\ \citenamefont {Verstraete}}]{ueda2026}%
  \BibitemOpen
  \bibfield  {author} {\bibinfo {author} {\bibfnamefont {A.}~\bibnamefont {Ueda}}, \bibinfo {author} {\bibfnamefont {V.~V.}\ \bibnamefont {Linden}}, \bibinfo {author} {\bibfnamefont {L.}~\bibnamefont {Lootens}}, \bibinfo {author} {\bibfnamefont {J.}~\bibnamefont {Haegeman}}, \bibinfo {author} {\bibfnamefont {P.}~\bibnamefont {Fendley}},\ and\ \bibinfo {author} {\bibfnamefont {F.}~\bibnamefont {Verstraete}},\ }\href {https://arxiv.org/abs/2510.26780} {\bibinfo {title} {Perfect particle transmission through duality defects}} (\bibinfo {year} {2026}),\ \Eprint {https://arxiv.org/abs/2510.26780} {arXiv:2510.26780 [hep-th]} \BibitemShut {NoStop}%
\bibitem [{Note6()}]{Note6}%
  \BibitemOpen
  \bibinfo {note} {We slightly abuse notation, using $A$ to refer both to the subgroup of $G$ as well as the direct sum $\DOTSB \bigoplus@ \slimits@ _{a \in A} U_a$ of all $A$ defects.}\BibitemShut {Stop}%
\bibitem [{\citenamefont {Choi}\ \emph {et~al.}(2022)\citenamefont {Choi}, \citenamefont {Córdova}, \citenamefont {Hsin}, \citenamefont {Lam},\ and\ \citenamefont {Shao}}]{Choi_2022}%
  \BibitemOpen
  \bibfield  {author} {\bibinfo {author} {\bibfnamefont {Y.}~\bibnamefont {Choi}}, \bibinfo {author} {\bibfnamefont {C.}~\bibnamefont {Córdova}}, \bibinfo {author} {\bibfnamefont {P.-S.}\ \bibnamefont {Hsin}}, \bibinfo {author} {\bibfnamefont {H.~T.}\ \bibnamefont {Lam}},\ and\ \bibinfo {author} {\bibfnamefont {S.-H.}\ \bibnamefont {Shao}},\ }\bibfield  {title} {\bibinfo {title} {Noninvertible duality defects in 3+1 dimensions},\ }\bibfield  {journal} {\bibinfo  {journal} {Physical Review D}\ }\textbf {\bibinfo {volume} {105}},\ \href {https://doi.org/10.1103/physrevd.105.125016} {10.1103/physrevd.105.125016} (\bibinfo {year} {2022})\BibitemShut {NoStop}%
\bibitem [{\citenamefont {Kaidi}\ \emph {et~al.}(2022)\citenamefont {Kaidi}, \citenamefont {Ohmori},\ and\ \citenamefont {Zheng}}]{Kaidi_2022}%
  \BibitemOpen
  \bibfield  {author} {\bibinfo {author} {\bibfnamefont {J.}~\bibnamefont {Kaidi}}, \bibinfo {author} {\bibfnamefont {K.}~\bibnamefont {Ohmori}},\ and\ \bibinfo {author} {\bibfnamefont {Y.}~\bibnamefont {Zheng}},\ }\bibfield  {title} {\bibinfo {title} {Kramers-wannier-like duality defects in 3+1d gauge theories},\ }\bibfield  {journal} {\bibinfo  {journal} {Physical Review Letters}\ }\textbf {\bibinfo {volume} {128}},\ \href {https://doi.org/10.1103/physrevlett.128.111601} {10.1103/physrevlett.128.111601} (\bibinfo {year} {2022})\BibitemShut {NoStop}%
\bibitem [{\citenamefont {Choi}\ \emph {et~al.}(2023)\citenamefont {Choi}, \citenamefont {Lu},\ and\ \citenamefont {Sun}}]{choi2023}%
  \BibitemOpen
  \bibfield  {author} {\bibinfo {author} {\bibfnamefont {Y.}~\bibnamefont {Choi}}, \bibinfo {author} {\bibfnamefont {D.-C.}\ \bibnamefont {Lu}},\ and\ \bibinfo {author} {\bibfnamefont {Z.}~\bibnamefont {Sun}},\ }\href {https://arxiv.org/abs/2310.19867} {\bibinfo {title} {Self-duality under gauging a non-invertible symmetry}} (\bibinfo {year} {2023}),\ \Eprint {https://arxiv.org/abs/2310.19867} {arXiv:2310.19867 [hep-th]} \BibitemShut {NoStop}%
\bibitem [{\citenamefont {Antinucci}\ \emph {et~al.}(2023)\citenamefont {Antinucci}, \citenamefont {Benini}, \citenamefont {Copetti}, \citenamefont {Galati},\ and\ \citenamefont {Rizi}}]{antinucci2023}%
  \BibitemOpen
  \bibfield  {author} {\bibinfo {author} {\bibfnamefont {A.}~\bibnamefont {Antinucci}}, \bibinfo {author} {\bibfnamefont {F.}~\bibnamefont {Benini}}, \bibinfo {author} {\bibfnamefont {C.}~\bibnamefont {Copetti}}, \bibinfo {author} {\bibfnamefont {G.}~\bibnamefont {Galati}},\ and\ \bibinfo {author} {\bibfnamefont {G.}~\bibnamefont {Rizi}},\ }\href {https://arxiv.org/abs/2308.11707} {\bibinfo {title} {Anomalies of non-invertible self-duality symmetries: fractionalization and gauging}} (\bibinfo {year} {2023}),\ \Eprint {https://arxiv.org/abs/2308.11707} {arXiv:2308.11707 [hep-th]} \BibitemShut {NoStop}%
\bibitem [{\citenamefont {van Beest}\ \emph {et~al.}(2023)\citenamefont {van Beest}, \citenamefont {Smith}, \citenamefont {Delmastro}, \citenamefont {Komargodski},\ and\ \citenamefont {Tong}}]{vanbeest2023}%
  \BibitemOpen
  \bibfield  {author} {\bibinfo {author} {\bibfnamefont {M.}~\bibnamefont {van Beest}}, \bibinfo {author} {\bibfnamefont {P.~B.}\ \bibnamefont {Smith}}, \bibinfo {author} {\bibfnamefont {D.}~\bibnamefont {Delmastro}}, \bibinfo {author} {\bibfnamefont {Z.}~\bibnamefont {Komargodski}},\ and\ \bibinfo {author} {\bibfnamefont {D.}~\bibnamefont {Tong}},\ }\href {https://arxiv.org/abs/2306.07318} {\bibinfo {title} {Monopoles, scattering, and generalized symmetries}} (\bibinfo {year} {2023}),\ \Eprint {https://arxiv.org/abs/2306.07318} {arXiv:2306.07318 [hep-th]} \BibitemShut {NoStop}%
\bibitem [{\citenamefont {Arias-Tamargo}\ \emph {et~al.}(2026)\citenamefont {Arias-Tamargo}, \citenamefont {Smith}, \citenamefont {Mouland},\ and\ \citenamefont {Hutt}}]{tamargo2026}%
  \BibitemOpen
  \bibfield  {author} {\bibinfo {author} {\bibfnamefont {G.}~\bibnamefont {Arias-Tamargo}}, \bibinfo {author} {\bibfnamefont {P.~B.}\ \bibnamefont {Smith}}, \bibinfo {author} {\bibfnamefont {R.}~\bibnamefont {Mouland}},\ and\ \bibinfo {author} {\bibfnamefont {M.~L. V.~C.}\ \bibnamefont {Hutt}},\ }\href {https://arxiv.org/abs/2605.13952} {\bibinfo {title} {Non-invertible symmetries and boundaries for two-dimensional fermions}} (\bibinfo {year} {2026}),\ \Eprint {https://arxiv.org/abs/2605.13952} {arXiv:2605.13952 [hep-th]} \BibitemShut {NoStop}%
\bibitem [{\citenamefont {Loladze}\ \emph {et~al.}(2026)\citenamefont {Loladze}, \citenamefont {Okui},\ and\ \citenamefont {Tong}}]{loladze2026}%
  \BibitemOpen
  \bibfield  {author} {\bibinfo {author} {\bibfnamefont {V.}~\bibnamefont {Loladze}}, \bibinfo {author} {\bibfnamefont {T.}~\bibnamefont {Okui}},\ and\ \bibinfo {author} {\bibfnamefont {D.}~\bibnamefont {Tong}},\ }\href {https://arxiv.org/abs/2508.21059} {\bibinfo {title} {Dynamics of the fermion-rotor system}} (\bibinfo {year} {2026}),\ \Eprint {https://arxiv.org/abs/2508.21059} {arXiv:2508.21059 [hep-th]} \BibitemShut {NoStop}%
\end{thebibliography}%

\appendix

\onecolumngrid

\section{Projected Dirac Fermion Calculations}
\label{app:dirac_cone}

Here we analyze the Gutzwiller projection of $N_f$ Dirac fermions by treating the $U(1)$ gauge field perturbatively. We use Euclidean signature throughout, with $\{\gamma^\mu, \gamma^\nu\} = 2\delta^{\mu\nu}$. The bare fermion propagator is $S_0(k,\omega) = \frac{-i(\gamma^j k_j + \gamma^0 \omega)}{k^2+\omega^2}$. In the momentum-time representation, this becomes $S_0(k,\tau) = \int \frac{d\omega}{2\pi} S_0(k, \omega)e^{i\omega \tau} = \frac{1}{2}e^{-k|\tau|}[-i \frac{\gamma^j k_j}{|k|} + \gamma^\tau \sgn(\tau)]$, while at equal times $S_0(k,0) = \frac{-i \gamma^j k_j}{2|k|}$. Notice that the propagator has a discontinuity at $\tau=0$. This discontinuity is important as it regulates certain IR divergences that could otherwise arise. We take the convention that $\int_q = \int \frac{d^dq}{(2\pi)^d}$ for momenta and $\int_\tau = \int_{-\epsilon}^\epsilon d\tau$. The gauge self-energy is given to leading order by the one-loop fermion bubble, 
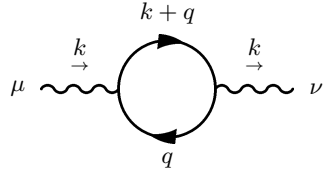
\begin{figure}[h]
\centering
\begin{fmffile}{diag_fermi}
\centering
\begin{fmfgraph*}(95,80)
       \fmfleft{i}
       \fmfright{o}
       \fmflabel{$\mu$}{i}
       \fmflabel{$\nu$}{o}
       \fmf{photon,tension=1,label={{$\underset{\rightarrow}{k}$}}}{v1,i}
       \fmf{photon,tension=1,label={{$\underset{\rightarrow}{k}$}}}{o,v2}
       \fmf{fermion,left,tension=0.4,label=$k+q$}{v1,v2}
       \fmf{fermion,left,tension=0.4,label=$q$}{v2,v1}
\end{fmfgraph*}
\end{fmffile}
\caption{Leading diagram contributing to $\Pi^{\mu\nu}(k)$.}
\end{figure}

\begin{align}
\begin{split}
\bar \Pi^{\mu\nu}(k) &= (-1)i^2 N_f \frac{1}{(2\epsilon)^2}\int_{q,\tau_1,\tau_2} \mathrm{Tr}[S_0(q+k,\tau_2-\tau_1) \gamma^\mu S_0(q,\tau_1-\tau_2) \gamma^\nu ]
\end{split}
\end{align}

The gauge field propagator is then given by Eq.~\ref{eq:gauge_prop}. We use the dressed propagator to compute the fermion self energy and dressed equal-time propagator

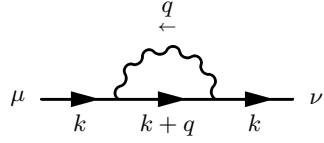
\begin{figure}[h]
\centering
\begin{fmffile}{onediagferm}
\centering
\begin{fmfgraph*}(95,80)
       \fmfleft{i}
       \fmfright{o}
       \fmflabel{$\mu$}{i}
       \fmflabel{$\nu$}{o}
       \fmf{fermion,tension=1,label=$k$}{i,v1}
       \fmf{fermion,tension=1,label=$k$}{v2,o}
       \fmf{fermion,tension=0.4,label=$k+q$}{v1,v2}
       \fmf{photon,left,tension=0.4,label=$\underset{\leftarrow}{q}$}{v1,v2}
\end{fmfgraph*}
\end{fmffile}
\caption{Leading diagram contributing to $\Sigma(k)$. }
\end{figure}
\begin{align}
\begin{split}
&\Sigma(k,\tau_1,\tau_2) =(i)^2\frac{1}{(2\epsilon)^2}\int_q \gamma^\mu S_0(k+q, \tau_2-\tau_1) \gamma^\nu \bar\Delta^{\mu\nu}(q, \tau_2, \tau_1) \\
&S(k,0) = S_0(k,0) + \int_{\tau_1, \tau_2} S_0(k, \tau_1) \Sigma(k, \tau_1, \tau_2) S_0(k, -\tau_2)\\
&\hspace{1.1cm} =S_0(k,0)\left[1 + 2\gamma_\psi \log (\frac{|k|}{\Lambda})\right] + \dots 
\end{split}
\end{align}
where we introduced a UV cutoff $\Lambda$ in the last line. This allows us to read off the anomalous dimension of the fermion from the loop integral. Lastly, we let $T^a$ be an $SU(N_f)$ generator and compute to leading order

\begin{figure}[h]
\centering
\begin{fmffile}{threeptrenorm}
\begin{minipage}{0.32\textwidth}
\centering
\begin{fmfgraph*}(100,90)
    \fmfstraight

    \fmfleft{i1,i2}
    \fmfright{o}

    \fmfforce{(0.00w,0.80h)}{i1}
    \fmfforce{(0.00w,0.20h)}{i2}
    \fmfforce{(0.28w,0.71h)}{v1}
    \fmfforce{(0.7w,0.58h)}{v3}
    \fmfforce{(1.00w,0.50h)}{o}

    \fmf{fermion,tension=1.2,label.side=left,label=$k$}{i1,v1}
    \fmf{fermion,tension=1.2,label.side=right,label=$k+q$}{v1,v3}
    \fmf{fermion,tension=1.2,label.side=left,label=$k$}{v3,o}
    \fmf{fermion,tension=1.1,label.side=left,label=$-k$}{o,i2}
    
    \fmf{photon,tension=0.28,label.side=left,left,label={$\underset{\leftarrow}{q}$}}{v1,v3}

    \fmfv{decor.shape=cross, decor.size=4thick}{i1}
    \fmfv{decor.shape=cross, decor.size=4thick}{i2}
    \fmfv{decor.shape=cross, decor.size=4thick}{o}
\end{fmfgraph*}
\end{minipage}
\begin{minipage}{0.32\textwidth}
\centering
\begin{fmfgraph*}(100,90)
    \fmfstraight

    \fmfleft{i1,i2}
    \fmfright{o}

    \fmfforce{(0.00w,0.80h)}{i1}
    \fmfforce{(0.00w,0.20h)}{i2}
    \fmfforce{(0.28w,0.29h)}{v1}
    \fmfforce{(0.7w,0.42h)}{v3}
    \fmfforce{(1.00w,0.50h)}{o}

    \fmf{fermion,tension=1.2,label.side=left,label=$k$}{i1,o}
    \fmf{fermion,tension=1.2,label.side=right,label=$-k+q$}{v3,v1}
    \fmf{fermion,tension=1.2,label.side=left,label=$-k$}{o,v3}
    \fmf{fermion,tension=1.1,label.side=left,label=$-k$}{v1,i2}
    
    \fmf{photon,tension=0.28,label.side=right,right,label={$\underset{\rightarrow}{q}$}}{v1,v3}

    \fmfv{decor.shape=cross, decor.size=4thick}{i1}
    \fmfv{decor.shape=cross, decor.size=4thick}{i2}
    \fmfv{decor.shape=cross, decor.size=4thick}{o}
\end{fmfgraph*}
\end{minipage}
\begin{minipage}{0.32\textwidth}
\centering
\begin{fmfgraph*}(100,90)
    \fmfstraight

    \fmfleft{i1,i2}
    \fmfright{o}

    \fmfforce{(0.00w,0.80h)}{i1}
    \fmfforce{(0.00w,0.20h)}{i2}
    \fmfforce{(0.28w,0.71h)}{v1}
    \fmfforce{(0.28w,0.29h)}{v3}
    \fmfforce{(1.00w,0.50h)}{o}

    \fmf{fermion,tension=1.2,label.side=left,label=$k$}{i1,v1}
    \fmf{fermion,tension=1.2,label.side=left,label=$-k$}{v3,i2}
    \fmf{fermion,tension=1.1,label.side=left,label=$k-q$}{v1,o}
    \fmf{fermion,tension=1.1,label.side=left,label=$-k-q$}{o,v3}
    
    \fmf{photon,tension=0.28,label.side=right,label={{$q \downarrow$}}}{v1,v3}

    \fmfv{decor.shape=cross, decor.size=4thick}{i1}
    \fmfv{decor.shape=cross, decor.size=4thick}{i2}
    \fmfv{decor.shape=cross, decor.size=4thick}{o}
\end{fmfgraph*}
\end{minipage}
\end{fmffile}
\caption{One-loop diagrams contributing to $\langle \psi(-k,0) (\bar{\psi} T^a \psi)(2k,0) \bar{\psi}(-k,0) \rangle$. }
\label{fig:three_diags}
\end{figure}
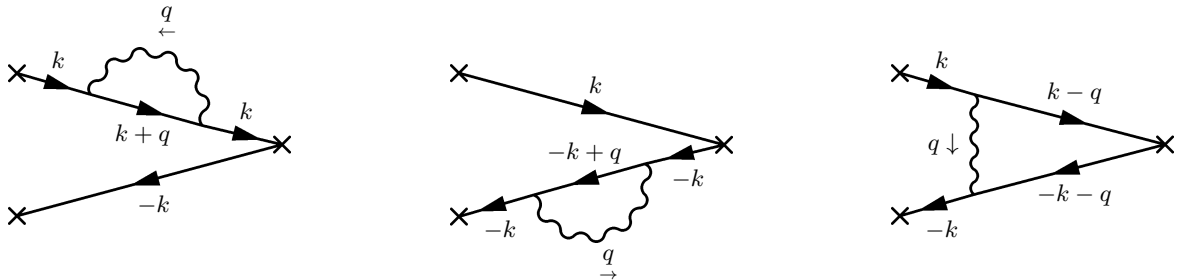

\begin{align}
\begin{split}
&\langle \psi(-k,0) (\bar{\psi} T^a \psi)(2k,0) \bar{\psi}(-k,0) \rangle \\
= S_0(k,0) T^a S_0(-k,0) &+\delta S(k, 0) T^a S_0(-k, 0) + S_0(k,0)T^a \delta S(-k, 0)\\
+ i^2 \int_q \int_{\tau_1, \tau_2} S_0(k, \tau_1) \gamma^\mu &S_0(k-q, -\tau_1) T^a S_0(-k-q, \tau_2) \gamma^\nu \Delta^{\mu\nu}(q, \tau_1, \tau_2) S_0(-k, -\tau_2) \\
= S_0(k) &T^a S_0(-k) (1 + (2\gamma_\psi + \gamma_{\bar{\psi} T^a \psi}) \ln(\frac{|k|}{\Lambda})) + \dots
\end{split}
\end{align}
where $\delta S(k,0) = S(k,0) - S_0(k,0)$. This allows us to read off the anomalous dimension of $\bar{\psi} T^a \psi$.

Here we show explicitly how this procedure works in $d=1$. We find that $\bar{\Pi}^{\mu\nu}(k) = \frac{N_f}{2\pi}|k|(2\delta^{\mu x}\delta^{\nu x} - \delta^{\mu\nu}) + O(\epsilon^3)$, giving $\bar\Delta^{\tau\tau}(k) = \frac{2\pi}{N_f |k|}$, $\bar\Delta^{xx}(k) = -\frac{2\pi}{N_f |k|}$. The corrected equal-time fermion propagator is 
\begin{align}
\begin{split}
S(k,0)&= S_0(k, 0)\left[1 + \frac{2\pi}{N_f} \int_q \frac{\Theta(-k (k+q))}{|q|}  + O(\epsilon)\right] \\
&=S_0(k,0) \left[1 - \frac{1}{N_f} \log (\frac{|k|}{\Lambda})\right]
\end{split}
\end{align}
giving $\gamma_\psi = -1/(2N_f)$. The vertex correction vanishes due to kinematic constraints and we find $\gamma_{\bar{\psi}T^a \psi} = 2\gamma_\psi = -1/N_f$, agreeing with the result from bosonization. 

In $d\geq2$ we find that there is no anomalous dimension generated and thus no correction to the free-fermion scaling deep in the IR. In $d=2$ the leading corrections to correlators scale as $\log \log (\frac{|k|}{\Lambda})$. 

\section{Projected Fermi Surface Calculations}
\label{app:proj_fermi}

We begin with the action describing the spinon Fermi surface with dispersion $E(k)$:

\begin{equation}
S_{0} = \int \frac{d^2k}{(2\pi)^2} \int_{-\infty}^\infty d\tau \sum_{i=1}^{N_f=2}\bar{\psi_{i}}(-k,\tau)[\partial_\tau + E(k)] \psi_{i}(k,\tau)
\end{equation}
For simplicity we set $E(k) = \frac{k^2}{2m} - \mu$. The fermion propagator is $G_0(k,\omega) = \frac{1}{-i\omega+E(k)}$. In momentum-time space, it becomes $G_0(k,\tau) = \int \frac{d\omega}{2\pi}e^{i\omega \tau}G_0(k,\omega) = e^{E(k) \tau}[\Theta(-\tau) \Theta(E(k)) - \Theta(\tau) \Theta(-E(k))]$, and at equal times $G_0(k,0) = \frac{1}{2}\sgn(E(k))$. Using the averaged gauge field description, the gauge-matter interaction is

\begin{align}
\begin{split}
S_{\mathrm{int}} = &- \frac{1}{2\epsilon}\int_{k,q}\int_{\tau}\left[i\bar{a}_\tau (-q) +  \frac{k_j+q_j/2}{m}\bar{a}_j(-q) \right]\bar{\psi}_{i}(k+q,\tau)\psi_i(-k,\tau) \\
&+ \frac{1}{2m(2\epsilon)^2} \int_{k,q_1,q_2}\int_\tau \bar{a}_j(-q_1)\cdot \bar{a}_j(-q_2) \bar{\psi}_{i}(k+q_1+q_2,\tau)\psi_i(-k,\tau)
\end{split}
\end{align}
In this section we focus on $d=2$. Then, to one-loop order the gauge self-energy is 

\begin{align}
\begin{split}
&\bar\Pi^{\tau\tau}(k) = \frac{N_f}{(2\epsilon)^2}\int_{q,\tau_1,\tau_2} G_0(q,\tau_2-\tau_1) G_0(k+q,\tau_1-\tau_2) = -\frac{N_f k_F}{2\pi^2}|k|\\
&\bar\Pi^{ij}(k) = \frac{N_F}{m(2\epsilon)^2} \delta^{ij}\int_{q,\tau} G_0(q,0) + \frac{N_f}{m^2(2\epsilon)^2}\int_{q,\tau_1,\tau_2} (q_i+\frac{k_i}{2})(q_j+\frac{k_j}{2})G_0(q,\tau_2-\tau_1)G_0(k+q,\tau_1-\tau_2)\\
&\hspace{1cm}= -\frac{N_f k_F^2}{8\pi m \epsilon}(P_T^{ij}+P_L^{ij}) - \frac{N_f k_F^3}{6\pi^2 m^2}(P_T^{ij} + 2P_L^{ij})|k|
\end{split}
\end{align}
to leading order in $|k|/k_F$. The corrected equal-time fermion propagator is

\begin{align}
\begin{split}
G(k,0) =G_0(k,0) &-\frac{1}{(2\epsilon)^2}\int_{\tau_1,\tau_2}\int_{q} G_0(k,\tau_1)G_0(k+q,\tau_2-\tau_1)\left[\bar\Delta^{\tau\tau}(q) + \frac{1}{m^2}(k_i + \frac{q_i}{2})(k_j + \frac{q_j}{2})\bar\Delta^{ij}(q)\right] G_0(k,-\tau_2)
\\ &- \frac{1}{m(2\epsilon)^2}\int_q \int_{\tau} G_0(k,\tau)\left[\sum_j\bar\Delta^{jj}(q)\right] G_0(k,-\tau)
\end{split}
\end{align}
We find that $\delta G(k,0)  =G(k,0)-G_0(k,0)$ is continuous with discontinuous derivative near $k_F$, scaling as $E(k) \log(\frac{E(k)}{E_0})$ (in addition to nonuniversal terms renormalizing $k_F$ and $m$) and thus only gives subleading corrections to correlation functions in the IR. For comparison, a genuine anomalous dimension would be sourced by a correction scaling as $\sgn(E(k)) \log(\frac{E(k)}{E_0})$. To analyze the behavior of fermion bilinears we must also compute the vertex correction:
\begin{figure}[h]
\centering
\begin{fmffile}{fermivertexrenorm}
\centering
\begin{fmfgraph*}(95,80)
    \fmfleft{i}
    \fmfright{o}

    \fmfforce{(0.00w,0.50h)}{i}
    \fmfforce{(1.00w,0.50h)}{o}
    \fmfforce{(0.50w,1.00h)}{v1}
    \fmfforce{(0.50w,0.00h)}{v2}

    \fmf{fermion,tension=1.2,label.side=left,label=$p+k$}{i,v1}
    \fmf{fermion,tension=1.2,label.side=left,label=$p+q+k$}{v1,o}
    \fmf{fermion,tension=1.2,label.side=left,label=$p+q$}{o,v2}
    \fmf{fermion,tension=1.2,label.side=left,label=$p$}{v2,i}
    
    \fmf{photon,tension=0.28,label.side=right,label={{$q \uparrow$}}}{v1,v2}

    \fmfv{decor.shape=cross, decor.size=6thick}{i}
    \fmfv{decor.shape=cross, decor.size=6thick}{o}
\end{fmfgraph*}
\end{fmffile}
\caption{Vertex correction for the spin-spin correlation function $\langle \vec{S}(k, \tau=0) \cdot \vec{S}(-k, \tau=0) \rangle$.}
\end{figure}
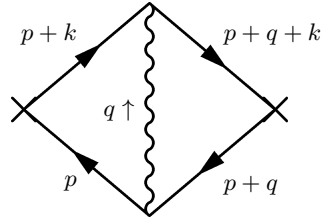

\begin{align}
\begin{split}
    V(k) = \frac{1}{(2\epsilon)^2}\int_{p,q,\tau_1,\tau_2}  &G_0(p+k,\tau_1) G_0(p+k+q,-\tau_1) G_0(p+q,\tau_2)G_0(p,-\tau_2)  \\
    \times&\left[\bar\Delta^{\tau\tau}(q) + \frac{1}{m^2}(p_i+k_i + \frac{q_i}{2})(p_j + \frac{q_j}{2})\bar{\Delta}^{ij}(q)\right]
\end{split}
\end{align}
Then, with $G(k) = \int_{\omega} G_0(k,\omega) + \int_{\omega_1, \omega_2} G_0(k,\omega_1) \Sigma(k,\omega_1,\omega_2) G_0(k,\omega_2) = G_0(k) + \frac{1}{4} \Sigma(k)$, we write $\delta G(k) = G(k) - G_0(k) = \frac{1}{4}\Sigma(k)$. The connected fermion bilinear correlation function becomes

\begin{align}
\begin{split}
    \chi(k) = \langle \vec{S}(k, 0) \cdot \vec{S}(-k, 0) \rangle_c =  -\text{tr}[\sigma^a \sigma^a] \left[\int_p G_0(p) G_0(p+k) + \int_p \delta G(p) G_0(p+k) + \int_p G_0(p) \delta G(p+k) - V(k)\right]
\end{split}
\end{align}

To understand the IR behavior we need only analyze the structure of the nonanalytic terms in $\chi(k)$ around $2k_F$, which can be understood via geometric arguments by analyzing the Fermi surface in patches. Defining $\delta = 2k_F-k$, the noninteracting part gives $\chi_0(k) \sim \Theta(\delta)\delta^{3/2}$, which gives the noninteracting real-space correlator $\chi_0(r) \sim 1/r^3$ with the usual $2k_F$ Friedel oscillations. Since $\Sigma(k)$ is continuous but has discontinuous derivative at $k_F$ the corrections from $\delta G$ are subleading, scaling like $\chi_{\Sigma} \sim \frac{1}{N_f}\Theta(\delta)\delta^{5/2}$, which contributes a $1/r^4$ term to $\chi(r)$. Lastly, the vertex correction gives $\chi_V \sim \frac{1}{N_f}\Theta(\delta) \delta^{3/2}$, renormalizing the amplitude of the leading $1/r^3$ power law but not giving any anomalous dimension. Thus the fermion bilinear scaling remains $\chi(r) \sim 1/r^3$ at long distances, and the IR scaling remains that of a Fermi liquid. 

\end{document}